\documentclass[preprint]{aastex63}

\hypersetup{linkcolor=red,citecolor=green,filecolor=cyan,urlcolor=magenta}
\usepackage[normalem]{ulem}
\usepackage{here}
\usepackage{color} 
\usepackage{multirow}
\renewcommand{\textbf}[1]{{#1}}
\newcommand{\jm}[1]{{}}
\newcommand{\jmold}[1]{{}}

\newcommand{\UVC}{\textit{u-v}~}

\newcommand{\SGR}{Sgr~A$^{*}$~}

\newcommand{\DEG}{^{\circ}}

\newcommand{\etal}{et al.}

\newcommand{\RS}{$R_{\rm S}$}

\newcommand{\FTMUAS}{$\sim 40~\mu \rm as$}
\DeclareRobustCommand{\erase}{\bgroup\markoverwith{\textcolor{red}{\rule[.5ex]{2pt}{0.4pt}}}\ULon}
\def\jm#1{{\bf[#1 -- JM]}}

\accepted{2022/05/05}
\submitjournal{ApJ}
\shorttitle{Jet and the SMBH of M\,87 observed with EHT}
\shortauthors{Miyoshi et al.}
\begin{document}
\title{The jet and resolved features of the central supermassive black hole \\ of M\,87 observed with EHT}
\correspondingauthor{Makoto Miyoshi}
\email{makoto.miyoshi@nao.ac.jp}
\author[0000-0002-6272-507X]{Makoto Miyoshi}
\affil{National Astronomical Observatory, Japan, 2-21-1, Osawa, Mitaka, Tokyo, Japan, 181-8588}
\author[0000-0003-2349-9003]{Yoshiaki Kato}
\affil{Computational Astrophysics Laboratory RIKEN, 2-1 Hirosawa, Wako, Saitama, 351-0198, Japan, e-mail: yoshiaki.kato@riken.jp}
\author[0000-0002-0411-4297]{Junichiro Makino}
\affil{Department of Planetology, Kobe University, 1-1 Rokkodaicho, Nada-ku, Kobe, Hyogo 650-0013, Japan, e-mail: makino@mail.jmlab.jp}
\nocollaboration{3}
\begin{abstract}
We report our independent image reconstruction of the M\,87 from the public data of the Event Horizon Telescope Collaborators (EHTC). Our result is different from the image published by the EHTC. Our analysis shows that (a) the structure at 230~GHz is consistent with those of lower frequency VLBI observations, 
(b) the jet structure is evident at 230~GHz extending from the core to a few mas, though the intensity rapidly decreases along the axis, and (c) the “unresolved core” is resolved into bright three features presumably showing an initial jet with a wide opening angle of $\sim 70\DEG$.

The ring-like structures of the EHTC can be created not only from the public data, but also from the simulated data of a point image. Also, the rings are very sensitive to the FOV size. The \UVC coverage of EHT lack $\sim40~\mu\rm as$ fringe spacings. Combining with a very narrow FOV, it created the $\sim40~\mu\rm as$ ring structure. We conclude that the absence of the jet and the presence of the ring in the EHTC result are both artifacts owing to the narrow FOV setting and the \UVC data sampling bias effect of the EHT array. Because the EHTC’s simulations only take into account the reproduction of the input image models, and not those of the input noise models, their optimal parameters can enhance the effects of sampling bias and produce artifacts such as the $\sim40~\mu\rm as$ ring structure, rather than reproducing the correct image.

\end{abstract}

\keywords{jets, accretion disks --- black hole physics --- galaxies: active --- galaxies: individual (M\, 87) --- interferometer: VLBI --- data calibrations --- data sampling bias}

\section{Introduction\label{Sec:Intro}}
~Supermassive black holes (SMBHs) at the centers of galaxies often have spectacular jets sharply collimated and extended to intergalactic scale. 
However, the mechanism of the generation of such jets by the black holes has been an enigma for over a century~\citep{Blandford2019}.

The SMBH of the elliptical galaxy M\,87, the first object of the astrophysical jet discovery~\citep{Curtis1918}, is the best place to study the origin of the jet because it has the largest apparent angular size for black holes with strong jets, due to the relatively small distance (16.7~Mpc;~\cite{Mei07}) and large mass 
($6.1 \pm~0.4 \times 10^9~M_{\odot}$;~\cite{G2011}), which implies that $1~R_{\rm S}$ = 7$~\mu\rm as$. 
The black hole with the largest apparent angular size, \SGR is present in our galaxy, but unfortunately, it has no jet and its activity is very low in comparison to that of a typical AGN. In addition, it is difficult to obtain high-resolution images of \SGR owing to its rapid time variability during VLBI observations~\citep{Miyoshi2019, Iwata2020}.

Observations of the core and jet of M\,87 have been performed in multiple wavelengths, from X-ray to radio
\citep{Biretta95, SBM1996, BSM99, Perlman1999, Perlman2001, Marshall2002, Wilson2002, MOJAVE, PW2005, Harris06, Madrid2007, WZ2009}.
Also, with high spatial resolution observations using VLBI of the SMBH of M\,87 have been performed in multiple frequencies up to 86~GHz~
\citep{Reid1989, Junor1999, LHE2003, Ly2004, CHS07, Kov2007, Ly2007, Walker2008, Hada2011, HE2011, AN2012, Giroletti2012, Hada2013, Nakamura-Asada2013, Asada2014, Hada2016, MLWH2016, Walker2016, Britzen2017, Hada2017,Kim2018}.
Using the "core shift" technique, the distance between the brightness peak of the core and the actual location of the SMBH has been estimated to be from 14 to 23~\RS~\citep{Hada2011}.
Observations with higher spatial resolution at 230~GHz should allow further exploration of the core and jet. 
Pioneering observations of EHT
\footnote{\url{https://eventhorizontelescope.org/}} 
were started at 2008~\citep{Doeleman2008}. 
\\

 In 2017, EHT attained sufficient sensitivity by including phased-ALMA in the array and equipping all stations with 32~Gbps recording systems. 
The EHTC reported their findings of a ring-shaped black hole shadow from the observational data~\citep{RefEHT1-6}. The ring diameter was approximately $42~\mu\rm as$, which is consistent with that expected from the measured mass of M\,87 SMBH ($6 \times 10^{9}~M_{\odot}$) using stellar dynamics~\citep{G2011}
\footnote{The M\,87 black hole mass is still controversial. 
A mass of $M_\mathrm{BH} = (3.5^{+0.9}_{-0.7})  \times 10^9\ M_\odot$ (68~$\%$ confidence) is obtained from gas dynamics~\citep{WBHS2013}.}.

We found three problems in the EHTC imaging results. 
First, although the EHT's intrinsic FOV (Field Of View) is large enough to cover both the core and the jet structure together, no jet structure has been reported by the EHTC. The M\,87 jet is powerful and has been detected in lower frequency VLBI observations.

There was no detailed description of the investigation of the jet structure in~\citep{RefEHT1-6}; in 2017, the EHT array achieved unprecedented sensitivity, so it is not surprising that many AGN experts have strong expectations for detecting new jet structures of M\,87.

Second, the ring diameter of the EHTC imaging
($d = 42\pm~3~\mu\rm as$;~\cite{EHTC1}) coincides with the separation between the main beam and the first sidelobe in the dirty beam (identical to point spread function (PSF)) of the EHT \UVC coverage for the M\,87 observations. 
In the EHTC paper, there is no description of the concrete structure of the dirty beam, such as sidelobes. Misidentification of sidelobes as real images is a common occurrence in radio interferometer observations with a small number of stations such as the EHT array. The EHTC do not seem to take such a risk into account (at least it is not clearly mentioned in their paper).
There is a possibility that the EHTC ring is a mixture of the real image and the residual sidelobes in the diffraction patterns.

 The last problem is the brightness temperature of the ring reported by the EHTC ($T_{\rm b} =6 \times 10^{9}~K$ at most ~from Figure 3 in \cite{EHTC1}
 
\footnote{
 The EHTC show several different "fiducial" images in their papers. We believe that the images shown in Figure 3 of \cite{EHTC1} are the FINAL "fiducial" images of the EHTC because \cite{EHTC1} is for reporting the scientific results about "the shadow of the supermassive black hole".}
 ), which is significantly lower than that of their previous M\,87 observations ($T_{\rm b}$~from~1.23~to~1.42 $\times 10^{10}$~K;~\cite{Akiyama2015}) despite having higher spatial resolutions.
 \footnote{
 The possibility of time variation in brightness temperature cannot be ruled out. The number of measurements is extremely small, and future observations are desirable.
 }
The 86~GHz Very Long Baseline Array (VLBA;~\cite{VLBA1993}) observations have shown that the core brightness temperature is $T_{\rm b} = 1.8 \times 10^{10}$~K~\citep{Hada2016}. 
\cite{Kim2018} also reported the brightness temperature is
$T_{\rm b} \sim (1 - 3) \times 10^{10}$~K at 86~GHz.
The spatial resolutions of both observations are lower than that of EHT ($\theta_{BEAM} > 100~\mu \rm as$), but they show higher brightness temperatures.
In any case, it is quite rare to observe a brightness temperature of less than $10^{10}~K$ for the M\ 87 core by VLBI.

In observations of very compact objects, if the spatial resolution is low, the measured brightness temperature could be underestimated because the solid angle of the emission region tends to be estimated larger than the actual size.
If the spatial resolution is higher, the measured brightness temperature can be expected to be higher because the solid angle of the emission region can be more accurately identified.
 The measured brightness temperature increases until the spatial resolution becomes sufficient to determine the fine structure of the compact object.
However, the measured brightness temperature may surely decrease once sufficient spatial resolution is achieved and the fine structure is recognized.
The EHTC observations show a ring diameter of about $40~\mu\rm as$, almost the same as the estimated source size in ~\cite{Akiyama2015}.
However, since it is a ring structure, the center of the image is darker, so assuming that the flux density is the same
\footnote{
The EHTC papers do not show the flux density of the ring image.}
, the highest-brightness part in the ring image should show a higher brightness temperature than that indicated by ~\cite{Akiyama2015}.
The lower brightness temperatures and/or flux densities in the images
obtained by the EHTC could be the result of the insufficient recovery of
the data coherence by improper calibrations.

~Because of these three problems we decided to reanalyze the data released by the EHTC
\footnote
{First M\,87 EHT Results: Calibrated Data\\ \url{https://eventhorizontelescope.org/for-astronomers/data}
\\
\url{http://datacommons.cyverse.org/browse/iplant/home/shared/commons_repo/curated/the EHTC_FirstM87Results_Apr2019}
~DOI:10.25739/g85n-f134
}.
Using the public data released by the EHTC, we succeeded in reconstructing the core and jet structure in M\,87. 

We have resolved the region containing the SMBH in M\,87 for the first time and found the structure of the core and knot separated by 
$\sim33~\mu\rm as$ ($550~\rm au$~or 4.7~\RS)
on the sky, which shows time variation.
This could be the scene of the initial ejection of the jet from the core.
We also found a feature to the west, 
$\sim83~\mu \rm as$
 away from the core.
These facts are important for identifying the jet formation mechanism from SMBHs. We need further observations to determine the nature of the features.

We also found emissions along the axis of the jet up to a point a few mas from the core, showing that the edges of the jet are brighter, similar to what was observed at low frequencies.

We first describe the observational data released by the EHTC in Section~\ref{Sec:obs+data}, our data calibration and imaging process in Section~\ref{Sec:ourreduction}, and our imaging results in Section~\ref{Sec:results}. 
Then, we investigate how the EHTC ring was created in Section~\ref{Sec:the EHTC-ring}.
In Appendix~\ref{Sec:UVsim}, we show that the EHT array cannot detect any feature whose size is larger than $30~\mu\rm as$.
As a supplement to Section~\ref{Sec:PSF},
Appendix~\ref{Sec:otherDbeams} shows the dirty beam (PSF) shapes of the EHT array for the M\,87 observations in two different types: natural weighting and uniform weighting. Both show the substructure with a scale of $\sim 40~\mu\rm as$.
In Appendix~\ref{Sec:MOD}, we show that the missing spatial Fourier components of \FTMUAS ~also affect the structure in our CLEAN map.

\section{Observational data\label{Sec:obs+data}}
 The observational data were recorded on 5, 6, 10, and 11 April 2017. 
 The EHT array consists of seven submillimeter radio telescopes located at five places across the globe, yielding a baseline length over 10000~km~\citep{RefEHT1-6}. 
For the observational details and the instruments, refer to the series of the EHTC papers~\citep{EHTC1,EHTC2,EHTC3,EHTC4,EHTC5,EHTC6}.  
The raw data archives have not been released by the EHTC yet, but they
released the calibrated visibility data with their recipe of the data
reduction procedure. We first analyzed the released EHT data sets of M\,87 using the standard VLBI data calibration procedure and imaging methods without referring to their data procedure.
 The data are time-averaged into 10~sec bins and are stored into 2 IF channels. 
 ~According to the header of the public FITS data the IF bandwidth is 1.856~GHz in each IF.
 Because of the removal of data of the strong calibrator source (3C 279), we could not perform the fringe search to correct the errors of station positions, clock parameters, and the receiving band-path calibration by ourselves. Therefore, our independent calibration was limited to the self-calibration method.\\

We checked the details of the data and noticed that the visibility values of the RR-channel and LL-channel are exactly the same. 
The headers of the EHTC open FITS data files contain two data columns labeled RR and LL, respectively; the FITS format data does indeed contain data columns labeled RR and LL.
We checked all the original public FITS data sets (there are 8 sets) and confirmed that the data in the RR and LL columns are the same in all the data sets; there are a total of 51119 pairs of RR and LL, and all the pairs have exactly the same real, imaginary, and weight values.

\begin{figure}[H]
\begin{center} 
\epsscale{0.5}
\plotone{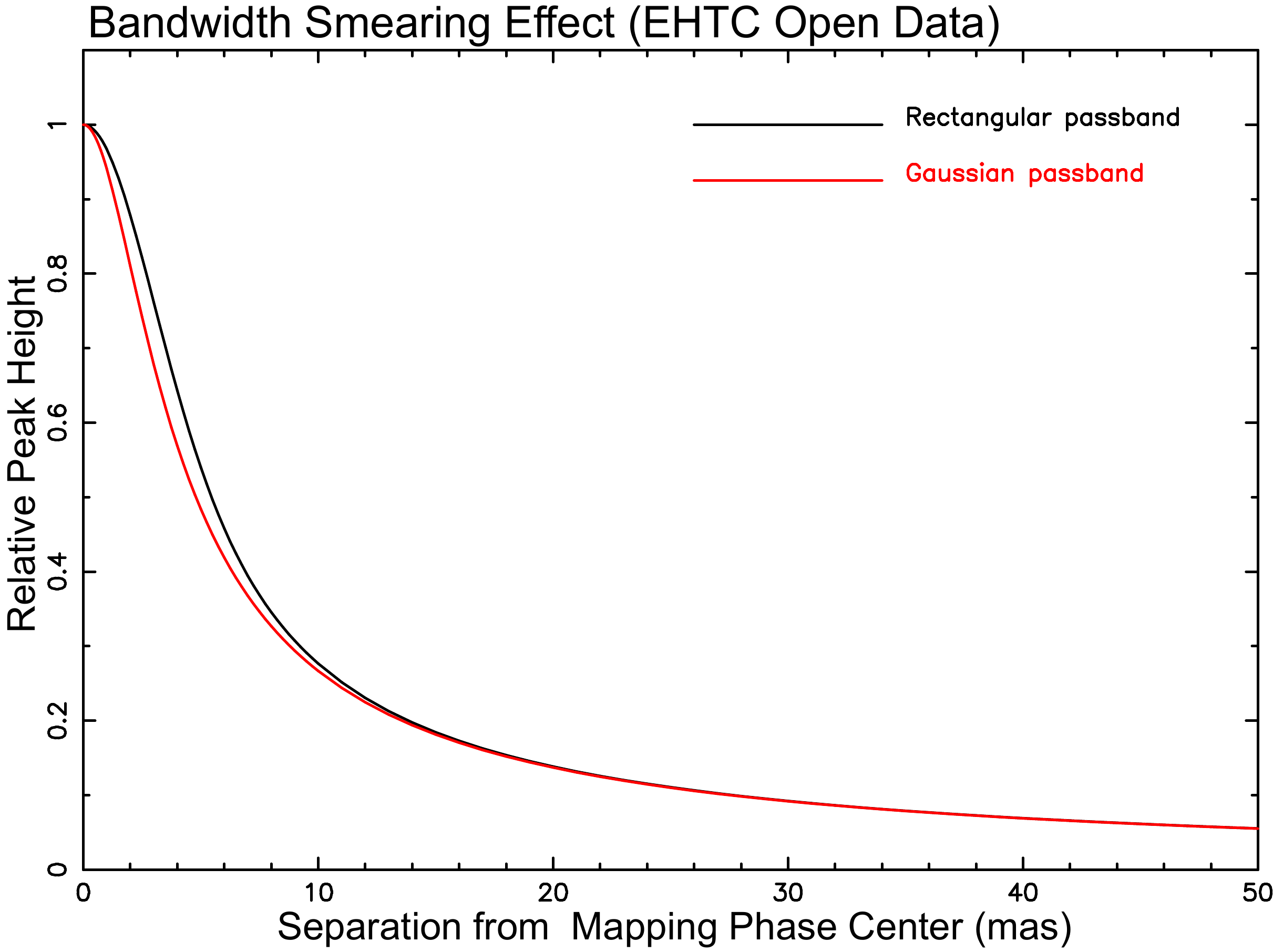}
\end{center} 
\caption{
The bandwidth smearing effect calculated explicitly for the  EHTC open data by following equations 6-75 and 6-76 in~\cite{TMS}.
Adapted synthesized beam size is $\theta_{b}=21.06~\mu as$, which is the geometric mean of major and minor axes of the beam shapes of the four observing days shown in Table 1 of~\cite{EHTC4}.
We substituted $\Delta\nu~=1.856~GHz$ for the bandwidth and $\nu_{0}=~229.071~GHz$ for the observing frequency.
}
\label{Fig:bws}
\end{figure}

We found in a document of the EHTC the following description:
\footnote{
README.md in \url{https://github.com/eventhorizontelescope/2019-D01-01}}
\begin{quotation}
The data are time averaged over 10 seconds and frequency averaged over all 32
intermediate frequencies (IFs). All polarization information is explicitly
removed. To make the resulting `uvfits` files compatible with popular
very-long-baseline interferometry (VLBI) software packages, the circularly
polarized cross-hand visibilities `RL` and `LR` are set to zero along with
their errors, while parallel-hands `RR` and `LL` are both set to an estimated
Stokes *I* value. Measurement errors for `RR` and `LL` are each set to sqrt(2)
times the statistical errors for Stokes *I*.   
\end{quotation}
In other words, the open data in the EHTC FITS format are not the visibility of either polarization, but the Stokes I, $V_{ij, I}=(V_{ij, RR}+V_{ij, LL})/2$,~\citep{EHTC3}, and the above-calculated values are stored in the columns of RR and LL.
This information is not included in the attached tables or files of the FITS data.
For this, the EHTC should have used the FITS format for intensity data instead of using that for dual polarization data.
Also, it means that the corrections made between the correlator output and the open data cannot be independently verified. 
\\
~EHTC's open data integrates the wide frequency band of 1.86~GHz into a single channel. Such wideband integration is extremely rare and unsuitable for public data because it results in loss of information over a wide field in the data due to the bandwidth smearing effect.
~The effect is similar to the peripheral light fall-off of optical camera lenses. Visibility data integrated in the frequency direction reduces the sensitivity in peripheral vision. This phenomenon occurs because originally independent (u, v) points are integrated in frequency domain. 
The further away from the center of the field of view (phase center), become larger the size of the PSF and lower the peak; the detection sensitivity in the peripheral vision becomes worse~\citep{TMS,Bridle89,Bridle99}.\\
Due to the bandwidth smearing effect, the peak of the PFS away from the phase center is suffered attenuation as shown in Figure~\ref{Fig:bws}.
In the case of the EHTC open data,
the ratios of  peak heights relative to that at the phase center are
$\sim~50~\%$ at a radius of 5~mas, and 
$\sim~27~\%$ at a radius of 10~mas.
Even at a radius of 20~mas from the center, the ratio is $\sim~14~ \%$.
If a component of sufficient intensity is present at an even far position from the center, it will be detected. We did not abandon such a possibility and set a wide field for imaging as explained in Section~\ref{Sec:ourreduction}.
\\

The coherence time of the obtained data has a significant impact on the data analysis and imaging results; the EHTC shows the atmospheric coherence time for all observations in the 2017 campaign~\citep{EHTC2}, but not for those limited to the M\,87 observations only.
 Therefore, we used the AIPS task COHER to check the coherence time of the visibility data.
Here, the coherence time is defined as the time when the amplitude becomes 
$1/e \sim 0.36$ by vector averaging.
The task COHER cannot identify the reasons for the coherence loss. In any case, the calculated coherence time implies the total amount of coherence loss that the data has suffered.
The coherence time $T_{cor} = 0.45 \pm 0.7 \rm ~min$ was obtained from the entire data set (average of all baselines).
However, the coherence time was not constant; the data from the first two days showed $T_{cor} = 0.54 \pm  0.91 \rm ~min$ and the data from the last two days showed $T_{cor} = 0.35 \pm 0.36 \rm ~min$. 
We took it as significant that $39\%$ of the total data showed 
$T_{cor} \sim 0.167\rm ~min~(\sim 10~sec)$.
Without any kind of calibrations, we do not expect to improve SNR by long time integration. We decided that no meaningful solution could be obtained by increasing the integration time (SOLINT, solution interval) in self-calibration. Therefore, we always set SOLINT to $0.15~min$ when performing self-calibration.

We used both data channels in their original form.
We found that our calibrations of the EHT data sets can be significantly improved and also obtained an improved solution for calibrations using the hybrid mapping method~\citep{RefHM1, RefHM2, RefHM3}. 
The observations were performed over four days. We succeeded in increasing the sensitivity by integrating  ~two days' data ~or all of them.
%

\section{Our data calibration and Imaging\label{Sec:ourreduction}}
In this section, we report on the procedures and results of data calibration and imaging using standard methods of VLBI data analysis for sources with unknown structures. 
In Section~\ref{Sec:hmp} we describe the hybrid mapping procedures used in this study. 
Section~\ref{Sec:1stp} describes how we identified the second feature from the first map, 
and 
Section~\ref{Sec:fin-stp} describes the process that followed. 
In Section~\ref{Sec:foi} , we present our final images. 
In Section~\ref{Sec:SN138AP}, we present a solution for self-calibration of both amplitude and phase, using the final image as a model to determine the quality of the EHT public data.
\subsection{Hybrid mapping process\label{Sec:hmp}}
In the analysis of VLBI data, the hybrid mapping method is widely used to obtain a calibration solution for the data and to reconstruct the brightness distribution.  Hybrid mapping, which consists of repeatedly assuming one image model, performing self-calibration, obtaining a trial solution for calibration, and improving the image model for the next self-calibration, is the only method that is essential for precise calibration of VLBI data~\citep{RefHM1, RefHM2, RefHM3}. VLBI systems are not so stable in phase and amplitude as connected radio interferometers. In addition, millimeter- and submillimeter-wave observations are more affected by atmospheric variations. Therefore, the hybrid mapping method is becoming more and more important in the calibration of high frequency VLBI data such as the EHT observations.
We performed a standard hybrid mapping process using the tasks CALIB and IMAGR in AIPS (the NRAO Astronomical Image Processing System
\footnote{\url{http://www.aips.nrao.edu/index.shtm}},~\cite{Greisen2003}).

\subsection{The first step in hybrid mapping process\label{Sec:1stp}}
\subsubsection{Solutions of self-calibration using a point source model\label{Sec:1st-solution}}
As a first step in this process, a single point source (located at the origin) was used as the first image model to obtain a solution for the visibility phase calibration from the self-calibration. The parameters used for the task CALIB are listed in Table~\ref{Tab:1}. 

As mentioned in Section ~\ref{Sec:obs+data}, the coherence time of EHT public data is very short. The solution interval (SOLINT) was set to 0.15 minutes. We set the $SNR~cutoff= 3$ for safety. This SNR cutoff value is larger than what many researchers use in the end. Solutions that did not meet the criteria (SNR cutoff) were flagged and abandoned. 

\begin{figure}[H]
\begin{center} 
\includegraphics[width=20cm,height=22cm]{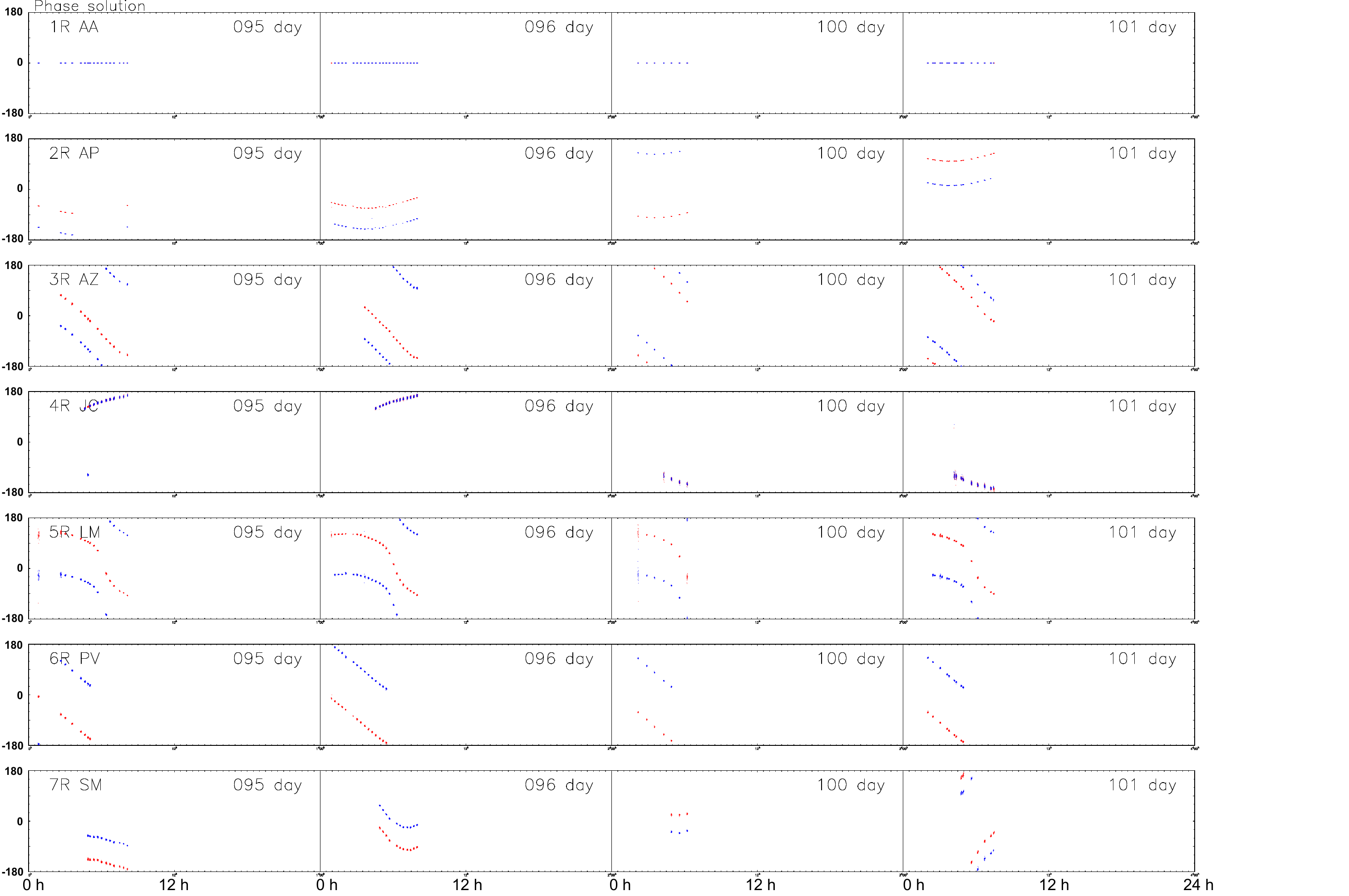}
\end{center} 
\vspace{-0.5cm}
\caption{Initial phase (only) solutions obtained by self-calibration using one-point model. 
The red dots are the solutions for IF 1 data, and the blue dots are those for IF 2 data. 
The solution for L (left-handed circular) polarization is not plotted here; the visibility data for LL is exactly the same as for RR, so the solution for L is the same as the solution for R in the Figure.
\label{Fig:mA}}
\end{figure}

Figure~\ref{Fig:mA} shows the phase solution for the first step. Because phase is a relative quantity, the phase of the ALMA station (AA) is used here as a reference. For all stations, the non-zero and time-varying phase values were calculated by self-calibration.  
The four stations, APEX (AP), SMT (AZ), LMT (LM), and Pico Veleta (PV), always show the same respective trends over the four days of observations, suggesting that errors in station positions remain (a sinusoidal curve with a period of one sidereal day is observed at all stations when there is an error in the position of the observed object). 

Another feature is the phase difference that occurs between IF1 and IF2, which is almost fixed for all stations except JCMT (JC) and AA (phase reference station), respectively.

If the EHT public data were sufficiently calibrated, the above two phenomena should not appear. 
In conclusion, the "calibrated" data published by the EHTC is not yet sufficiently calibrated. In order to obtain reliable images, the EHT's public data needs to be further calibrated.

%
\begin{table}[H]
\begin{center}
\begin{tabular}{lr} \hline
Parameters        & \\ \hline \hline
 SOLTYPE &'L1'\\
 SOLMODE &'P' (phase only) \\
 SMODEL &1,0 (1~Jy single point) \\
 REFANT &1 (ALMA)\\
 SOLINT (solution interval) &0.15 (min)\\
 APARM(1)&1\\
 APARM(7) (SNR cut off) &3\\ \hline
\end{tabular}
\end{center}
\caption{Parameters of CALIB for the first self-calibration.} \label{Tab:1}
\end{table}

%
\subsubsection{The first CLEAN map\label{Sec:ByFCLNMap}}
\begin{figure}[H]
\begin{center} 
\epsscale{1.2}
\plotone{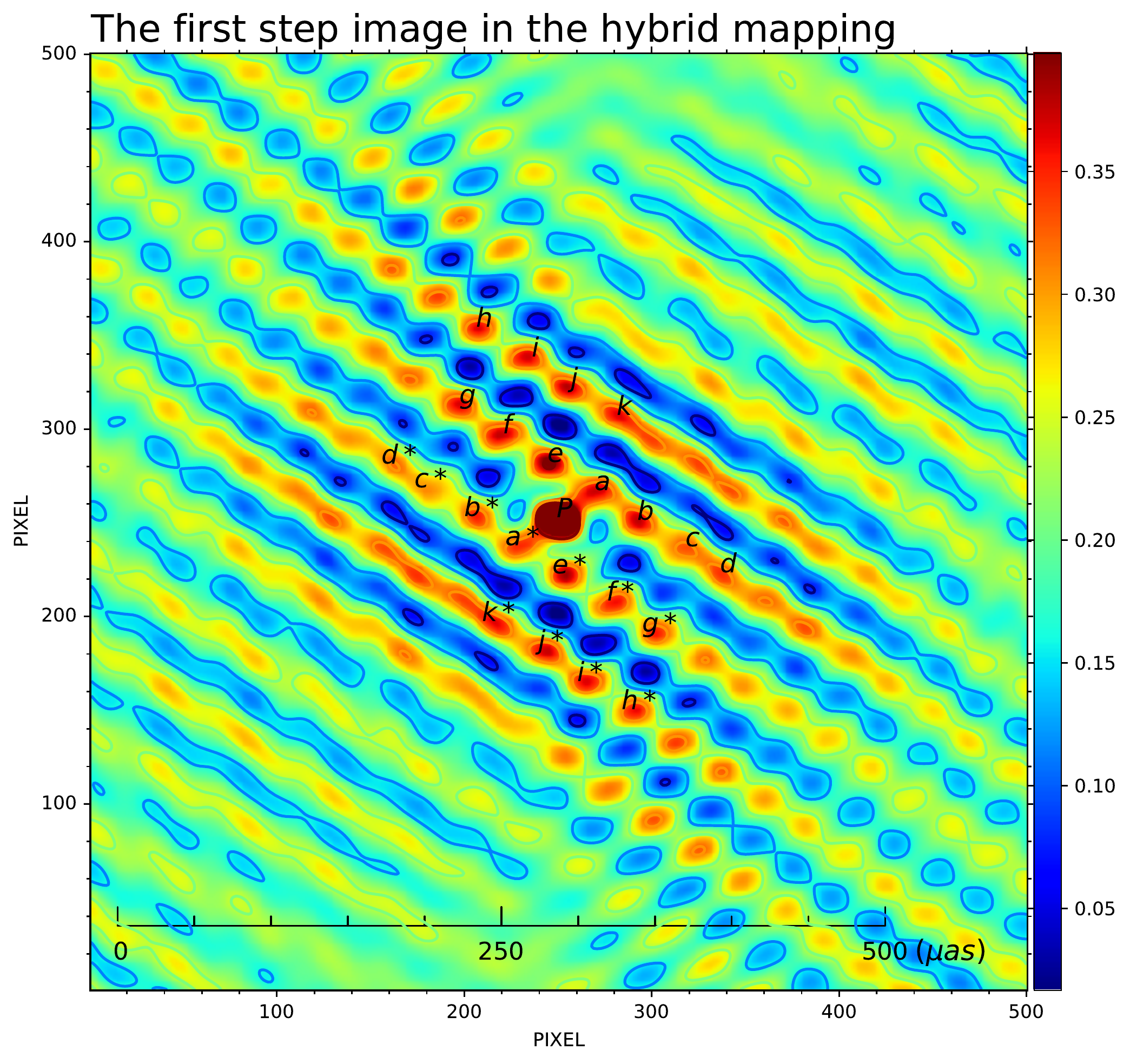}
\end{center} 
\caption{First step image of our hybrid mapping process. 
The restored beam size is about $20~\mu\rm as$, and the center $600~\mu\rm as$ square of the map is magnified. Contour lines are drawn at every 20$\%$ level of the peak brightness of $3.981~\times10^{-1}$ (arbitrary units are used in the figure, not Jy/beam). One pixel corresponds to $1.22011~\mu\rm as$.
}
\label{Fig:mB}
\end{figure}

Figure~\ref{Fig:mB} shows the CLEAN~\citep{Clark1980,Hogbom1974} image obtained from the data after applying the first phase calibration solution shown in Figure ~\ref{Fig:mA}. The parameters of the imaging of IMAGR are shown in Table~\ref{Tab:2}.
(In all figures showing the imaging results, the x-axis indicates relative Right Ascension and the y-axis relative Declination.)  
The purpose of this imaging is to find the second brightest component following the central brightest peak. 
As will be explained in Section~\ref{Sec:uvs}, despite the fact that the EHTC has involved the largest number of stations ever, the \UVC coverage of the EHT array is formed by only 7 stations, or actually 5 stations if we exclude the very short baselines. The synthesized beam (dirty beam) is not as sharply shaped as the dirty beams of multi-element interferometers such as ALMA and VLA. 
It is not easy to find the complex brightness distribution of the observed sources from a tentative map composed of such a scattered dirty beam (PSF). Therefore, we performed CLEAN, specified by the parameters shown in Table~\ref{Tab:2}. 
This method is effective when the structure of the observed source is not point symmetric. We set the loop gain (GAIN) to 1.0 and extracted all of the brightest peak from the dirty map in the first CLEAN subtraction. Next, this component was replaced by a sharp Gaussian restoring beam and combined with the brightness distribution of the remaining dirty map. The image in Figure~\ref{Fig:mB} was created in this way.

This has the effect of removing the bright but scattered PSF shape of the brightest point that dominates the dirty map, and clarifying the presence of the second brightest component in the image. Note that if the data is not properly calibrated and the actual PSF corresponding to a point source differs from the theoretically calculated PSF shape, the brightness distribution caused by the brightest point may remain in the afterimage. However, such remaining brightness distribution also shows a point-symmetric structure with respect to the location of the brightest point (practically the same location as the center of the map), and does not contribute to the asymmetric structure of the image. Therefore, if there is an asymmetric structure in the image, it is not related to the brightest component, but is due to another bright point source. So, by searching for the asymmetric structure, we can find the second component of the observed source.

This image (Figure~\ref{Fig:mB} shows its central $600~\mu as$ square) has a nearly point-symmetric structure with respect to the center of the map. 
The overall feature is a series of multiple ridges in the $PA = 55 \DEG$ direction. This structure is due to the non-uniformity of the \UVC  coverage. 
In addition to the central P, there are several other bright features. 
The peak brightness of these features is shown in Table~\ref{Tab:SymP}.
Features a, b, c, d, e, f, g, h, i, j, and k have corresponding features located at their symmetry points 
(denoted as a$^{*}$, b$^{*}$, c$^{*}$, d$^{*}$, e$^{*}$, f$^{*}$, g$^{*}$, h$^{*}$, i$^{*}$, j$^{*}$, and k$^{*}$).  
Curiously, the features located in the upper right from the center are always brighter than the symmetric features located in the lower left, i.e., the brightness ratio is greater than one.  
This may indicate the existence of a large-scale asymmetric brightness distribution in the observed object, extending from the center to the upper right. This is roughly consistent with the M\,87 jet propagation direction $PA = -72~\DEG$ \citep{Walker2018}. 
Examining the brightness ratio of each pair, we find that the pair c~\&~c$^{*}$ ($ratio = 1.146$) is the largest, 
followed by the pair a~\&~a$^{*}$ ($ratio = 1.122$).
Regarding the absolute value of brightness, the brightness of a ($2.396 \times 10 ^ {-2}$ Jy/Beam) is larger than that of c ($1.982 \times 10 ^ {-2} $ Jy/Beam). 
In addition, there is a ridge extending from the central bright point (P) toward feature a. The direction of this ridge ($PA = -45~\DEG$) is completely different from the direction of multiple ridges seen in the entire image, and no other ridge shows the same direction. Based on these characteristics, we decided to continue the hybrid mapping by adopting two points for the next image model. In other words, we chose to use a model with a point at each of the two locations, the center and the location of feature a.

\begin{table}
\begin{center}
\begin{tabular}{lr} \hline
Parameters & \\ \hline \hline
DOCALIB & 2 \\
CELLSIZE& $1.22011\times 10^{-6}$,~$1.22011\times 10^{-6}$ (asec)\\
FLDSIZE & 8192, 8192 (pix)\\
ROBUST& 0\\
NITER & 1\\
GAIN & 1.0\\
\hline
\end{tabular}
\end{center}
\caption{Parameters of IMAGR for the first trial imaging.}\label{Tab:2}
\end{table}
%

\small{
\begin{table}[H]
\begin{center}
\begin{tabular}{cc|cc|cc} \hline
Name&Brightness (Jy/beam)&Name&Brightness (Jy/beam)&Ratio&Order\\ \hline \hline
P &$8.736\times 10^{-2}$&  &  &  &  \\ 
a &$2.396\times 10^{-2}$& a$^{*}$&$2.136\times 10^{-2}$& 1.122  & 2 \\
b &$2.438\times 10^{-2}$& b$^{*}$&$2.187\times 10^{-2}$& 1.115  & 3 \\
c &$1.982\times 10^{-2}$& c$^{*}$&$1.729\times 10^{-2}$& 1.146  & 1 \\
d &$2.099\times 10^{-2}$& d$^{*}$&$1.889\times 10^{-2}$& 1.111  & 4 \\
e &$2.780\times 10^{-2}$& e$^{*}$&$2.534\times 10^{-2}$& 1.097  & 5 \\
f &$2.420\times 10^{-2}$& f$^{*}$&$2.257\times 10^{-2}$& 1.072  & 7 \\
g &$2.368\times 10^{-2}$& g$^{*}$&$2.205\times 10^{-2}$& 1.074  & 6 \\ 
h &$2.334\times 10^{-2}$& h$^{*}$&$2.282\times 10^{-2}$& 1.023  & 9 \\
i &$2.364\times 10^{-2}$& i$^{*}$&$2.337\times 10^{-2}$& 1.012  &10 \\
j &$2.475\times 10^{-2}$& j$^{*}$&$2.404\times 10^{-2}$& 1.030  & 8 \\
k &$2.328\times 10^{-2}$& k$^{*}$&$2.319\times 10^{-2}$& 1.004  &11 \\
\hline
\end{tabular}
\end{center}
\caption{
Peak brightness of the features that appeared in the first step image. The 11 features near the center are shown. P is the peak in the center that was replaced by the restoring beam. Features other than P are as shown in the brightness distribution of the residual dirty map.
}\label{Tab:SymP}
\end{table}
}
%
\subsection{Our hybrid mapping process\label{Sec:fin-stp}}
After obtaining the image models for the two points, more than 100 iterations, including trials and errors, were performed in the hybrid mapping process. Most of the CLEAN images were run with the parameters listed in Table~\ref{T:3}.

As described in~\cite{EHTC4}, care must be taken in the choice of FOV, as incorrect restrictions will result in incorrect image structures. Considering the well-known structure of M\,87, we restricted the imaging region by eight BOXes where emission could be detected. 
For the self-calibration, the selected CLEAN components were used as the next imaging model and the parameters in Table~\ref{Tab:1} were used. By repeating the phase-only self-calibration in this way, we were able to find better images and calibration solutions.
This is because the method of simultaneously solving the amplitude solution with hybrid mapping has the risk of going in the wrong direction. Self-calibration of phase-only can be done with a certain degree of safety. Even if the wrong model is used and a completely wrong solution is obtained, the closed phase is automatically preserved and convergence to a completely wrong image can usually be avoided.
On the other hand, the amplitude solution from self-calibration can be infinitely large or small by using the wrong image model. It is safe to try the self-calibration of the amplitude solution after a confident correct image model is obtained from the phase self-calibration.
%

%
%
\begin{table}[H]
\begin{center}
\begin{tabular}{lr} \hline
Parameters & \\ \hline \hline
ANTENNAS & 0 (all) \\
DOCALIB & 2 \\
CELLSIZE& $1.5\times~10^{-6}$, $1.5\times~10^{-6}$ (asec)\\
UVRANGE& 0, 0 (no limit)\\
FLDSIZE & 16384, 16384 (pix)\\
ROBUST& 0\\
NITER & 40000\\
GAIN &  0.05 or~0.005\\
FLUX &-1.0 (Jy)\\ 
BMAJ, BMIN & 0.00002, 0.0002, and 0 (asec)\\
BPA & 0 ($\DEG$) \\
RASHIFT & $-9.25\times~10^{-3}$ (asec)\\
DECSHIFT& $8.15\times~10^{-3}$ (asec)\\
NBOXES & 8\\
 BOX(1) & -1, 912, 1329, 2185 (pix)\\
 BOX(2) & -1, 820, 2049, 2729 (pix)\\
 BOX(3) & -1, 912, 3105, 3129 (pix)\\
 BOX(4) & -1, 1184, 4481, 3689 (pix)\\
 BOX(5) & -1, 1792, 6321, 4473 (pix)\\
 BOX(6) & -1, 2448, 8753, 5673 (pix)\\
 BOX(7) & -1, 2848, 11537, 7001 (pix)\\
 BOX(8) & -1, 2800, 13377, 7833 (pix)\\ \hline
\end{tabular}
\end{center}
\caption{Typical parameters of IMAGR for our hybrid mapping process.
The imaging area is $24.576~\rm mas$ square, but it is limited by the BOX setting. "$Flux = -1.0$" means that the terminating condition of CLEAN subtraction is the first occurrence of a negative maximum value.
}
\label{T:3}
\end{table}
\subsection{Our final image\label{Sec:foi}}
\begin{figure}[H]
\begin{center} 
\epsscale{1.1}
\plottwo
{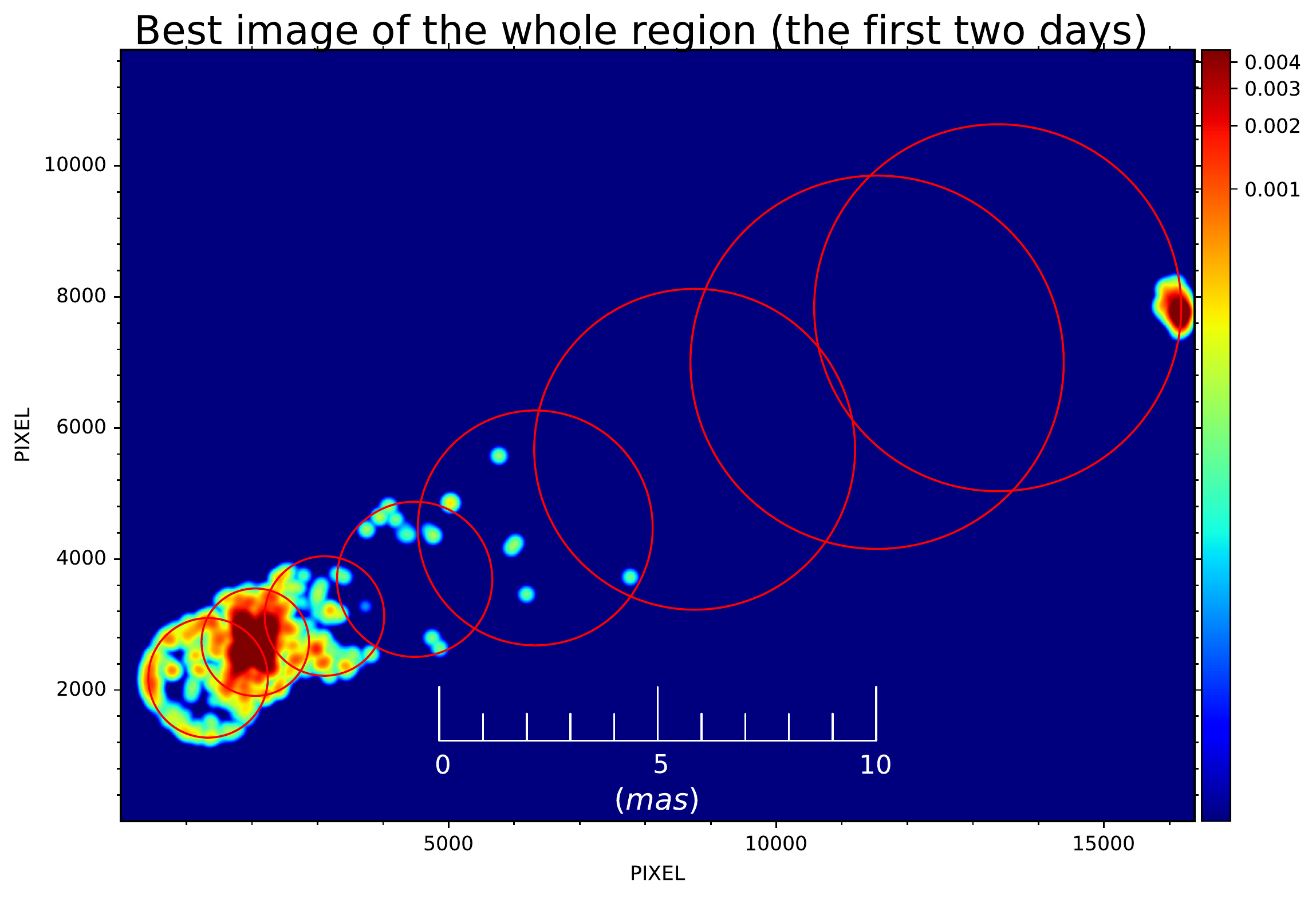}
{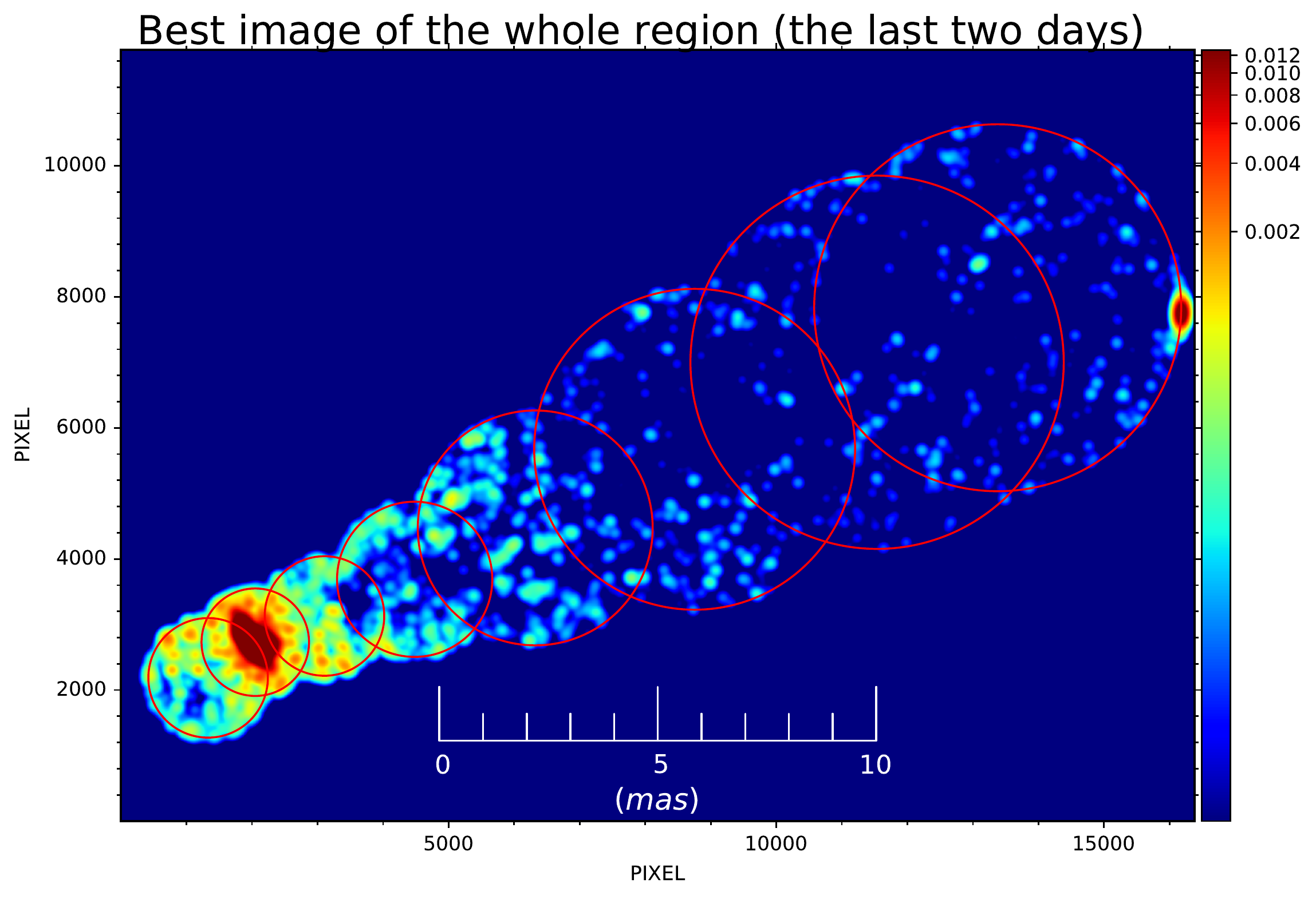}
\end{center} 
\vspace{-4mm}
\caption{
The best images obtained from the analysis. 
The left panel shows the best image for the data of the first two days. 
The right panel shows the best image for the data of the last two days. 
In both images, the grouped-CLEAN components are convolved with a circular Gaussian restoring beam of HPBW $ = 200~\mu\rm as$.  
A logarithmic pseudo-color is used to express the large differences in brightness distribution (arbitrary unit).
The eight red circles indicate the BOX area used.
The bright spot seen on the right (west) edge is not real brightness distributions (see the content in this Section~\ref{Sec:foi}).
}
\label{Fig:sA2}
\end{figure}

In the latter half of the imaging process, we selected several candidates for the final image by comparing the difference in closure phase between the image and the real data. Furthermore, we created several image models using the CLEAN components of the candidate images to find the best image with the minimum closure residuals. 
Since we found that the source structure has time variation, we divided the data into the data of the first two days and the data of the last two days. As a result, we obtained the best images as shown in Figure~\ref{Fig:sA2}. 
The data from the first two days was a composite of 
seven~raw CLEAN maps, and the data from the last two days was a composite of nine raw CLEAN maps. Both consist of CLEAN components only. In the images of the first two days, adjacent CLEAN components within $2~\mu \rm as$ are merged into one component. In the images of the last two days, adjacent CLEAN components within $5~\mu \rm as$ are merged into one component.

Since the eight BOXes cover a large area, the resulting CLEAN component contains three types of emission: real emission, associated diffraction (sidelobes), and false acquisitions. For example, the bright emission on the right edge is not real. Such unreal bright spots often appear when the VLBI data is analyzed and imaged. When such unreal brightness appears, there may actually be strong emissions outside the BOXes. We produced a large image (30 arcsec field of view) using very short baseline (SM-JC and AA-AP) visibility data, but could not detect any new strong features.

 To get a more complete image, we need to select only the real ones from these CLEAN components and perform CLEAN imaging again with each narrow BOX to cover the selected ones.  However, since our data did not have enough \UVC coverage and quality to select the correct ones among the CLEAN components, we gave up the task of extracting the CLEAN components this time.
 Nevertheless, the quality of the final image does not seem to be too bad. The closure residuals of the resulting image show a small variance comparable to the EHTC ring image. 
The residual of our map for the first two days of data is 
$-4.90~\pm~37.93\DEG$
(the residual of the EHTC ring is $-0.73~\pm~45.33\DEG$) and the residual of our map for the last two days of data is 
$1.79~\pm~42.11\DEG$
(the residual of the EHTC ring is $4.22~\pm~45.74\DEG$). Here, we integrated the 5-minute data points and calculated the closure phase of all triangles.
For more information on closure phase residuals, see Section~\ref{Sec:SCimage-vis}.
 The EHTC ring images for comparison were generated using the EHTC-DIFMAP pipeline.
It should be emphasized that the final image clearly contains an unrealistic CLEAN component, but still shows the same level of closure phase residuals as the image of the EHTC ring.

We also used this final image model to attempt self-calibration of the amplitude and phase for the solution, performed CLEAN, and obtained new images. However, the residuals of the closure phase in the new image were not improved. Therefore, we terminated the hybrid mapping without amplitude calibration. The images shown in Figure~\ref{Fig:sA2} are our best final images. 
The two upper panels in Figures~\ref{Fig:sA3}
and the panels in Figure~\ref{Fig:afmas} are partial extracts from the final images. 
Figure~\ref{fig:Fig-a} shows the full image of the last two days of data.

\subsection{Solutions of self-calibration both amplitude and phase using a final image model\label{Sec:SN138AP}}
\begin{figure}[H]
\begin{center} 
\includegraphics[width=20cm,height=22cm]{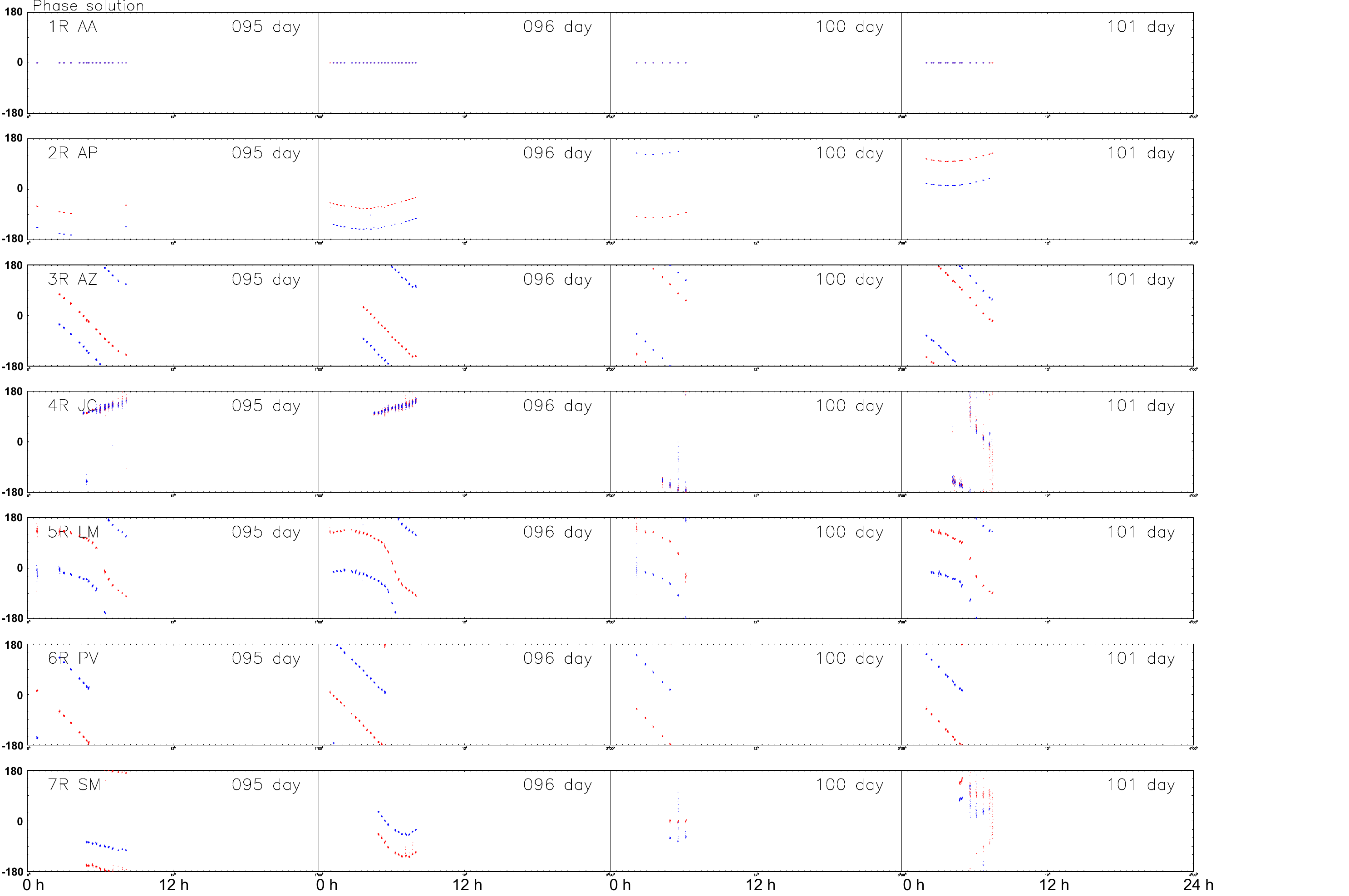}
\end{center} 
\vspace{-0.5cm}
\caption{Phase solutions obtained from self-calibration with A\&P mode using CALIB in AIPS. As the image model, all of the grouped CLEAN components of the last two days' image were used. The red points show the solutions for IF 1 data, the blue ones show those for IF 2 data.
\label{Fig:mD}}
\end{figure}

\begin{figure}[H]
\begin{center} 
\includegraphics[width=20cm,height=22cm]{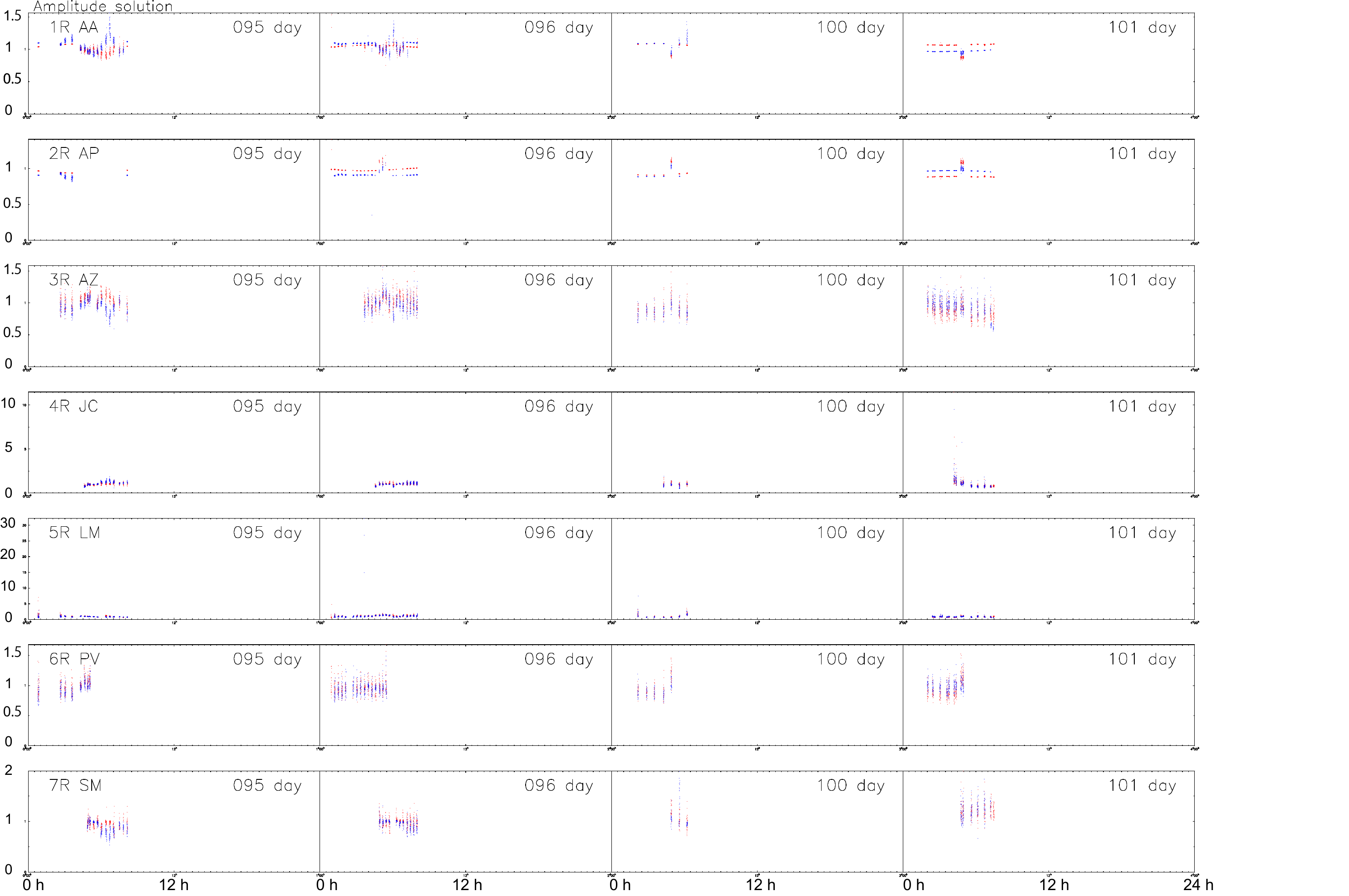}
\end{center} 
\vspace{-0.5cm}
\caption{Amplitude solutions obtained from self-calibration with A\&P mode using CALIB in AIPS. As the image model, all of the grouped CLEAN components of the last two days' image were used. The red points show the solutions for IF 1 data, the blue ones show those for IF 2 data.
\label{Fig:mE}}
\end{figure}

Here we show solutions of the self-calibration for both amplitude and phase using one of the final images, although we did not apply it to the data calibration. The reason we dare to show the unused solutions here is that we believe this study provides insight into the quality of EHT  public data and the reliability of our images. 
We performed our self-calibration using CALIB in the AIPS task in "A\&P" mode with the image (CLEAN components). Other parameters of CALIB are the same as in the first self-calibration (Table~\ref{Tab:1}). Figures~\ref{Fig:mD} and \ref{Fig:mE} show the self-calibration solutions (the total amount of solutions to be applied for data calibration). 

Figure~\ref{Fig:mD} shows the phase solution. Compared to the initial phase solution shown in Figure~\ref{Fig:mA}, there is no significant change in the overall perspective. 
This can be attributed to the fact that the brightness distribution of the observed source is concentrated in the center and does not deviate significantly from the self-calibration solution assuming a single point source.
However, there are offsets in the phase changes of JC, LM, PV, and SM stations. In addition, the phases of the two stations in Hawaii, JC and SM, show a larger phase dispersion than the solution of self-calibration assuming a point source on the 100 and 101 days' data. 
Although small in comparison, the phases of JC and SM on 95 and 96 days' data also show phase scatters on the same hours.

 The amplitude solution is shown in Figure~\ref{Fig:mE}.
In general, errors in amplitude are due to noise in the atmosphere and in the receiving system.
In addition, changes in aperture efficiency depending on the elevation angle of antenna often cause systematic errors in amplitude.
These effects can be measured by an auxiliary method.
%

For large aperture antennas, gain loss due to offset tracking of the target source from the narrow main beam angle may occur, which is difficult to calibrate.

Furthermore, coherency (phase stability) loss is observed due to the
variations in each station clock and atmospheric variations, which is more difficult to measure correctly than other error factors.


Amplitude solutions for AA, AP, PV and SM stations are within $50\%$ fluctuation ($\sim 1.0 \pm 0.5$). Such values are often found in amplitude solutions of most of the self-calibrations of VLBI data. 
On the other hand, JC and LM stations occasionally show large amplitude solutions reaching 10 and 30, respectively.
 The JCMT station shows a large amplitude value at $T~\sim 4.25~$hour on the last observation day (101 day).
On the other hand, for the LMT station, the amplitude is large for several times as follows.\\
~(a) $T \sim 1~hour$ and $T \sim 2.6~hour$  on the first observing day (095 day)\\
~(b) $T \sim 1~hour$  on the second observing day (096 day)\\
~(c) $T \sim 2.25~hour$ and $T \sim 6.25~hour$  on the 3rd observing day (100 day).\\
The EHTC did not show the overall calibrations to be applied, but noticed the sudden large-amplitude errors at the LMT station (Figure 21 in \cite{EHTC4}). 
\\
%

 These large amplitude solutions may have implied that the resultant image is significantly wrong. 
For comparisons, we examined the solutions of self-calibration in the case of the EHTC ring image. 
Consequently, we found that the self-calibration solutions by the EHTC ring image also demonstrate large amplitudes occasionally, similar to those of our image (Section~ \ref{Sec:EHTSNPLT}). 
Therefore, if such large amplitudes found in self-calibration solutions are negative signs against the resultant image quality, 
the results obtained by both the EHTC and our work should be rejected. 
The EHTC is concerned and considering the fact that some stations require large-amplitude corrections during data analysis.
EHTC then analyzed the data from 3C~279, which was observed with  M\,87, and obtained consistent imaging results from all imaging methods. At the same time, the EHTC found that the amplitude correction was consistent with that obtained with M\,87 (\cite{EHTC4}).
The amplitude corrections they found are also consistent with those we showed above. In other words, it is natural to consider that such a large amount of amplitude variation actually occurred.
To add, the fact that the EHTC obtained a non-ring structure from the 3C~279 data and that the amount of error corrections the EHTC obtained at that time were consistent with those obtained from the M\,87 data does not mean that the ring image of the EHTC is the correct image of M\,87.
The large amplitude solutions from the self-calibrations indicate that the "calibrated" data released by the EHTC are not of high quality with respect to the amplitude. 

%
%
\section{Imaging results\label{Sec:results}}
~In this section, we describe the brightness distribution obtained in our final image models.
Unlike the EHTC result, we could not detect any ring structure but
found that the emissions at 230~GHz come not only from the narrow central region less than $128~\mu\rm as$ in diameter (the EHTC's FOV), but also from the jet region. 
We found a core-knot structure at the center and weak spot-like features along the M\,87 jet stream though the reliability must be discussed.

In Section~\ref{Sec:resultsCC}, we show the structure of the central region.
In Section~\ref{Sec:Jet}, features seeming to belong to jet are presented.
In Section \ref{Sec:Reliability}, we investigate the reliability of our final image
from three 
points of views: the attainable sensitivity (Section~\ref{Sec:Sensitivity}), the robustness of the main features (Section~\ref{Sec:ROFI}),
and the self-consistency of our imaging (Section~\ref{Sec:SCimage-vis}), where we compare with those of the EHTC.
%
\subsection{The core\label{Sec:resultsCC}}
\begin{figure}
\begin{center} 
\epsscale{0.96}
\plotone{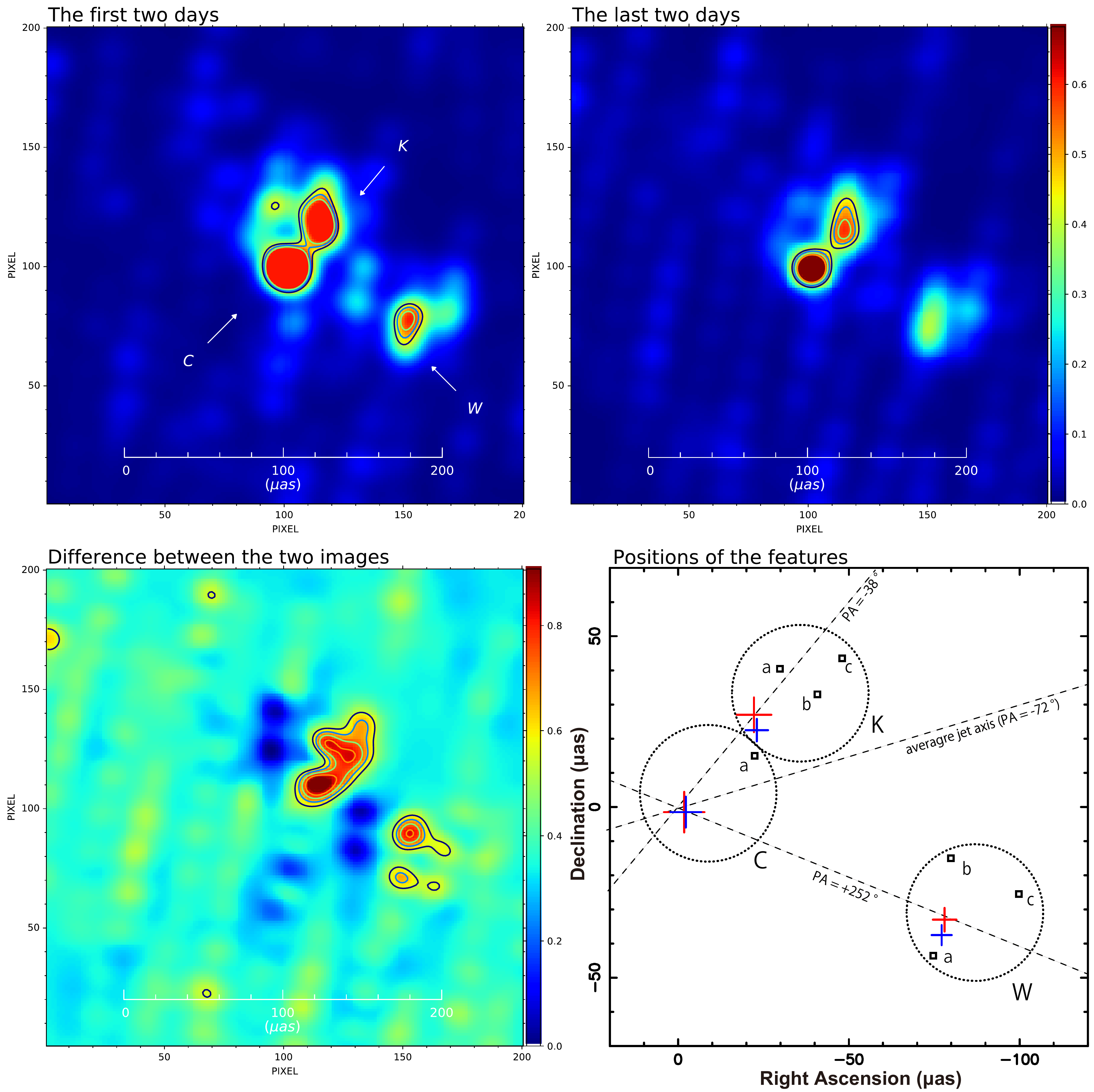}
\end{center} 
\vspace{-4mm}
\caption{
Images of the central core region: 
The top left panel is the image obtained from the data of the first two days.
The upper right panel is that obtained from the data of the last two days.
In both images, the CLEAN components are grouped and convolved with the restoring beam of a circular Gaussian with HPBW of $20~\mu \rm as$.
The brightness distribution is shown in pseudo-color (arbitrary unit).
As shown in the upper left panel, three features, C(core), K(knot), and W(west component), were detected.
The lower left panel shows the difference between the last image and the first image, i.e., the time variation of the brightness distribution during 5 days.
The lower right panel shows the positions of the C, K, and W peaks.
The red crosses are for the first two days, and the blue crosses are for the last two days.
The size of the crosses is ten times the size of the error bars in position.
The small squares also indicate the location of the large increase in intensity.
\label{Fig:sA3}}
\end{figure}
%
\small{
\begin{table}[H]
\begin{tabular}{rrrrr}
\hline \hline
                             &\multicolumn{2}{c}{\textbf{the first two days'}} & \multicolumn{2}{c}{\textbf{the last two days'}} \\ \cline{2-5}
\textbf{Peak position}&\multicolumn{1}{c}{R. A. ($\mu \rm as$)}& \multicolumn{1}{c}{$\delta$ ($\mu \rm as$)} & \multicolumn{1}{c}{R. A. ($\mu \rm as$)}& \multicolumn{1}{c}{$\delta$ ($\mu \rm as$)}\\ \hline
Core & \multicolumn{1}{r}{$~-1.8\pm0.6$} & $~-1.5\pm0.6$& $~-2.3\pm0.5$& $~-1.5\pm0.5$\\
Knot & \multicolumn{1}{r}{$-22.2\pm0.5$} & $~27.0\pm0.5$& $-23.1\pm0.3$& $~22.5\pm0.3$\\
West & \multicolumn{1}{r}{$-78.1\pm0.3$} & $-33.0\pm0.3$& $-77.2\pm0.3$& $-37.5\pm0.3$\\ \hline

\multicolumn{3}{r}{}     & \multicolumn{1}{c}{$\Delta$ R. A. ($\mu \rm as$)} & \multicolumn{1}{c}{$\Delta\delta$ ($\mu \rm as$)}\\ \hline
Core & \multirow{3}{*}{}& \multicolumn{1}{l}{\multirow{3}{*}{}}        & $-0.5$ & $~~0.0$\\ 
Knot &                      & \multicolumn{1}{l}{    }                 & $-0.9$ & $~-4.5$\\ 
West &                      & \multicolumn{1}{l}{    }                 & $ 0.9$ & $~-4.5$\\ \hline \hline

\textbf{Position of intensity increase} & \multicolumn{2}{r}{} & \multicolumn{1}{r}{R. A. ($\mu \rm as$)} & \multicolumn{1}{r}{$\delta$ ($\mu \rm as$)} \\ \hline
Core~area~a&\multicolumn{2}{l}{}  & $-22.4\pm0.4$ & $~15.0\pm0.4$\\ 
Knot~area~a&\multicolumn{2}{l}{}  & $-29.8\pm0.7$ & $~40.5\pm0.7$\\ 
~~~~~~~~~~~b&\multicolumn{2}{l}{} & $-40.8\pm0.3$ & $~33.0\pm0.3$\\ 
~~~~~~~~~~~c&\multicolumn{2}{l}{} & $-48.0\pm0.4$ & $~43.5\pm0.4$\\ 
West area~a&\multicolumn{2}{l}{}  & $-74.7\pm0.7$ & $-43.5\pm0.7$\\ 
~~~~~~~~~~~b&\multicolumn{2}{l}{} & $-79.9\pm0.5$ & $-15.0\pm0.5$\\ 
~~~~~~~~~~~c&\multicolumn{2}{l}{} & $-99.8\pm0.7$ & $-25.5\pm0.7$\\ \hline \hline

\textbf{Integrated Intensity} & \multicolumn{2}{r}{(mJy)}& \multicolumn{2}{r}{(mJy)} \\ \hline
Core & \multicolumn{2}{r}{$55.6\pm5.2$} & \multicolumn{2}{r}{$66.1\pm4.7$} \\
Knot & \multicolumn{2}{r}{$33.5\pm2.7$} & \multicolumn{2}{r}{$44.9\pm2.3$} \\
West & \multicolumn{2}{r}{$22.5\pm1.2$} & \multicolumn{2}{r}{$30.2\pm1.4$} \\ \hline \hline

\textbf{Brightness temperature}& \multicolumn{2}{r}{(K)}& \multicolumn{2}{r}{(K)} \\ \hline
Core & \multicolumn{2}{r}{$>1.0\times10^{10}$}& \multicolumn{2}{r}{$>1.2\times10^{10}$}\\
Knot & \multicolumn{2}{r}{$>6.0\times10^{~9}$}& \multicolumn{2}{r}{$>8.1\times10^{~9}$}\\
West & \multicolumn{2}{r}{$>4.0\times10^{~9}$}& \multicolumn{2}{r}{$>5.4\times10^{~9}$}\\ \hline \hline

\textbf{Grouped CLEAN components} & \multicolumn{2}{r}{(mJy)}& \multicolumn{2}{r}{(mJy)}\\ \hline
$F_{GCC}> 0.1 mJy$ & \multicolumn{2}{r}{$707.4~(n=1151)$} & \multicolumn{2}{r}{$1032.6~(n=1657)$}\\
All                & \multicolumn{2}{r}{$767.8~(n=2824)$} & \multicolumn{2}{r}{$1154.6~(n=7844)$}\\  \hline \hline
\end{tabular}
\caption{
Properties of main features: positional offsets from the map phase center in $\mu \rm as$, flux densities in~mJy, and minimum brightness temperatures in Kelvin calculated with the assumption that the emission area is $15~\mu \rm as$ in diameter.
Positions of intensity increase in features are also shown. 
At the bottom, the sum-up intensities and the numbers of the grouped CLEAN components in the whole images are shown.
}
\label{Tab:Fs}
\end{table}
}
%

In the central core region, we could not find the ring structure reported by the EHTC, but found a core-knot structure.
Figure~\ref{Fig:sA3} shows the images of the central region ($300~\mu \rm as$ square).
As noted in Section \ref{Sec:foi}, since the data calibration is not yet complete, 
our final images show the sidelobe structure around actual features. This is a common phenomenon in synthesis imaging with radio interferometers with only a few stations.
 The images in Figure~\ref{Fig:sA3} show that ``the unresolved VLBI core" 
  in M \,87  
 has finally resolved into substructures. 

 The high spatial resolution of the EHT array clearly shows the presence of two bright peaks, i.e., the core and knot structure.
The core is indicated by ''C" and the knot by ''K" in the upper left panel of Figure~\ref{Fig:sA3}.
 In addition, we found a feature, ''W", located west about 
$83~\mu\rm as$ ~apart from the core C. 
 The flux densities from the obtained CLEAN components are 
 $F_{C}\sim 60~mJy$ for the core (C), 
 $F_{K}\sim 40~mJy$ for the knot (K), and 
 $F_{W}\sim 25~mJy$ for the west feature (W).
In this observation, the solid angle of features was not so clear.
Here, we assume that the solid angle of the emission is $15~\mu\rm as$ in diameter, and calculate the brightness temperatures (lower limit).
The average brightness of feature C is $T_{\rm b} =1.1~\times 10^{10}$~K.
Feature K has a brightness of 
$T_{\rm b} =7.1~\times 10^{9}$~K.
~Feature W is $T_{\rm b} =4.7~\times 10^{9}$~K.
Thus, we have detected central features with brightness temperatures higher than the EHTC ring (up to $T_{\rm b} \sim 6\times 10^{9}$~K).
The solid angle assumed here is the maximum size of a single, smoothed object that the EHT array can detect in the 230~GHz (Appendix \ref{Sec:UVsim}).
Therefore, the actual brightness temperature is likely to be much higher. If the solid angle of the emission is $5 ~\mu \rm as$ in diameter, the brightness temperature of core C reaches $10^{11}~K$.
If this is the case, the brightness temperature is an order of magnitude higher than the previous measurement cases~\citep{Kim2018, Hada2016, Akiyama2015}. This is mainly because the size of the emitting region has been identified as smaller due to the higher spatial resolution.
%

 High brightness temperatures were often detected from some active galactic nuclei~\citep{Horiu2004, Homan2011}, and can be explained by the Doppler boosting effect of relativistic motion of jet approaching toward us. Previous observations found no high velocity movement in the M\,87 central core.
 Therefore, the brightness temperatures are not due to such Doppler boosting effects. 
If they actually reflect the physical temperatures, they can be explained easily by the simple RIAF (Radiatively Inefficient Accretion Flow) disks~\citep{KFM2008, Nakamura1997}.
Our observational results are consistent with those of previous studies, supporting the existence of the RIAF disk in the M\,87 core~\citep{DiMatteo2003}.\\

There is a clear difference between the two images observed over the five days.
According to \cite{EHTC3}, they found a change in the closure phase between data sets from the first two days and the last two days. 
In other words, there was a clear time variation. 
However, the EHTC could not clearly identify from the structure of that ring where that change occurred~\citep{EHTC4}. 
We identified the change in the closure phase as due to a change in the 
core knot structure (features C and K).
In particular, the change in the position of feature K was also seen in the trial images during the hybrid mapping process.\\

Assuming that features are single components, we fitted a Gaussian brightness distribution to each feature and measured the central position and displacement over five days.
Relative to the position of feature C, the change in position of feature K is 
$\Delta\alpha =-0.4~\mu \rm as$,~$\Delta\delta =-4.5~\mu \rm as$ in 5~days, and the proper motion is 
$0.33~\rm mas/yr~(v \sim 0.1~c)$.
~Feature K appears to be approaching feature C as if showing an inflow motion. 
However, if we look at the differences in the brightness distributions shown in the lower left of the Figure~\ref{Fig:sA3}, we can see that the changes in the brightness distribution of feature K occur in three places, all~at the north end of feature K. 
In the latter measurement of the position of feature K, feature K appears to be moving south because the brightness distribution of feature C affects the measurement; K is moving north on the line of $PA = -38 \DEG$ as a whole. The position of feature C has hardly changed, but the location of "a" where the intensity increased is at the northwest side of feature C.
In other words, the structure of features C and K and their time variation  can be interpreted as an outflow emanating in the direction of $PA = -38\DEG$ from the origin.
%
There has never been a measurement of the motion of a knot so close to the central core. 
\\

In comparison, it is difficult to interpret what feature W is. 
Three hypotheses are presented below.
\begin{enumerate}
\item
\textbf{Gravitational lensing image.}
Feature W is morphologically similar to feature K, and the pattern of brightness variation is also similar.
This can be attributed to the formation of the gravitational lens image of feature K due to the strong gravitational field of the SMBH in M\,87. 
Assuming that the position of SMBH is approximately equal to the position of feature C (the distance between the core of M\,87 and SMBH is $41~\mu \rm as$~or 6~\RS;~\cite{Hada2011}), feature W would be located (or at least projected) at 12~\RS~from the SMBH.
There is a possibility that the radio waves emitting from feature K to the far side, orbiting in the strong gravity field of the SMBH, and being changed the propagation direction, come towards us (Black Hole Echo;~\cite{Saida}).
If feature W is such a lensing image caused by the strong gravity field of the SMBH, it should be the image of the backside of feature K, so it is most likely a mirror image of feature K. However, the shape of feature W does not look like such an image.
Needless to say, there are many possibilities for a gravitational lensing due to a strong gravity field, so detailed calculations are required to deny it completely; however, the possibility that feature W is a gravitational lensing image is not very high.

\item
\textbf{Another central black hole.}
Feature C is the primary SMBH of M\,87, and feature W is a secondary SMBH orbiting the primary SMBH.
If there is a binary SMBH in M\,87, it can be permanently observed with the EHT array.
Based on these observations, we calculated two possible orbits.
\begin{enumerate}
\item 
The proper motion of feature W 
($\mu = 0.34~\rm mas/yr,~\rm v \sim 0.09~c$) is assumed to be due to a circular orbit motion, and its orbital radius is assumed to be the separation 
$R =83~\mu \rm as$~from feature C. 
In this case, the orbital period T is $\sim 1.5$~years.  
If the real radius is 
$R = 1.4\times~10^{3}~au$,
the mass of central object $M_{c}$ is only $1.2\times~10^{6}~M_{\odot}$.
~Since the estimated mass is too small as compared to those of the previous M\,87 studies, this assumption must be rejected.
\item
We assume that the observed proper motions and structure change of feature W are only due to changes in surrounding matters, and that the measured proper motions of feature W have nothing to do with its orbital motion.
In other words, we assume here that no change in the position of the center of gravity of feature W is observed.
Also, it is assumed that feature C has an SMBH with a mass of $6\times 10^9~M_{\odot}$ and that the orbital radius of feature W is the distance  between features C and W.
The distance between them is $83~\mu\rm as$ ($1.4\times10^{3}$ au or $11.9~R_{\rm S}$).
~It is consistent with the 86 GHz core size of $\sim 80~\mu \rm as$ at 86~GHz observed in 2014~\citep{Hada2016}, suggesting that the two features C and W are not transient.
Also sinusoidal oscillations of the position angle of the jet were observed with a period of roughly 8 to 10 years~\citep{Walker2018}. 
If the two features C, and W compose a binary of black holes, its orbital motion can cause such jet oscillations. Certainly, the apparent separation of approximately 
$1.4\times10^{3}$ ~au~is too short to explain the observed period of the jet oscillation. However, if the real distance is longer by a factor of about 
3.42,~which is the correction factor of the viewing angle of the jet axis from us ($\sim 17~\DEG$), the orbital period of the binary can be $\sim 10$ years. 
\end{enumerate}
\item
\textbf{Unstable initial knot.} Feature W is another knot moving toward a different direction. The jet of M\,87 is known to have a wide opening angle at scales well below $1\rm~mas$ \citep{Junor1999}. Furthermore, \cite{Walker2018} found evidence from 43~GHz observations that the initial opening angle $\theta_{app}$~is $\sim70~\DEG$. We found the angle $\angle KCW$ is $70~\DEG$, and further that the line of the average jet axis ($PA = -72\DEG$) divides its angle almost evenly into $34\DEG,~ 36\DEG$. Furthermore, the lines CK and CW extend in the directions of $PA = -38\DEG$ and $PA =252\DEG$, respectively. These directions are very similar to the ridges observed at 43~GHz from where \cite{Walker2018} measured the initial opening angle. We guess that not only feature K, feature W is also an initial knot that has just emerged from the core, and still, the shape is very unstable and shows changes significantly.
\end{enumerate}
Adopting the most conservative hypothesis, feature W, like feature K, can be understood to be a knot that represents the initial jet structure.\\

 As we will discuss in Section~\ref{Sec:ROFI}, the core-knot structure (features C and K) is robust in the sense that it can be obtained with different imaging parameters. On the other hand, the ~$40~\mu\rm as$ ring of the EHTC is sensitive to BOX parameters and can be easily destroyed, even if it can be created as shown in Section~\ref{Sec:staERING}. Due to the robustness of the core-knot structure, we consider it to be a real structure.
On the other hand, the feature W is sensitive to the BOX size, so its detection is not as reliable as it could be. However, the structure corresponding to feature W had already appeared in the first imaging results (Section~\ref{Sec:ByFCLNMap}). 
That is, feature c in Table~\ref{Tab:SymP} is in a similar position to feature W and also shows the largest asymmetry. Also, if we run the EHTC-DIFMAP pipeline without its BOX setting, an emission feature appears at the position close to feature W (lower right panel of Figure~\ref{fig:the EHTCDIFMAP1}). These suggest that the feature W is also a real structure.
%

\subsection{The jet\label{Sec:Jet}}
Here, we show the overall brightness distribution (Section~\ref{sec:overall}) and that of so-called the jet launching region, which is a few mas away from the core (Section~\ref{Sec:afew-mas}).
\subsubsection{The overall structure\label{sec:overall}}
\begin{figure}[H]  
\begin{center} 
\plotone{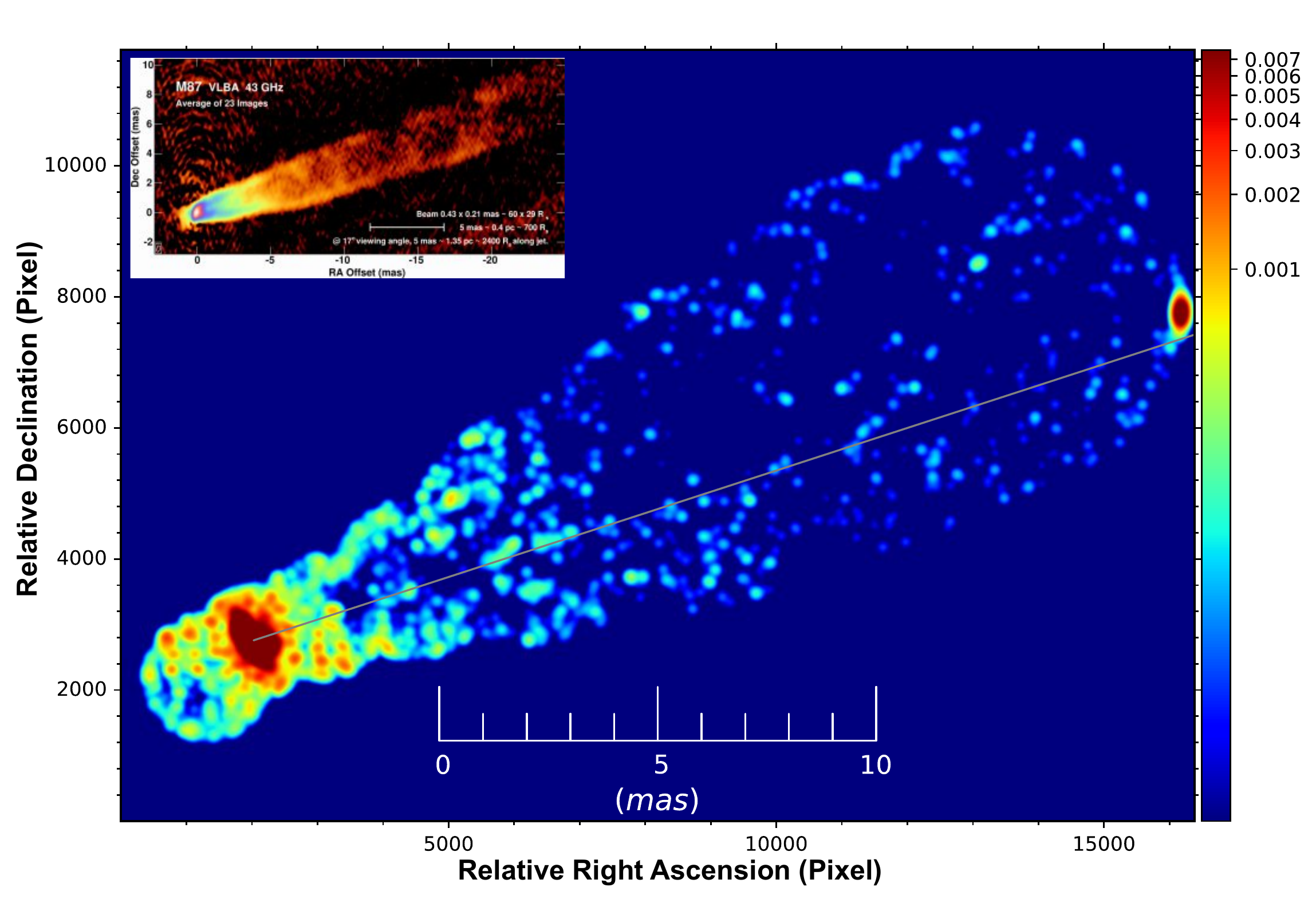}
\end{center} 
\caption{The overall structure of M\,87 we obtained (image from the data of the last two days).
The grey line shows the average direction of the jet axis ($PA = -72\DEG$ from~\cite{Walker2018}). 
The restoring beam is $200~\mu \rm as$ circular Gaussian, which is much larger than the default beam in order to make the emission obvious.
The logarithmic pseudo-color (arbitrary unit) is used to enhance the darker parts of the image.
The image consists only of the grouped CLEAN components obtained.
For comparison, the average image at 43~GHz from VLBA observations taken from Figure 1 in~\cite{Walker2018} is shown in the inset.
The bright spot seen on the right (west) edge is not real brightness distribution (see Section~\ref{Sec:foi}).
}
\label{fig:Fig-a}
\end{figure}
\begin{figure}[H]
\begin{center} 
\epsscale{0.6}
\plotone{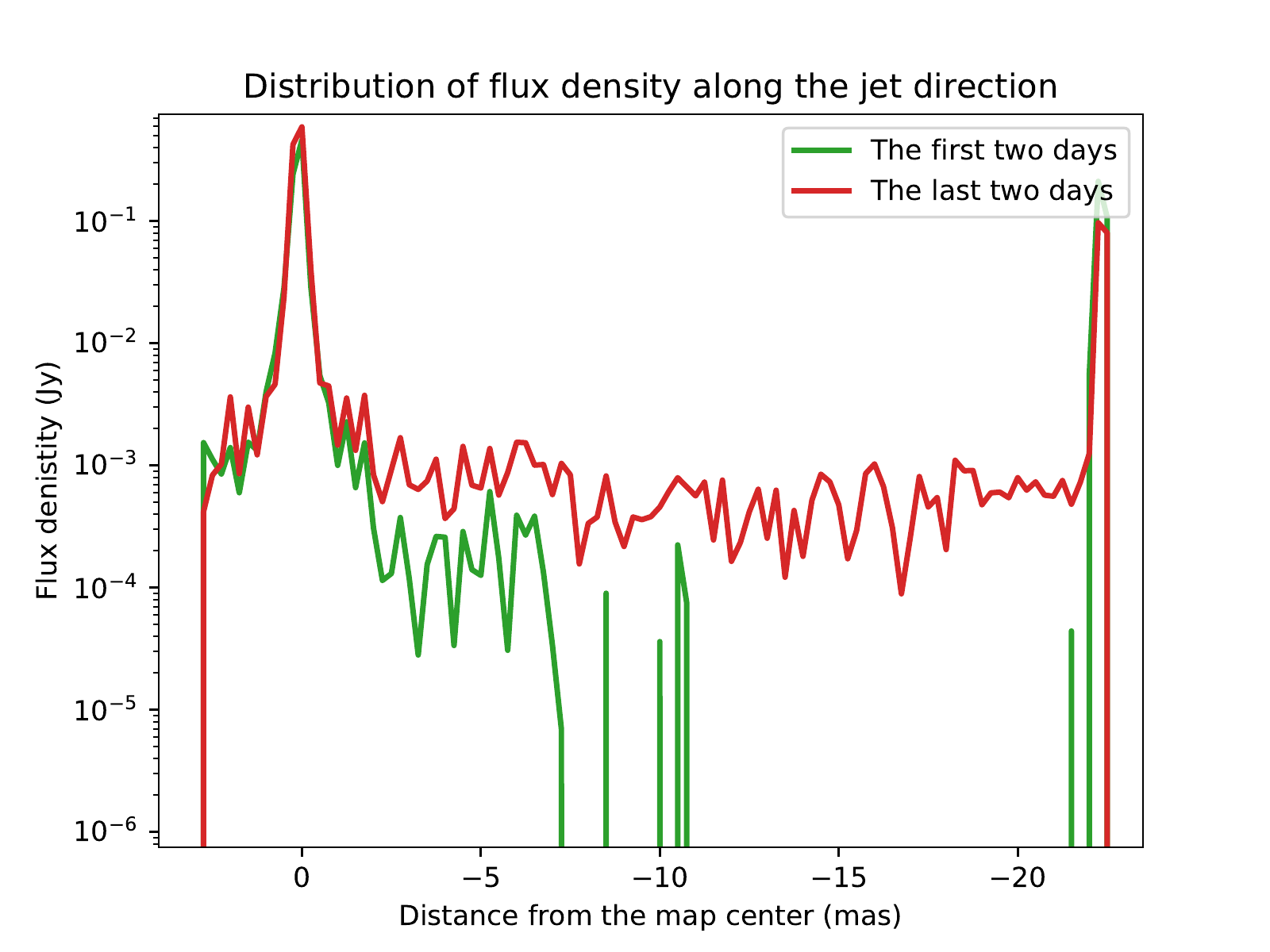}
\end{center} 
\caption{
Flux density distribution 
corrected for the band width smearing effects.
~Here we show the sum of the flux densities of the CLEAN components obtained in each small region.
The horizontal axis shows the distance along the average direction of the jet axis ($PA = -72\DEG$ from~\cite{Walker2018}) from the map center (near the peak of the core).
The binning intervals are 0.25~\rm mas.
The vertical axis shows the sum of the flux densities in~\rm Jy (logarithmic scale).
The green line shows the flux density distribution for the image from data of the first two days, and the red line shows the flux density distribution for the image from data of the last two days.
The peaks seen on the right edge are not real flux densities (see Section~\ref{Sec:foi}).
\label{Fig:A}}
\end{figure}
%
Figure \ref{fig:Fig-a} shows the overall structure of the M\,87 we obtained. In order to make the emission obvious, we used a restoring beam of $200~\mu\rm as$ circular Gaussian, 10 times larger than the default beam size.
As already mentioned, it is found that the emission at 230 GHz comes not only from the central point source, but also from other regions.

The EHTC either assumed or concluded that there is no bright source outside the narrow region ($128~\mu\rm as$ in diameter) where the ring was found.
 However, we found that the emission was not from such a narrow range, but from a wide range of more than a few milli-arcseconds (mas). This is consistent with the results of VLBI observations at 43~GHz and 86~GHz.

Our final image shows a similar structure to the average image in the 43 GHz band (the inset of Figure~\ref{fig:Fig-a}).
There are two main similarities.

First, as in the 43~GHz image, our image shows that the "jet" has an extended structure leading to the core.
Then, up to a few mas away from the core along the jet axis, both edges are bright, as in the previous observations.

Second, the brightness distribution of the jet in the 230-GHz is consistent with the trend of those obtained from lower-frequency observations.
The core is vastly brighter than the jet structure.
Within the radius of 0.25~mas ($250~\mu \rm as$) from the center, 63~\% (the first two days' image) and 75~\% (the last two days' image) of all obtained flux densities are concentrated. However, features C, K, and W (several tens~mJy at most, see Table~\ref{Tab:Fs}) do not occupy them, rather the flux densities are distributed over a wider area. In contrast, the EHTC rings have a total of about 500~mJy that is contained entirely within a diameter of only a few tens of $\mu \rm as$.
\\
The results of this observation at 230~GHz show that the brightness in the jet region is orders of magnitude lower than that in the core region. In addition, the decay of the flux density along the jet is more rapid at 230 GHz than in the lower frequency observations.
Compared to the peak luminosity of the core, the relative intensities are 
$6.6\times~10^{-2}$~at 0.25~mas from the core, 
$9.9\times~10^{-3}$~at 0.5~mas, and 
$2.3\times10^{-3}$~at 1~mas (Figure~\ref{Fig:A}).
While, in the observation at 43~GHz, the decreases of intensity are
$2.8\times~10^{-1}$ at 0.25~mas from the core peak, 
$8.5\times~10^{-2}$ at 0.5~mas, and 
$2.5\times~10^{-2}$ at 1~mas (from the upper panel of Figure 6 in \cite{Walker2018}).
At 1~mas position, the intensity of the jet structure is 2.5~$\%$ of the core peak at 43~GHz, 
however the intensity at 230 GHz is only ~0.2~$\%$ ~of the core peak,
 namely the intensity of the jet structure is greatly attenuated at 230~GHz.
However, with respect to the structure of the brightness and intensity distribution of both edges, the trend is in great agreement with the previous results of the M\,87 jet.

The total flux density measured by the EHTC \citep{EHTC3} was 1.12~and 1.18~Jy, for the first two days and the last two days, respectively.
In contrast, the total flux density of the CLEAN component in our analysis is 767.8 and  1154.6~mJy,~respectively.
That is, there are missing flux densities of 353~and ~25.4~mJy,~respectively
\footnote{The sum of the flux densities we obtained as CLEAN components is larger than that of the EHTC ring. 
We made the ring according to their open procedure and measured its flux density ($\sim500~\rm mJy$).}.

The difference between the total flux density of our image and the single-dish flux density is most likely due to the presence of extended emission somewhere, which the present EHT array cannot detect. As shown in Appendix~\ref{Sec:UVsim}, the EHT array cannot detect an smoothed emission feature (like Gaussian brightness distributions) with size larger than $30~\mu \rm as$.

\subsubsection{The Jet launching region\label{Sec:afew-mas}}
In this section, we present the structure within a few mas from the core. We have found emission belonging to the jet component that was not detected by the EHTC.

\begin{figure}[H]
\begin{center} 
\epsscale{1.15}
\plotone{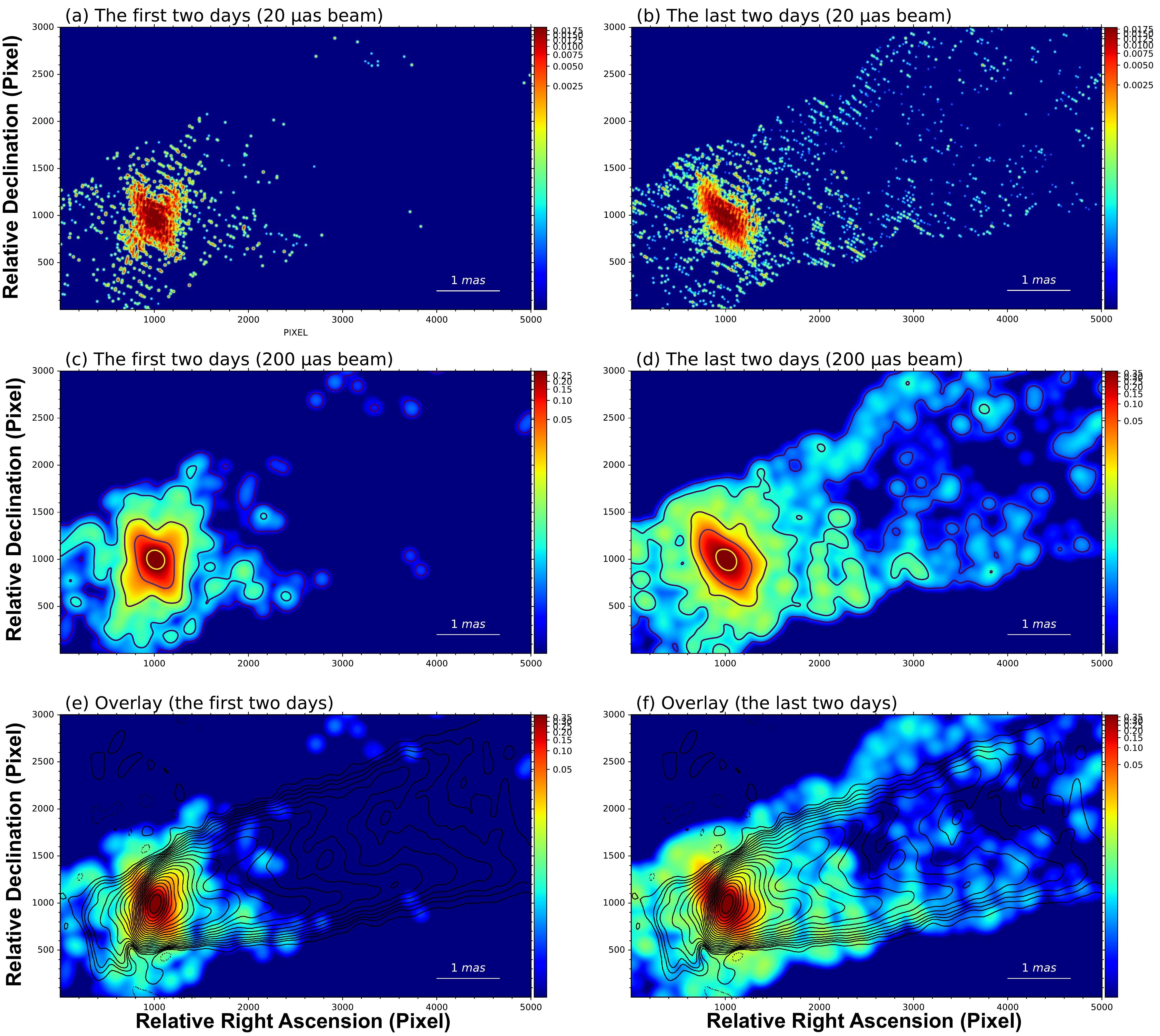}
\end{center} 
\caption{Images of the core and the jet launching region:
Panels (a), (c), and (e) on the left show images of data from the first two days.
The right panels (b), (d), and (f) show the images of data from the last two days.
The logarithmic pseudo-color (arbitrary unit) is used to represent the large differences in brightness distribution.
The upper panels (a) and (b) are composed by a circular Gaussian restoring beam with HPBW of $20~\mu \rm as$.
The middle panels (c) and (d) are composed by a circular Gaussian restoring beam with HPBW of $200~\mu \rm as$.
The levels of contour lines in the panels (c) and (d) are $10^{-5}$, $10^{-4}$, $10^{-3}$, $10^{-2}$, and $10^{-1}$ of the peak brightness.
The lower two panels, (e) and (f), show overlaid ones with the VLBA averaged image at 43~GHz.
 The contour lines show the VLBA averaged image at 43~GHz taken from Figure 3 in~\cite{Walker2018}, and the levels of contour lines are -0.3, 0.3, 0.6, 0.85, 1.2, 1.7~mJy~beam$^{-1}$. Their restoring beam shape is $0.43\times$ 0.21~mas with $PA = -16\DEG$. Peak positions of the two images are used for the alignment of the two images.
\label{Fig:afmas}}
\end{figure}

In Figure~\ref{Fig:afmas}, the brightness distribution in this region is represented in three ways.
The logarithmic pseudo-color (arbitrary unit) is used to represent the large differences in brightness distribution.
The upper panels (a) and (b) are composed by restoring beams of a circular Gaussian with HPBW of $20~\mu \rm as$, which corresponds to the size of the spatial resolution of the EHT array for M\,87 observations.
The middle panels (c) and (d) are composed by restoring beams of circular Gaussian with HPBW of $200~\mu \rm as$. In order to facilitate comparison with previous results, the beam size is close to the spatial resolution of previous lower frequency observations (43,~86~GHz).
Panels (e) and (f) in the bottom row show the image by the large restoring beam overlaid with the 43 GHz averaged image by the VLBA \citep{Walker2018}.  Note that the 43~GHz image is time-averaged over 17 years, so knot-like features have been averaged out.
It can be seen that the brightness distribution at 230~GHz is consistent with that at 43~GHz.
The left panels (a), (c), and (e) show images of the data from the first two days.
Panels (b), (d), and (f) on the right show images of the data from the last two days.
Obviously, they are different from each other. However, the differences seen in the regions of a few mas cannot be attributed to the intrinsic variability of the source that occurred during the five days. Rather, it seems to be mainly dependent on the observational conditions.\\

The emission areas of our 230~GHz results are consistent with that of the 43~GHz average image. They also show that both edges of the jet are brightened, which phenomena have been observed so far.
Based on our data analysis, it seems that the detection of emissions in the range of several mas from the core of the M\,87 has been successful to some extent.

\subsection{Reliability of our final images\label{Sec:Reliability}}

As mentioned in Section ~\ref{Sec:ourreduction}, our calibration method was limited to self-calibration because the public the EHTC data do not contain raw data.

We also had to give up on the amplitude self-calibration because the closure phase residuals were not reduced as compared to the case when only phase calibration was performed.
Therefore, the calibration is not yet fully complete.
As clues to the reliability, we describe the properties of the final images from
three~aspects: detection limit (Section~\ref{Sec:Sensitivity}), 
robustness (Section~\ref{Sec:ROFI}),
and the self-consistency of our imaging (Section~\ref{Sec:SCimage-vis}), where our images show better self-consistency than those of the EHTC.
\subsubsection{From sensitivity\label{Sec:Sensitivity}}
\begin{figure}[H]  
\begin{center} 
\plotone{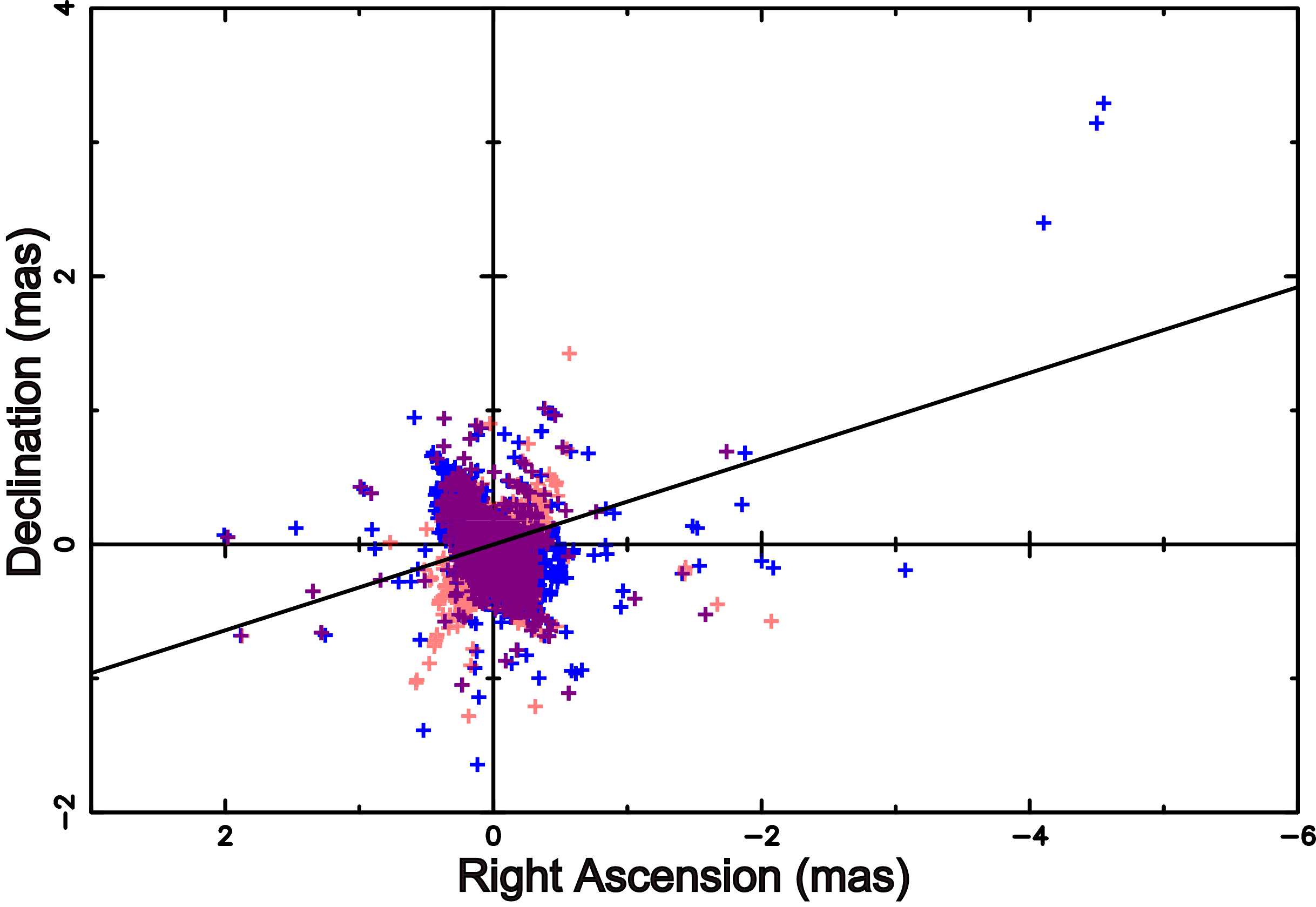}
\end{center} 
\caption{
Distributions of the grouped CLEAN components with flux densities larger than 0.1~mJy. Red dots are from the image of the first two days, blue dots are from the image of the last two days.
The sloped line indicates the average direction of the jet axis ($PA = -72\DEG$ from~\cite{Walker2018}).
}\label{Fig:lt01mjy}
\end{figure}

\cite{EHTC3} shows that the typical sensitivity of a baseline connected to ALMA is $\sim 1~mJy$. We estimate that this sensitivity is for an integration of about 5 minutes.
For an on-source time of 2 days, the attainable sensitivity reaches close to $0.1~mJy$ (ALMA-LMT baseline, $S/N~=~4$, assuming a point source).
It is difficult to estimate the practical sensitivity of the synthesized image of an interferometer composed of antennas with different performances, such as the EHT array. However, it is unlikely that the image sensitivity will not be worse than the baseline sensitivity noted above.


Here, we consider $0.1~mJy$ to be a reliable detection limit for our final images. Figure~\ref{Fig:lt01mjy} shows the distribution of the grouped CLEAN components with flux densities larger than $0.1~mJy$.
Almost all of the~components are concentrated within a few mas of the core.
(The remaining components are located in a false bright spot created outside the range of this figure, about $20~mas$~west of the center.)
~The image from the core to a few mas along the average jet axis seems to be reliable in terms of detection limits.

A large number of grouped CLEAN components with flux densities larger than $0.1~mJy$ are found in our final images.
The number of grouped CLEAN components with flux densities larger than $0.1~mJy$ is
1151 from the images of the first two days (with a sum of flux density of $707.4~mJy$), and 1657~from the images of the last two days (with a sum of flux density of 
$1032.6~mJy$) (Table~\ref{Tab:Fs}).

\subsubsection{The robustness of our final image\label{Sec:ROFI}}

In this section, we discuss another property of our final images: the robustness of the image structure.

If the data is not yet completely calibrated, 
BOX technique is effective.
As well as the EHTC, we also used the BOX technique to limit the imaging area (FOV). This technique has the potential to produce good images even if the calibration is incomplete.
On the other hand, it may be creating structures that do not actually exist. In fact, the bright spot on the right side of our final image (Figure ~\ref{fig:Fig-a}) is such an example.
Therefore, care must be taken when using the BOX technique, because a false structure will be created in the BOX area, and the real structure outside of the BOX area will be removed from the image.\\

\begin{figure}[H]
\begin{center} 
\epsscale{1.05}
\plottwo{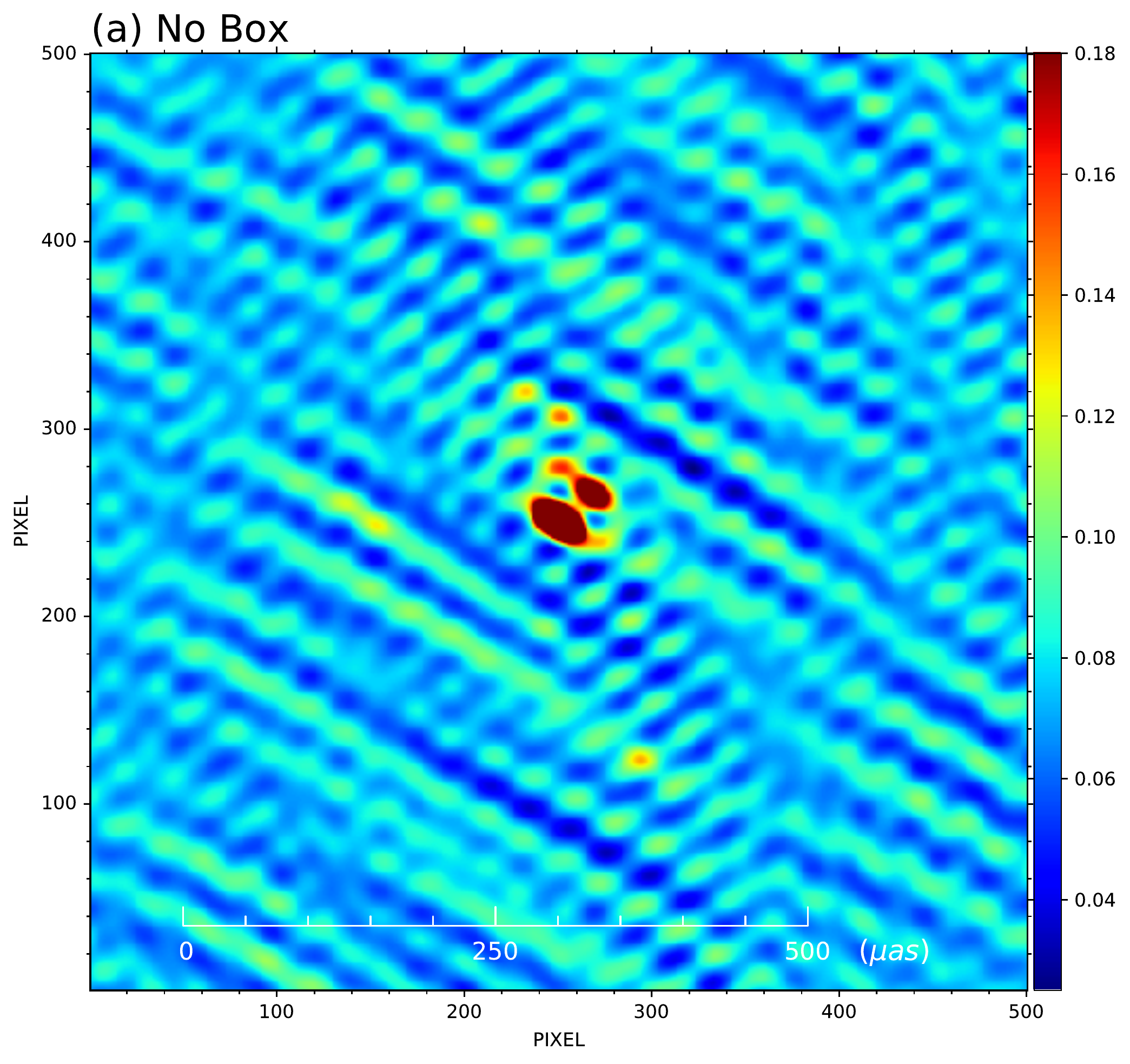}{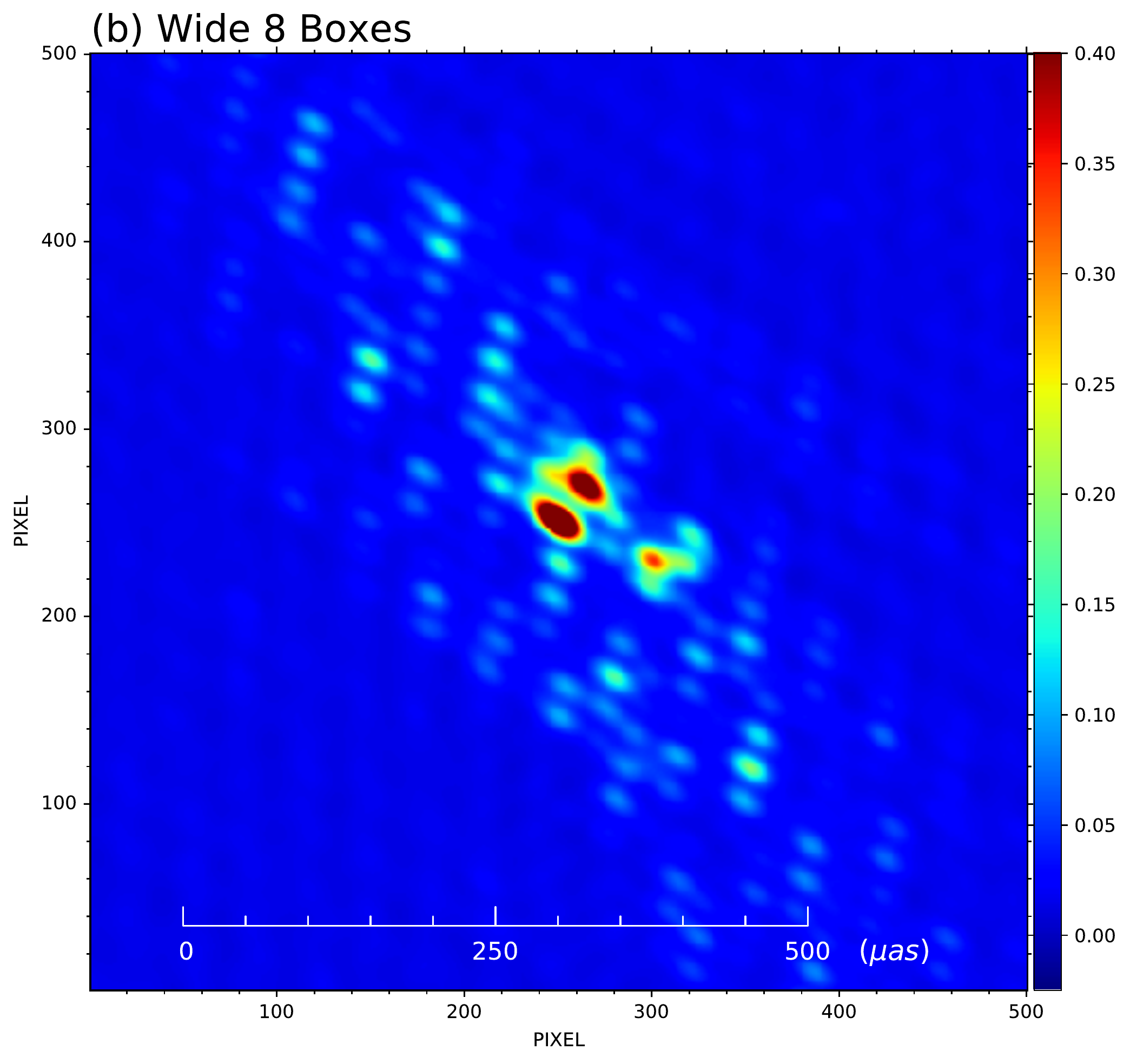}
\plottwo{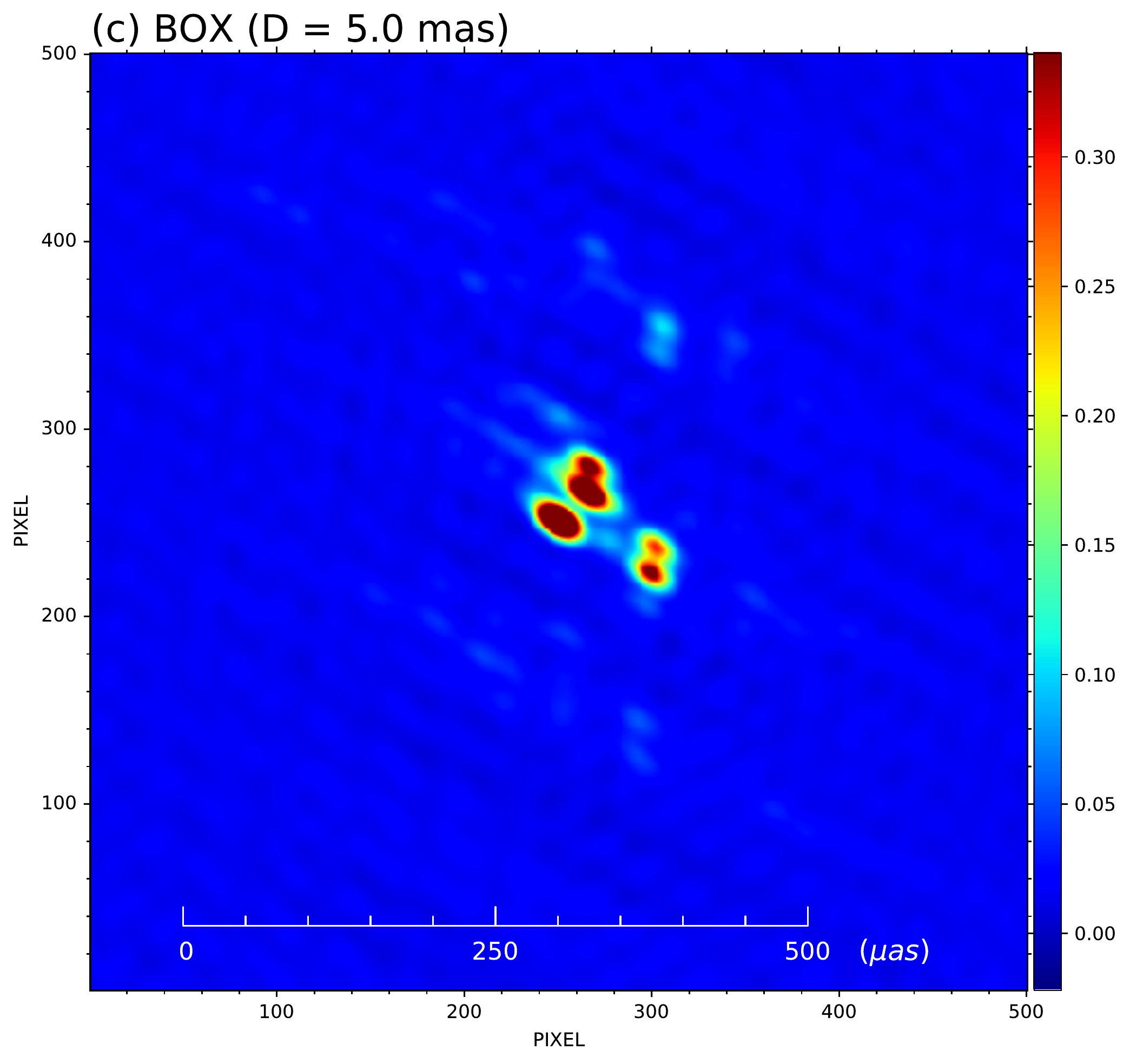}{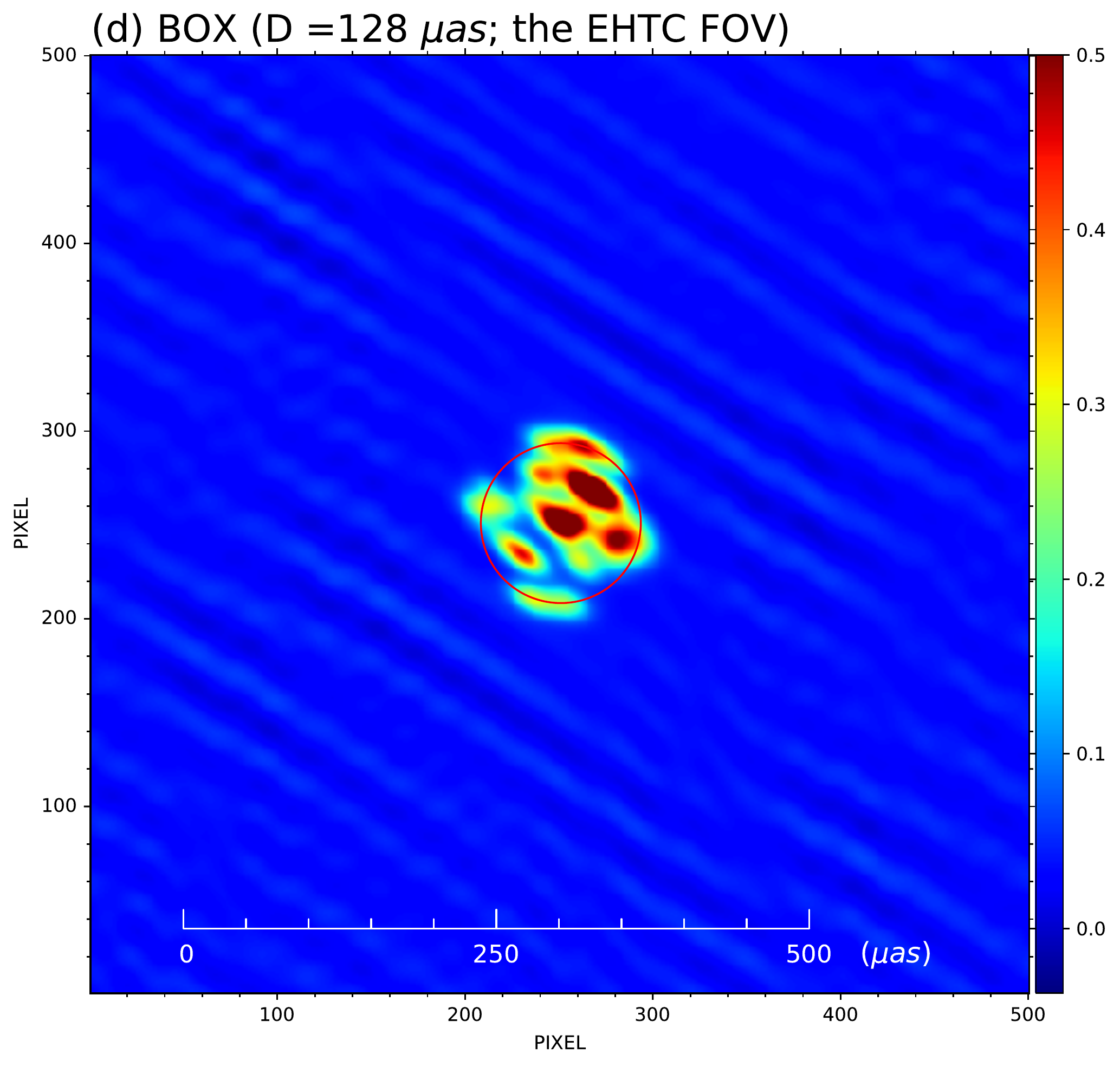}

\end{center} 
\vspace{-0.5cm}
\caption{Comparison of images obtained by changing the size of Box.
Panel (a) is the image without BOX (the FOV is $24.576~\rm mas$ square).
Panel (b) shows the image with the same 8 BOXes that we used to obtain our final images.
Panel (c) is the image with a small BOX (circle with diameter $D=5~\rm mas$ is used).
Panel (d) is the image with a very narrow BOX (circle with diameter $D = 128~\mu \rm as$ that corresponds to the FOV the EHTC used). 
These four CLEAN images were produced using data of the entire four days.
}
\label{fig:Fig-add1}
\end{figure}

 ~We examine how the image is affected when we change the BOX parameters. In other words, we investigate the stability of the image.
We compare the images obtained by changing the size of the BOX.
The panels in Figure~\ref{fig:Fig-add1} show the four cases.
The upper left panel (a) is the image without BOX (the FOV is $24.576~\rm mas$ square).
The top right panel (b) shows the image with the same 8 BOXes that we used to obtain our final images.
The lower right panel (c) is the image with a small BOX (circle with diameter $D=5~\rm mas$ is used).
The lower right panel (d) is the image with a very narrow BOX (circle with diameter $D = 128~\mu \rm as$ that corresponds to the FOV the EHTC used). 
These four CLEAN images were produced using data of the entire four days.
%

In all panels, the emission can be seen at the positions of features C (core) and K (knots).
On the other hand, feature W disappears in the case where the BOX setting is omitted (no BOX case).
In the case of the EHTC FOV, no emission is seen at the position of feature W because the position of feature W is outside the BOX setting.

Without the BOX setting, the S/N of the image is degraded.
From the comparison between panel (a) and the other panels, we can see that the BOX setting compensates for the lack of calibration and improves the image quality. Thus, presence or absence of the BOX setting seems to have an effect on the image quality.
Another noteworthy point is that in the case of the very narrow BOX setting (panel d), several different bright spots newly appear in the BOX.
Moreover, some of them are located at the boundaries of the BOX. In such a case, other actual brightness distributions could exist outside the BOX setting.

Since feature W disappears in the CLEAN image without the BOX setting. Feature W is considered to be less reliable than features C and K.
As mentioned at the end of the Section~\ref{Sec:resultsCC}, there are other reasons to consider that feature W is a real existence.
%
\subsubsection{
Self-consistency of our imaging as compared to those of the EHTC images
}\label{Sec:SCimage-vis}
\begin{figure}[H]
\epsscale{1.0}
\begin{center} 
\plotone{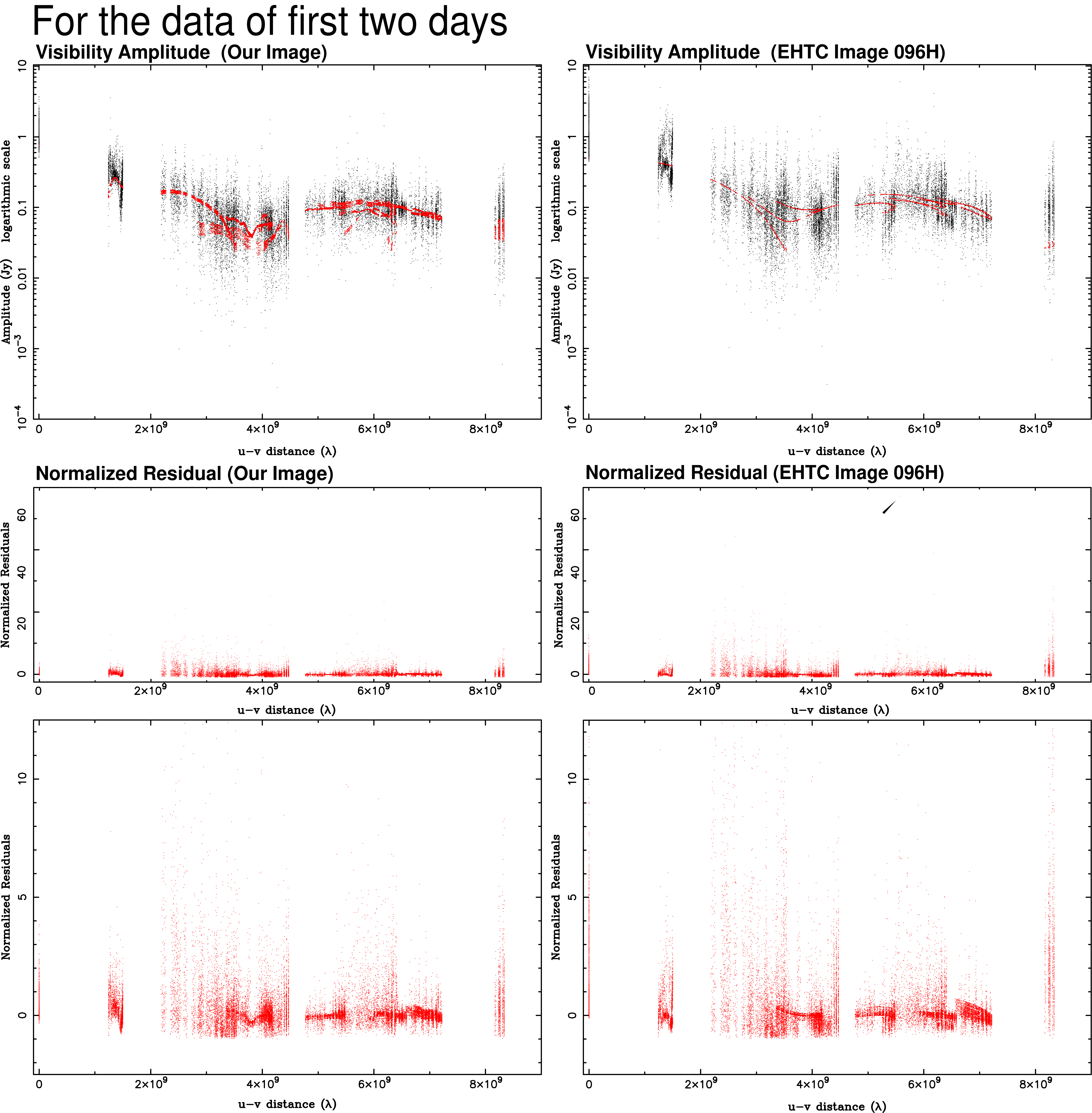}
\end{center} 
\vspace{-0.5cm}
\caption{
Relation between the visibility amplitude and u-v distance for the first two days of data.
The left panels are for our image, and the right panels are for the EHTC ring image model with the EHTC DIFMAP pipeline using 096H data 
(April 6).
Every dot is a raw visibility point for a 10-second integration.
The top panels show the plots of visibility amplitude versus u-v distance. 
The black dots in them show those calibrated by self-calibration solutions using the image model.
The red dots are those from the Fourier-transformation of the image model.
The middle and bottom panels show normalized amplitude residuals.
The middle panel shows all data points. The vertical axis scale is very large due to data points with very large values, as indicated by black lines.
The bottom panels show a magnified view of the range of normalized residuals from -5 to +15. In all cases, the minimum value of the normalized residuals is greater than -1.
}
\label{fig:Vis-amp1}
\end{figure}

\begin{figure}[H]
\epsscale{1.0}
\begin{center} 
\plotone{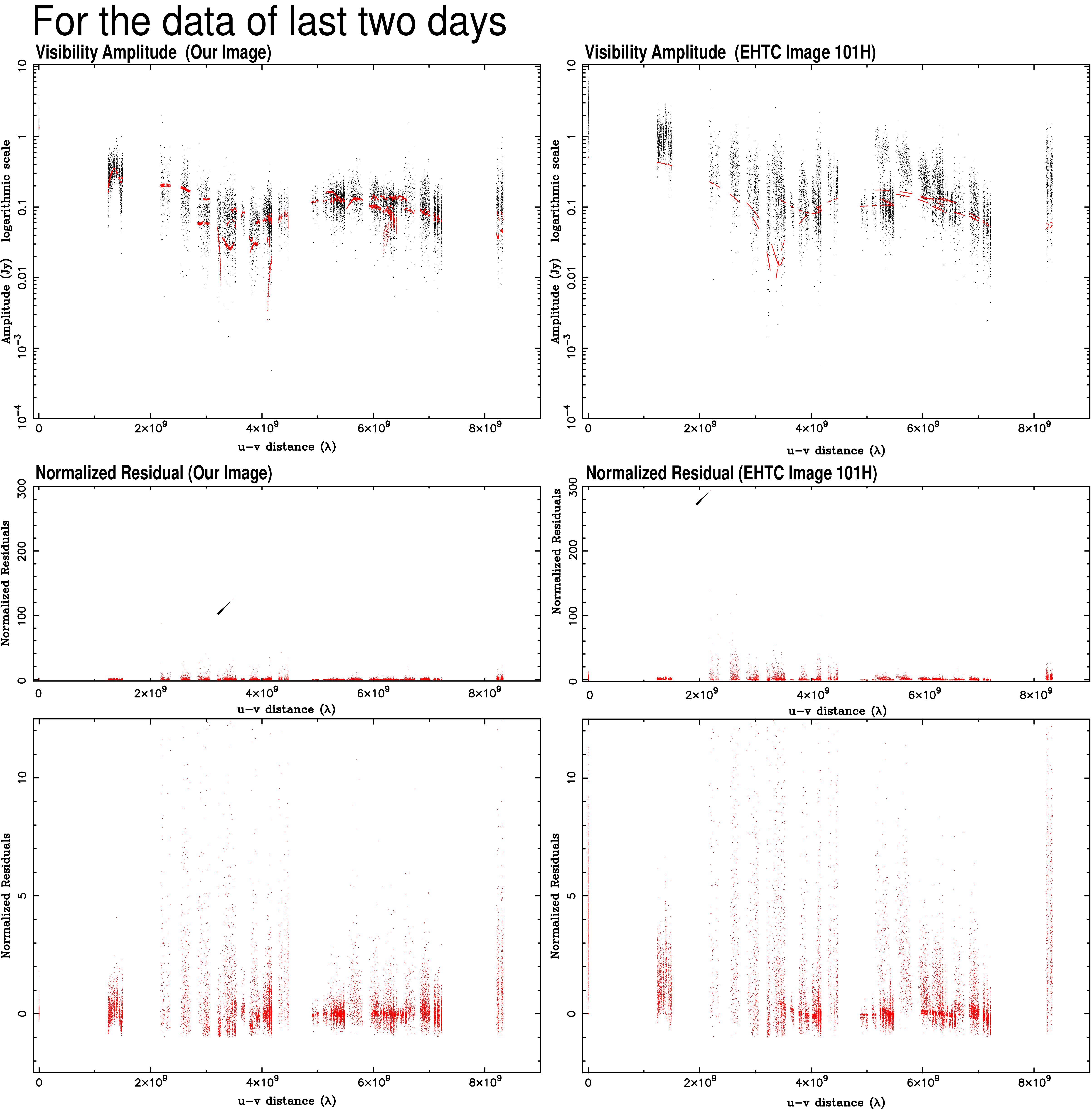}
\end{center} 
\vspace{-0.5cm}
\caption{
Relation between the visibility amplitude and u-v distance for the last two days of data.
The left panels are for our image, and the right panels are for the EHTC ring image model with the EHTC DIFMAP pipeline using 101H data 
(April 11). 
Every dot is a raw visibility point for a 10-second integration.
The top panels show the plots of visibility amplitude versus u-v distance. 
The black dots in them show those calibrated by self-calibration solutions using the image model.
The red dots are those from the Fourier-transformation of the image model.
The middle and bottom panels show normalized amplitude residuals.
The middle panel shows all data points. The vertical axis scale is very large due to data points with very large values, as indicated by black lines.
The bottom panels show a magnified view of the range of normalized residuals from -5 to +15. In all cases, the minimum value of the normalized residuals is greater than -1.
}
\label{fig:Vis-amp2}
\end{figure}

\begin{figure}[H]
\begin{center} 
\epsscale{1.15}
\plotone{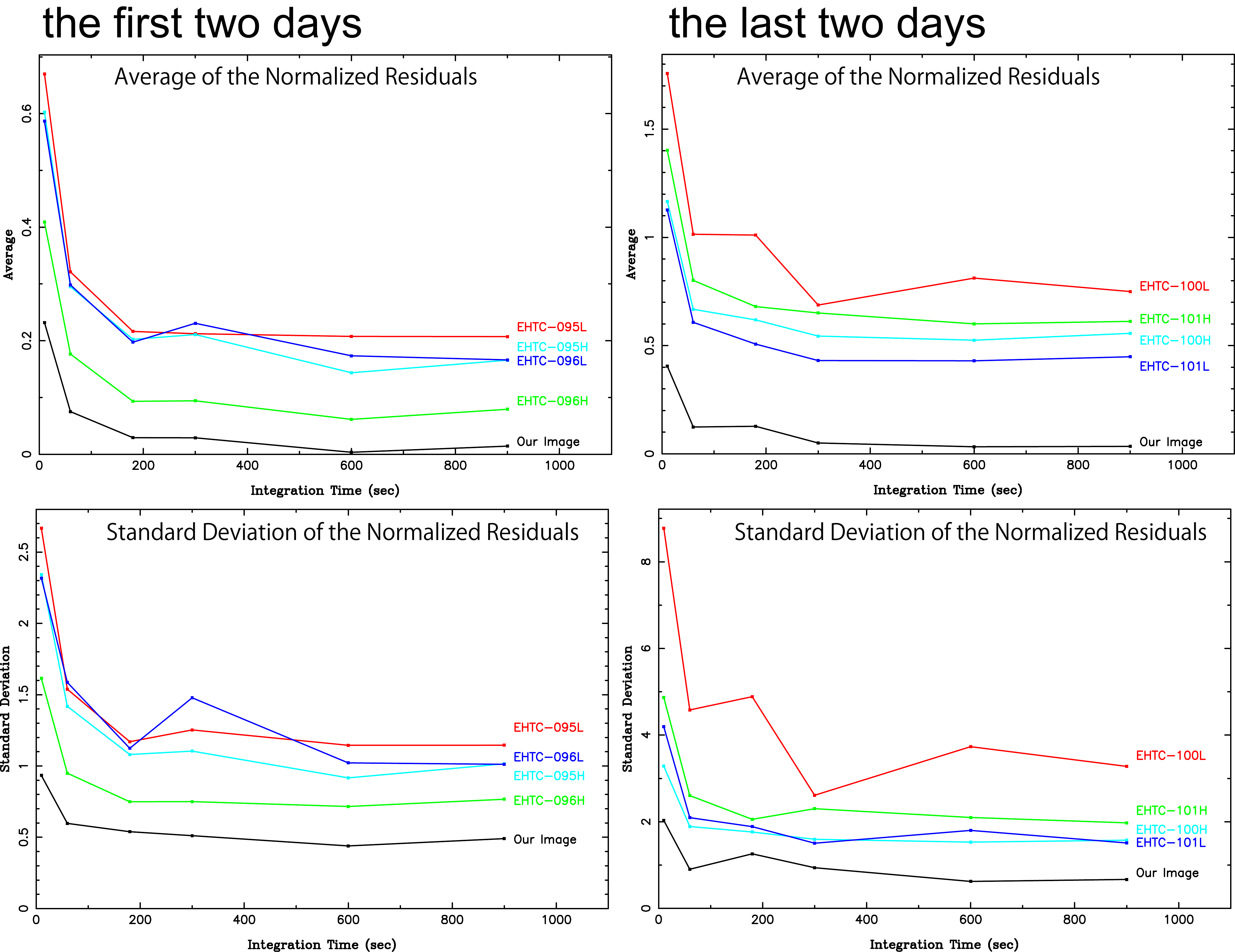}
\end{center} 
\vspace{-0.5cm}
\caption{
Statistics of normalized residuals.
The upper panels show the average values of the normalized residuals.
The lower panels show the standard deviations.
The left panels are the data of the first two days, and the right ones are the data of the last two days.
The black line shows the case of our final image, and other color lines show the cases of the EHTC images.
}
\label{fig:Vis-amp-sta}
\end{figure}

\begin{figure}[H]
\begin{center} 
\epsscale{0.9}
\plotone{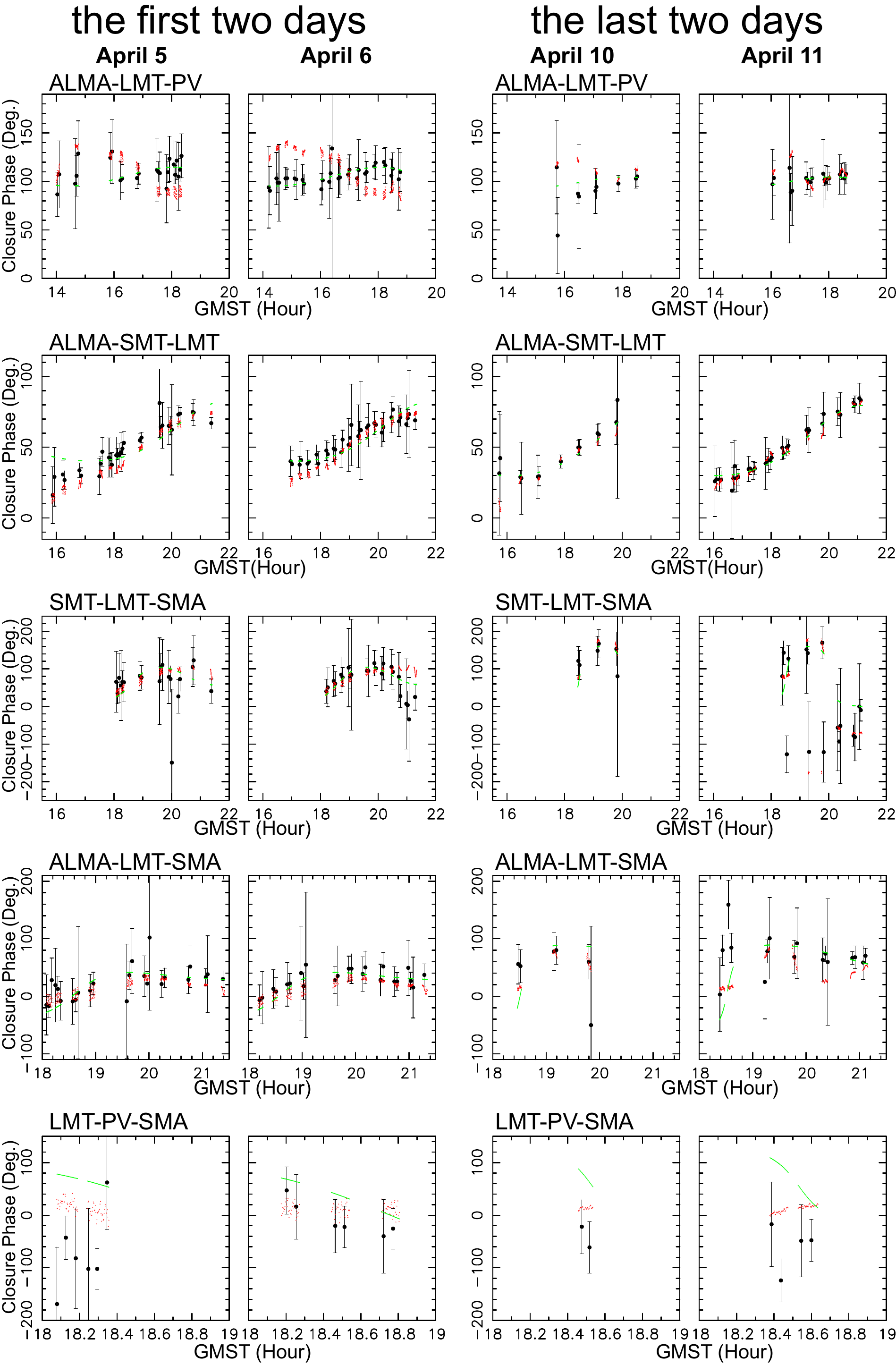}
\end{center} 
\vspace{-0.75cm}
\caption{
Closure phase variations of five triangles (ALMA-LMT-PV, ALMA-SMT-LMT, SMT-LMT-SMA, ALMA-LMT-SMA, and LMT-PV-SMA).
Closure phases of the observed data  are shown by black dots with $\pm 3~\sigma$ error bars, which are obtained from 5 minutes integration.
Red dots show those from our image models. 
Green dots show those from EHTC ring image models
(the EHTC 096H image for the first two day's data, the EHTC 101H image for the last two day's).
}
\label{fig:exam-closVis}
\end{figure}

At the end of this section on image reliability, we present the degree of matching between the visibility and the image model. Here, we compare the results with those of the EHTC ring.
\begin{enumerate}
\item Relations between the visibility amplitude and \UVC distance (projected baseline length):
The amplitude of visibility obtained by inverse Fourier transforming the image model is compared with those of the observed visibility data.
Figures~\ref{fig:Vis-amp1},~and \ref{fig:Vis-amp2} correspond to Figure 12 in \cite{EHTC4}.
This kind of comparison of visibilities is often performed to check the reliability of an image.
However, here, the observed visibility data are calibrated by self-calibration solutions using the image model.
Therefore, it is important to note that the amplitudes of the observed visibility data and those from the image model are no longer independent of each other. 
What can be safely determined from this comparison is the internal consistency of the imaging and calibration process.
Figure~\ref{fig:Vis-amp1} shows the data from the first two days, and Figure~\ref{fig:Vis-amp2} shows the data from the last two days.
The top row of each shows the variation of the visibility amplitude with respect to the projected baseline length. The red dots are those of the image model.
The middle and bottom panels show the normalized residual amplitudes between the image model and the calibrated observation data.
The plotted points are the calibrated raw data that have not been time-integrated, and we can see that the scatter is much larger than seen in their Figure 12 (\cite{EHTC4}), where the time-integrated points have been plotted.
We can see that the average and standard deviation of the normalized residuals of our final image are much smaller than any of the EHTC ring images
in Figure~\ref{fig:Vis-amp-sta}.
As an example, we show the normalized residual values for $t = 180~sec$ integration. \\
For the data of the first two days,
our image shows $\rm NR_{ours} = 0.030 \pm 0.539$, while
the EHTC images show $\rm NR_{EHTC} = 0.148 \pm 0.933$.\\
For the data of the last two days,
our image shows  $\rm NR_{ours} =0.127 \pm 1.259$, while
the EHTC images show $\rm NR_{EHTC} =0.589 \pm 2.370$.
Here, in the case of EHTC, we used the simple averages of those values for the four EHTC images.
~One thing that interests us is the large discrepancy in amplitude of the EHTC ring image cases at the longest baseline lengths over $8\times 10^{9} \lambda$. It is three times larger than those of our final image cases.
Since the EHTC ring images are very compact, if the images are really correct, the amplitude residuals at the longer baseline should become small at least.
Another is the amplitudes at the very shorter baselines nearly $\rm zero~\lambda$. They contain the components of the extended structure which are resolved by high spatial resolution by EHT, so it is not surprising even if they do not match.
Our images reproduce the amplitudes of the very short baselines well, but the differences are more significant in the cases of the EHTC rings. In our cases, the normalized residuals are 4 at most, but in the cases of the EHTC rings, they are distributed widely in the range of 0 to 15, which is not surprising since the EHTC rings are compact and have no extended components. 
However, in the Figure 12 in \cite{EHTC4}, the maximum is 4, as if the result shows good self-consistency. The EHTC Figure also shows the same results for the normalized residuals at the longest baselines. 
This is not consistent with our own analysis.
Perhaps a different integration time of the data may cause this apparent discrepancy. (There is no explanation for the integration time of the data points in Figure 12 in \cite{EHTC4}).
~Since the scatter of data points is affected by thermal noise, its value changes depending on the integration time of the data. Therefore, we examined the amounts of normalized residuals by changing the integration time.
Figure~\ref{fig:Vis-amp-sta} shows the average and standard deviation of the obtained normalized residuals.
It can be seen that at any integration time, our final image always shows smaller values than that of the EHTC ring image.
The averages of the 10-second integrations and integrations over 180 seconds differ by a factor of 3, which explains the discrepancy above.\\
The diagram of the visibility amplitude and \UVC distance 
shows that our final images, both the first and the last days, show a better consistency of imaging and calibration processing compared to the cases of the EHTC ring images.
The diagrams indicate that our images, not those of EHTC, are supported by the data.
\item Closure phase variations:
 Following Figure 13 of \cite{EHTC4}, we show the closure phases of the observed data and those derived from image models for some triangles in Figure~\ref{fig:exam-closVis}. We added the closure phases of ALMA-LMT-SMA and LMT-PV-SMA to the three triangles (ALMA-LMT-PV, ALMA-SMT-LMT, SMT-LMT-SMA) shown by the EHTC.
Closure phase is a quantity that is free from systematic phase errors and reflects the observed source structure. All panels in Figure~\ref{fig:exam-closVis} show large phase variations, which correspond not to time variation in the structure of the observed source, but to time variation of the shape of the triangle composed by the three stations as seen from the observed source. The green dots are the closure phase corresponding to the EHTC ring image, and the red dots correspond to our image. The dots of our image (green dots) appear to be more aligned with the observed data than those of the EHTC ring image. Our image is more complex than the EHTC ring, resulting in short-term small closure phase variations.\\
The three panels from the top right toward the bottom correspond to the panels shown in Figure 13 of~\cite{EHTC4}.
In the case of the two triangles ALMA-LMT-PV and SMT-LMT-SMA, our results are consistent with those of EHTC.
However, our results for the closure phase in ALMA-SMT-LMT triangle differ from those of EHTC.
In our case, the closure phase shows an increase from $+25\DEG$ to $+85\DEG$, while that of the EHTC shows a decrease from $-25\DEG$ to $-80\DEG$. Both the first and last values and the amount of change are opposite in positive and negative.
All triangles were examined, but none were identical to the closure phase variation shown by the EHTC for the ALMA-SMT-LMT triangle.
Our closure phase values are from the AIPS task, CLPLT. All triangles were examined, and none showed the similar variation that the EHTC showed for the triangle. Also, there are no significant closure phase discrepancies between the real and model data. There seems nothing wrong with the CLPLT calculations. \\

In our analysis, there is no clear difference in closure phase matching between our images and the EHTC rings.
A notable difference is in the case of the LMT-PV-SMA triangle, where the closure phase in the EHTC rings is beyond $\pm 3~\sigma$ error bars, whereas in our final images, it manages to fall within it.

The values of the closure phases also change depending on the integration time of the data; however, even when the integration time is changed, the residuals of either of them do not become overwhelmingly small.
~Figure~\ref{fig:res-closure-sta} shows the statistics of the closure phase residuals for all triangles.
As far as the closure phase residuals are concerned, between our images and the EHTC rings, there is no significant difference.
Our image of the core-knot structure also shows the same magnitude of closure  phase residuals as those of the EHTC ring image. 
As an example, we show the standard deviations of the closure phase residuals at $t = 180~sec$ integration.
For the data of the first two days,
our image shows $\sigma_{ours} = 40.5\DEG$, while
the EHTC image shows $\sigma_{EHTC} = 38.5\DEG$.
For the data of the last two days,
our image shows $\sigma_{ours} = 43.2\DEG$, while the EHTC image shows $\sigma_{EHTC} = 43.7\DEG$.

As for closure phase residual, there is no significant difference between ours and the EHTC rings.
If we claim that the EHTC ring image is correct due to the closure phase residual, then our images are also correct at the same time.\footnote{
The correct image satisfies the closure phase of the observed data. However, the opposite is not always true. Moreover, there are numerous image models that show small closure phase residuals.}
\\
If these residuals are due to thermal noise, they should decrease inversely proportional to the root square of the integration time T, but as Figure~\ref{fig:exam-closVis} shows, they do not decrease in that way. 
It means that both images of ours and the EHTC's still have differences from the true image.
\end{enumerate}

%
\begin{figure}[H]
\begin{center} 
\plotone{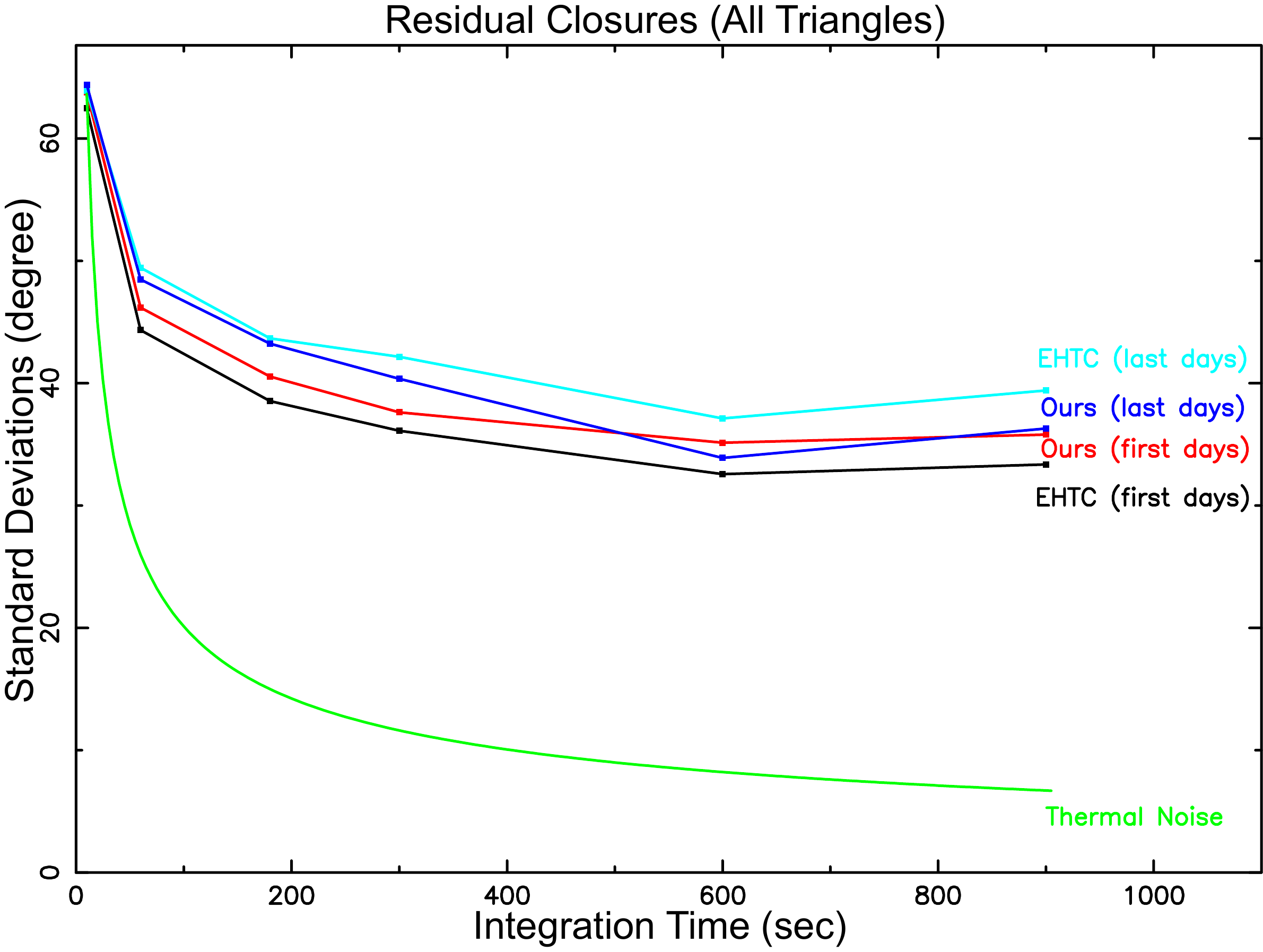}
\end{center} 
\vspace{-0.5cm}
\caption{
Standard deviations of the closure phase residuals from all triangles.
The case of the EHTC image (for the first two days) is shown by black line, that for the last two days by cyan line.
The cases of our final images are shown by red line for the first days and blue line for the last days.
The green line shows in the case that the residuals are due from thermal noise.
}
\label{fig:res-closure-sta}
\end{figure}
%
%
\section{Why the EHTC found their \texorpdfstring{$\sim 40~\mu \rm as$} ~~ring?\label{Sec:the EHTC-ring}}
In Sections~\ref{Sec:ourreduction} and~\ref{Sec:results}, we tried to reconstruct the image from the EHT public data using the hybrid mapping process. Our final images contain a core-knot structure at the center and features along the jet axis towards the outside. 
It is consistent with those obtained by 43~GHz or 86~GHz observations. 
On the other hand, three of the EHTC imaging teams all obtained $\sim 40~\mu\rm as$ rings similar to each other, but no jet structure. In this section, we show the evidences that the EHTC ring is an artifact.

The essential reason why the ring image was obtained by all the EHTC imaging teams is the limited \UVC coverage of the EHT array for M\,87, 
namely the data sampling bias, though the EHTC realized 230~GHz VLBI observations on a scale that has never been accomplished before.
In addition, the very narrow FOV settings of the EHTC strongly help to create $\sim 40~\mu\rm as$ ring shape from the EHT \UVC data sampling bias.

~First, in Section \ref{Sec:uvs}, we discuss the nature of the \UVC coverage of EHT for the M\,87 observations. The spatial Fourier components for the fringe spacing of $\sim 40~\mu\rm as$ are lacking.
~Second, in Section \ref{Sec:PSF}, we discuss how the dirty beam
(PSF) of the M\,87 EHT observations is affected by this lack of
the spatial Fourier components for the fringe spacing of $\sim 40~\mu\rm as$.
~Third, in Section~\ref{Sec:byDmap}, we show the dirty map is greatly affected 
by the PSF shape in the case of the EHTC data.
~Fourth, in Section~\ref{Sec:simData}, we show that even from the
simulated visibility data of a point source we can create  $\sim 40~\mu\rm as$ rings.
This means that the \UVC coverage of the EHT array for M\,87 can create the $\sim 40~\mu\rm as$ ring regardless of the real structure of the observed object.~
In other words, the EHTC result is indistinguishable from an artifact.
~Fifth, in Section~\ref{Sec:pipeline}, we investigate one of their open procedures for imaging demonstration. The EHTC used three methods for their imaging. We investigated their DIFMAP\citep{Shepherd1997} pipeline, which is the closest to traditional procedures for VLBI, and found an improper point.
In the EHTC-DIFMAP pipeline, they set a very narrow FOV setting by using BOX technique.
When we ran the EHTC-DIFMAP pipeline without the BOX, the $40~\mu\rm as$ ring disappeared. Instead, a core-knot structure appeared. We also checked the output of the simulated data for other shape model images and found that the EHTC-DIFMAP pipeline did not reproduce the input model images correctly. 
In other words, the EHTC-DIFMAP pipeline does not prove the correctness of the EHTC ring image from the data.
~In addition, we estimate the amount of calibrations that the EHTC performed
during their imaging process (by the DIFMAP method), and compare them with those we did in our analysis.
We found that a large amount of additional ”calibrations” are required to make the $40~\mu\rm as$~ring (Section~\ref{Sec:EHTSNPLT}). 
~Also, we discuss the robustness of the EHTC $\sim 40~\mu\rm as$ ring structure in Section~\ref{Sec:staERING}. The structure is very sensitive to the imaging parameters. If we change the BOX size larger, the ring image changes significantly.\\

Despite their "isolated image analysis" and surveys involving large-scale simulations, the EHTC have obtained artificial ring images. 
Finally, we explain in Section~\ref{Sec:EHTCL4} why their objective survey has produced artifacts.
They determined their optimal imaging parameters through large-scale simulations. However, they did not take into account the sampling bias that tends to produce the \FTMUAS ~structure.
 The EHTC focused only on the reproducibility of the input image models and not on the simultaneous reproducibility of the input error models. It cannot be ruled out that their optimal parameters may have had the property of creating the EHTC's \FTMUAS ~ring by not performing proper calibration and imaging, but rather by enhancing the sampling bias effect.
 The facts shown in this section are strong evidence that the EHTC $\sim 40~\mu \rm as$ ring image is a forcibly created artifact.
 In Section~\ref{Sec:reasons}, we  summarize the reasons why the EHTC obtained the artifact image unintentionally.

\subsection{\textit{u-v}~coverage of the EHT array for M\,87\label{Sec:uvs}}

\begin{figure}[H]
\begin{center}
\epsscale{1.}
\plotone{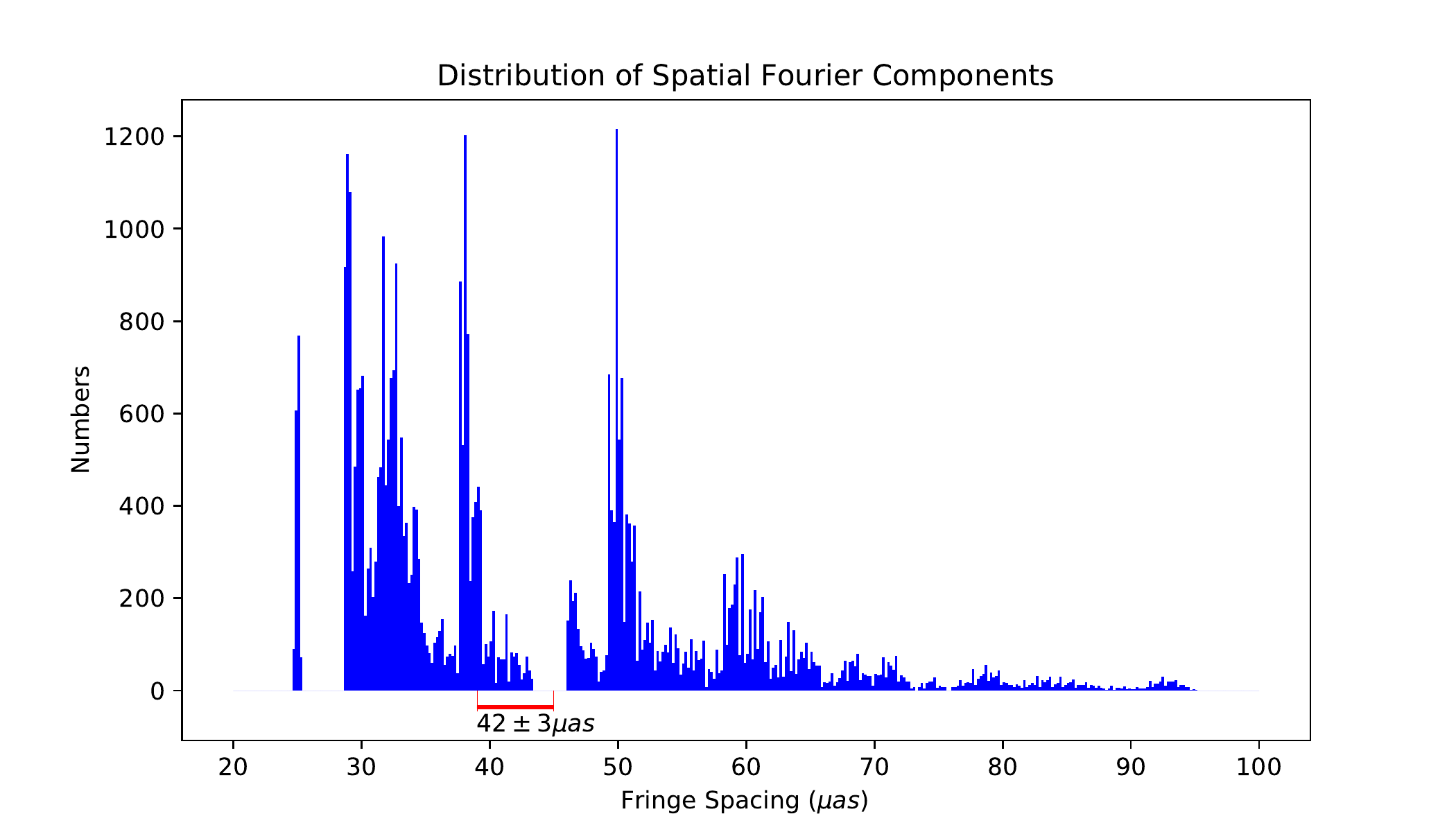}
\end{center}
\caption{Distribution of the sampling data (all data of the four days from all baselines). The x-axis shows the fringe spacings of sampled visibility data in $\mu\rm as $ unit. The y-axis shows the number of sampled data. Here, samples larger than the fringe spacing of $100~\mu\rm as$ are omitted. The red line segment indicates the range of the diameter of the ring measured by the EHTC ($d = 42 \pm 3~\mu\rm as $).
}\label{Fig:uvs}
\end{figure}

We here investigate the \UVC sampling of the EHT array for M\,87 observations.
The \UVC coverage itself is shown in the EHTC paper I~\citep{EHTC1}. 
It shows a good \UVC coverage as an interferometer of practically 5 stations and 10 baselines.
~We looked at the number of data samples from another point of view.
Figure~\ref{Fig:uvs} shows the distribution of spatial Fourier
components (fringe spacings) sampled by EHT for M\,87 during the four observational
days. The number of sampled data is plotted against the fringe
spacings in $\mu\rm as$ unit.
We can see that there are no sample for ranges $d = 26-28~\mu\rm as$, $d=
44-46~\mu\rm as$, and $d = 95-100~\mu\rm as$
\footnote{
Fringe spacings larger than $100~\mu \rm as$ are omitted in the histogram.  We are interested in the samplings contributing to very high spatial resolutions less than $100~\mu\rm as$.
}.
The number of sampled data for the size of the EHT ring ($d =
42\pm~3~\mu\rm as$) is quite small, and for spacing between $d = 44
-46~\mu\rm as$, there are no sampled data.

The few numbers of sampling around $42~\mu\rm as$ fringe spacings should affect the imaging performance. In the following subsections, we investigate how this lack of spatial Fourier components can affect the imaging result.

\subsection{Substructures in the dirty beam\label{Sec:PSF}}
The EHTC show no description of the dirty beam structure.
We, then, examine the dirty beam calculated from the \UVC coverage. 
Dirty beam is the term used in the field of radio interferometer and its meaning is the same as that of PSF (Point Spread Function) in optics. 
In other words, it is the diffraction image of a single point when the data calibration is perfect.
If we can obtain all spatial Fourier components, the dirty beam becomes a two-dimensional $\delta$-function. 

\begin{figure}[H]
\epsscale{0.9}
\plotone{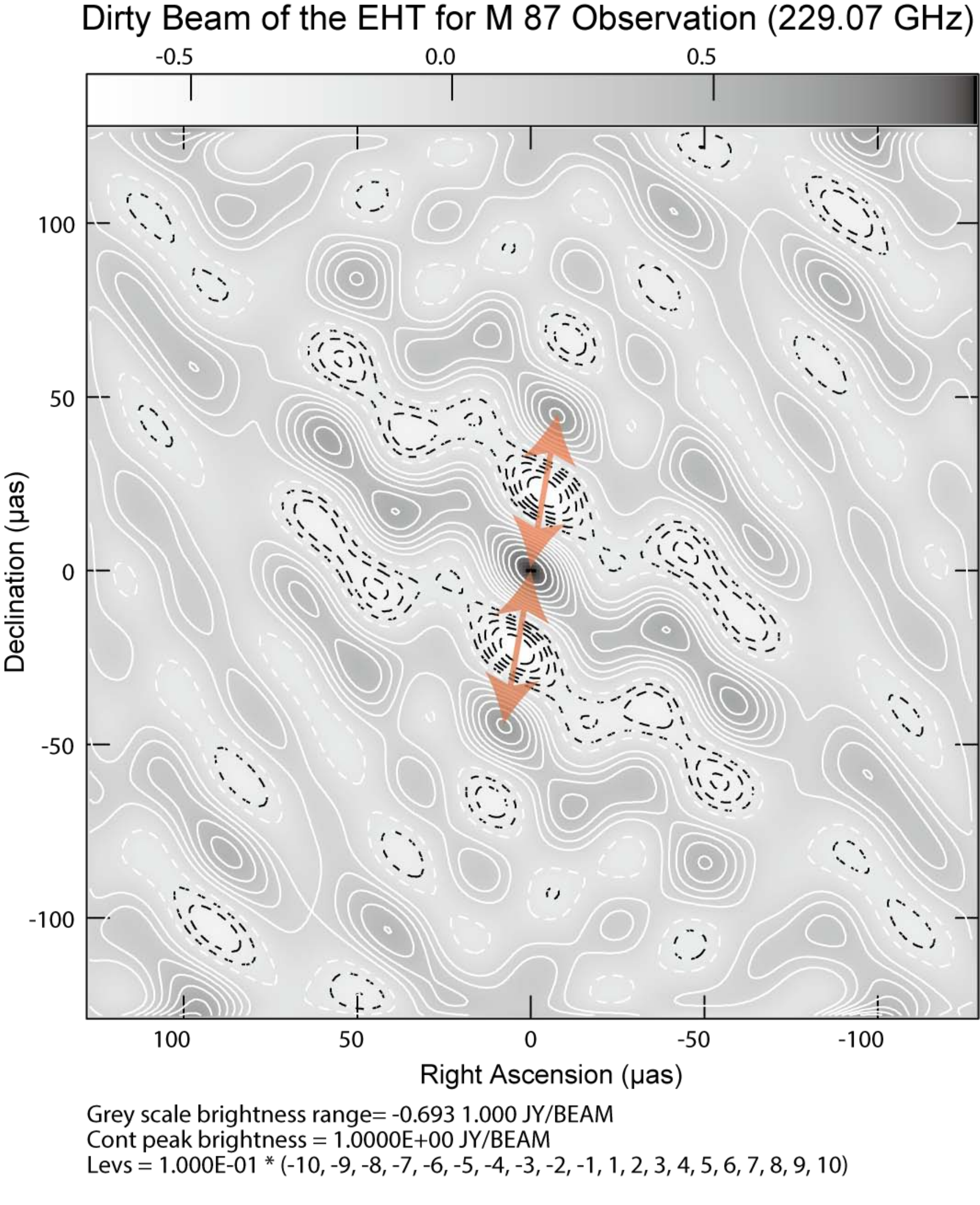}
\caption{Point spread function (dirty beam) of EHT on the first day observation of M\,87. The contour levels are at every $10~\%$ points of the peak from $-100$ to $100~\%$. Positive levels are shown by white lines, and negative levels by black dotted lines. The white dotted lines are for zero levels. 
The grayscale ranges from the minimum to the maximum brightness of the dirty beam image.
The first sidelobe levels reach more than 70~\% of the height of the main beam.
We put two red arrows with a length of $45~\mu\rm as $  between the main beam and the first   sidelobes. For the dirty beam image, we set the parameter $ROBUST = 0$ in IMAGR (its default setting).\label{Fig:uvd}}
\end{figure}

In practice, we can obtain only a limited sample of spatial Fourier components, and thus the dirty beam has a complex shape.
Figure~\ref{Fig:uvd} shows the dirty beam of the EHT array on the first day observation of M\,87.
The FWHM of the main beam is approximately 20$~\mu\rm as$.
We can see a point-symmetric pattern around the main beam. This pattern includes peaks lower than the main beam. 
Such peaks are often called ``sidelobes'' in radio astronomy. 
We can see that the distances between the main peak and the first peaks of sidelobes are about 45$~\mu\rm as$.
\footnote{
Further, the first sidelobe levels reach more than 70~\% of the height of the main beam.
The sidelobe level is still extremely high even when changing \UVC weights.
Natural and uniform weighting cases, both show that the first sidelobe levels reach more than 60~\% of the height of the main beam
(see Appendix \ref{Sec:otherDbeams}).
}
The separation between them corresponds to the radial range for which the spatial Fourier components are missing, and is very close to the diameter of the EHTC ring ($42~\pm~3~\mu\rm as$). Thus, it is important to clarify whether or not this PSF structure has affected the ring image or not.
In addition, Appendix \ref{Sec:otherDbeams} shows other PSF shapes by different weightings of \UVC points. Even in these PSF shapes, the separations between the main beam and adjacent sidelobes do not change much. They also show substructures with $\sim 40~\mu\rm as$ spacings.

\subsection{The relation between the dirty map convolved by the PSF shape and the ring structure\label{Sec:byDmap}}
~Here, we show what happens if we are not very careful in using the
dirty map when applying the hybrid mapping process. 
A dirty map is an image created by simply performing an inverse
Fourier transform on the obtained spatial Fourier
components. Therefore, it is influenced by the data sampling bias,
that is, the structure of PSF is convoluted into the dirty
map. Furthermore, if the data calibration is insufficient, it will be
reflected in the dirty map, and 
the obtained structure in such a dirty map can be far from the actual image.
Therefore, it is dangerous to perform self-calibration using the dirty map as the image model.

It is certain that in the case of multi-element radio interferometers like ALMA, or VLA, the dirty beams have sharp main beams and very low sidelobes. Only in such cases, it is not so dangerous to estimate the true image by the dirty map.
While,
in the case of VLBI observations, the dirty beam is comparatively dirty, 
usually, we do not estimate the true image from them and do not use them as the model image for self-calibration.

\begin{figure}[H] 
\epsscale{1.15}
\gridline{\fig{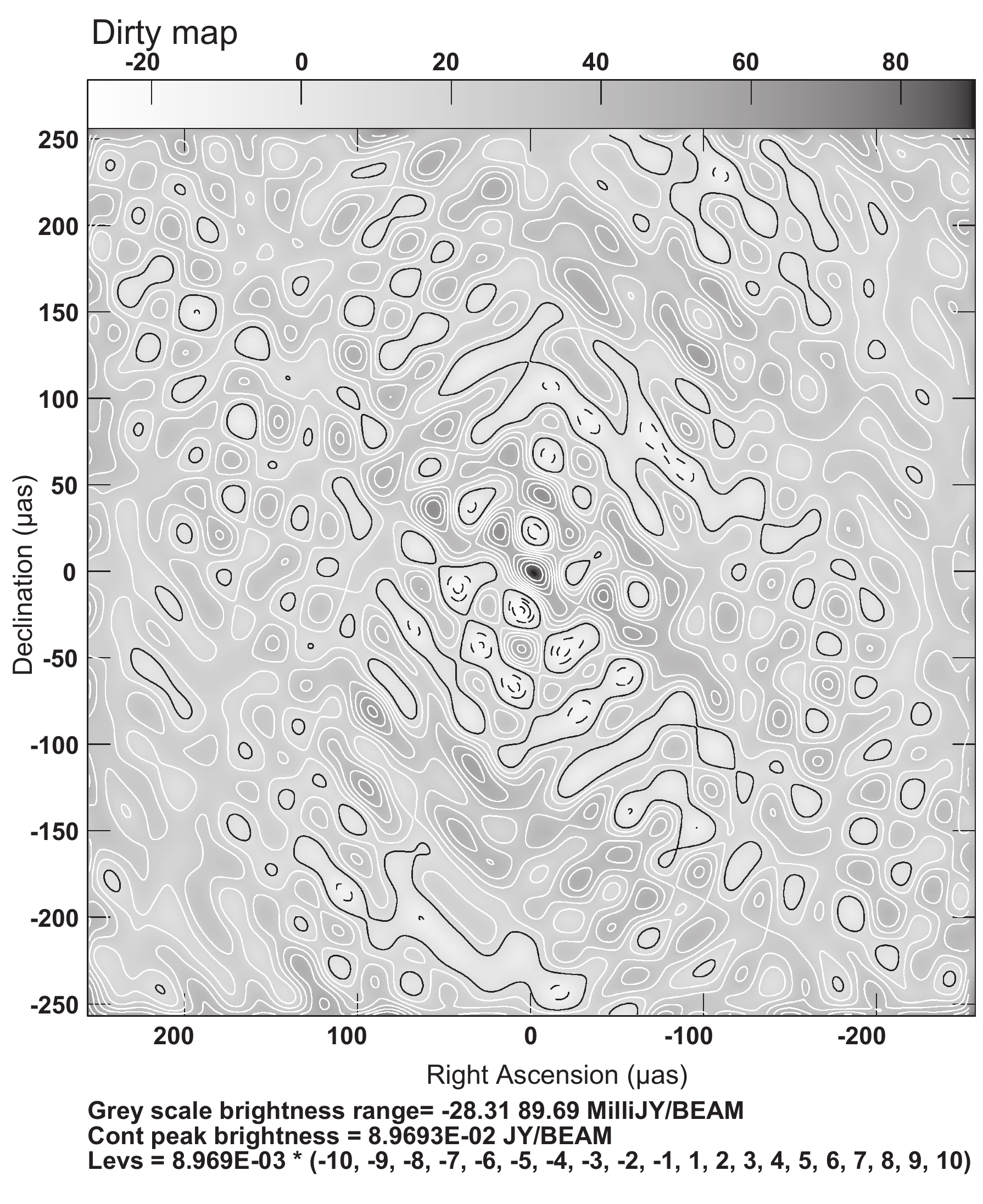}{0.4\textwidth}{}\fig{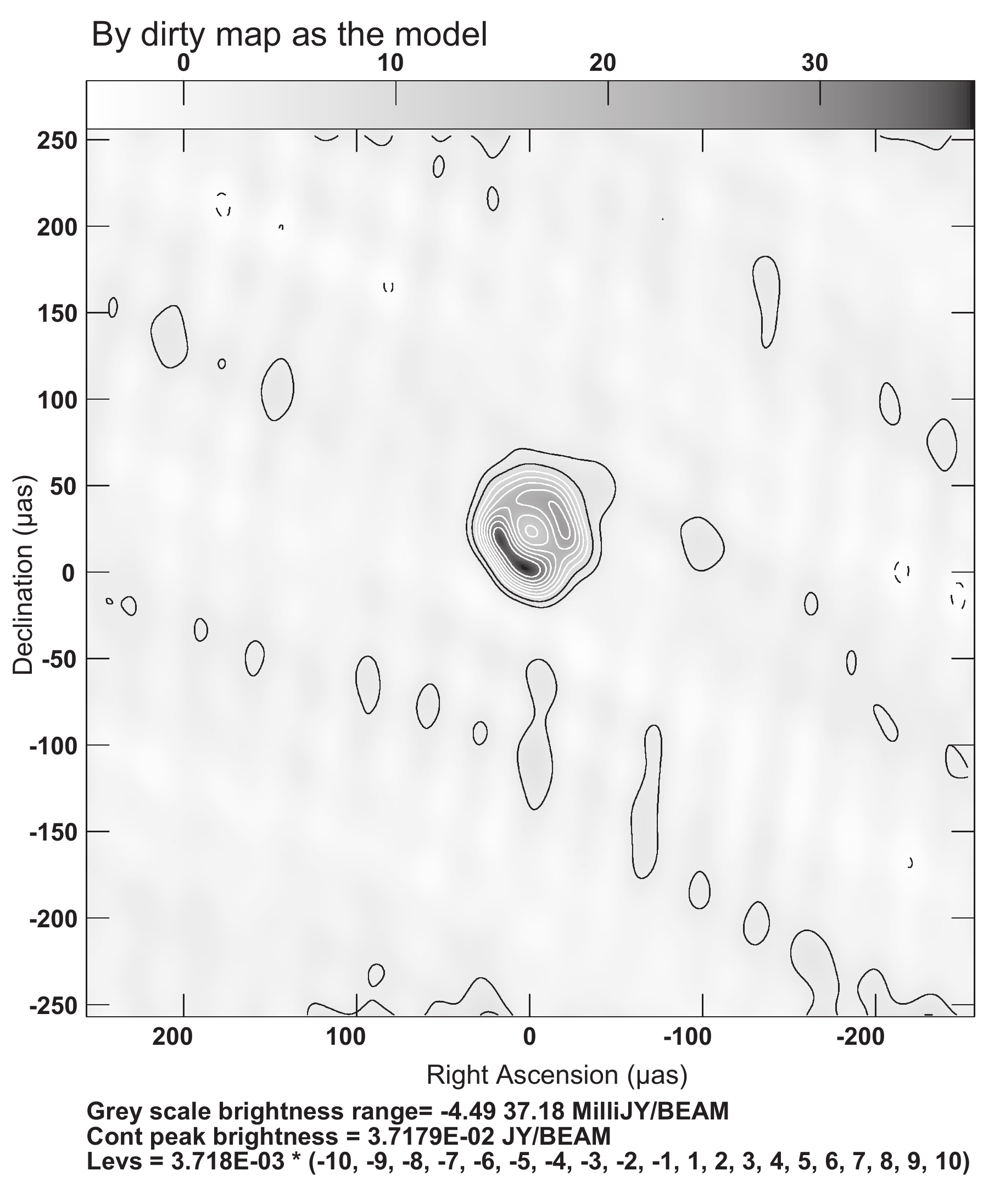}{0.4\textwidth}{}}
\gridline{\fig{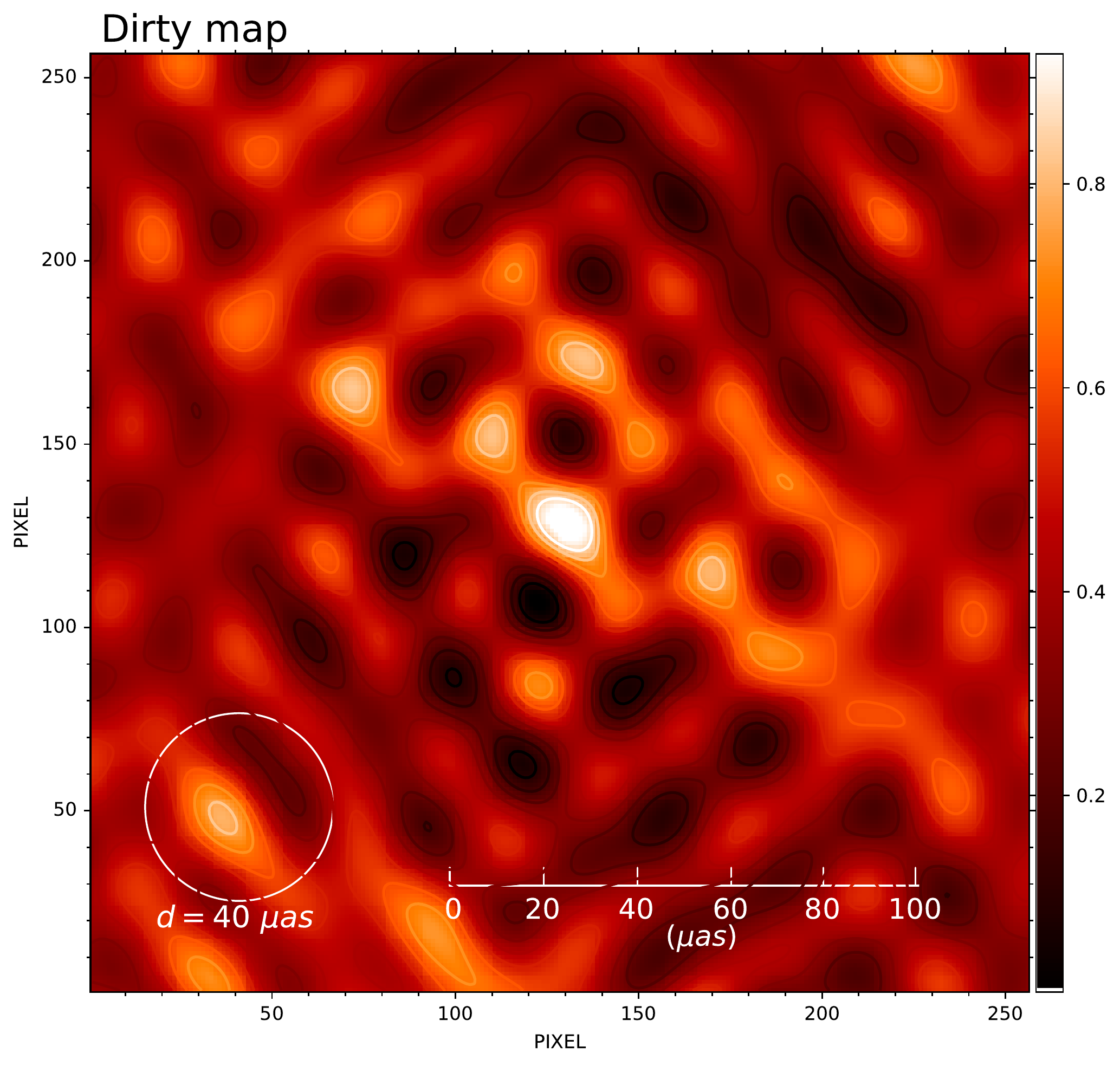}{0.44\textwidth}{}\fig{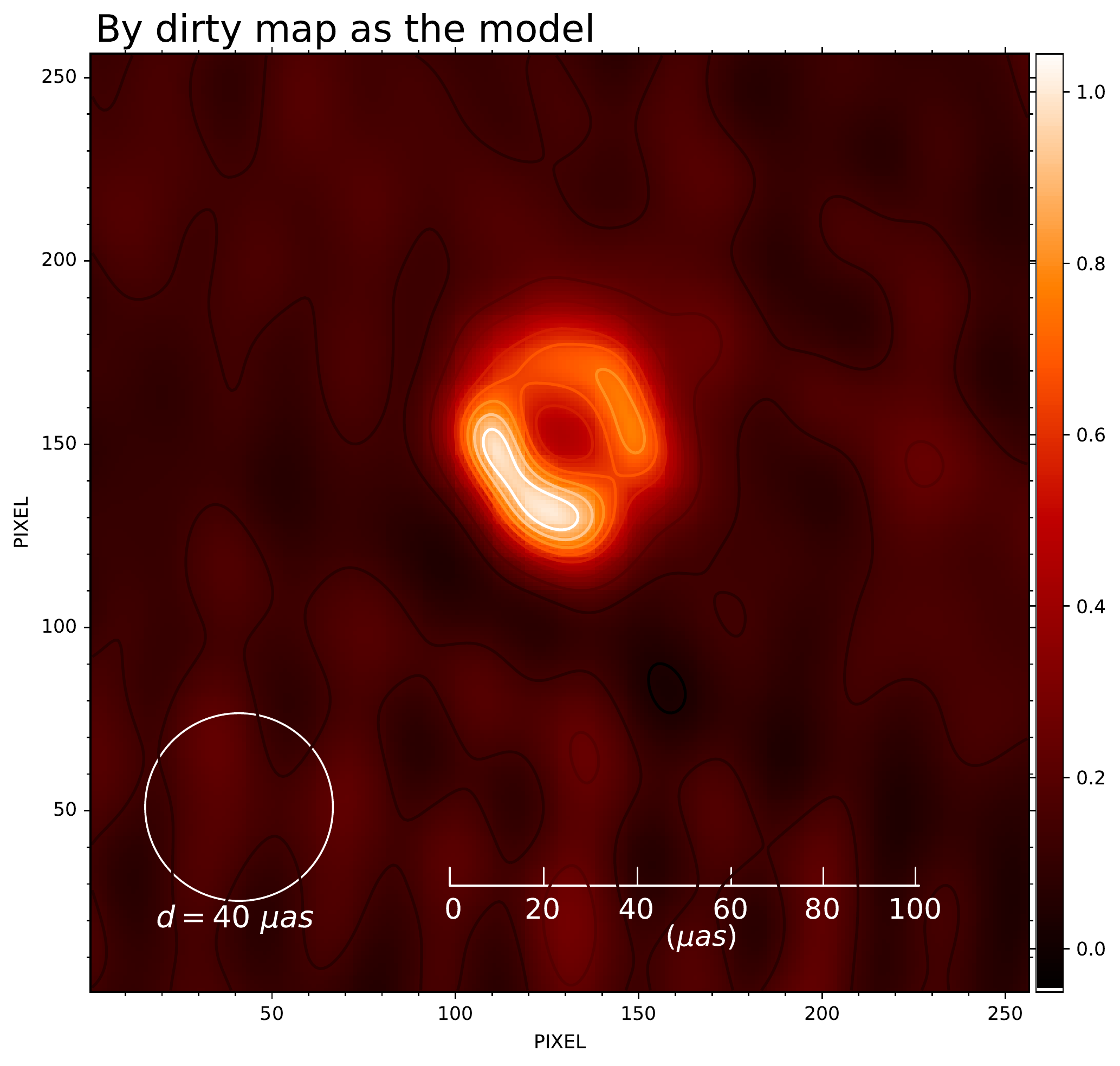}{0.44\textwidth}{}}
\caption{
Ring image obtained using the dirty map.
The left panels show the dirty map of the first day data from the EHT observation.\label{Fig:dmap}
The right panels show the image resulting from self-calibration of the central part of the dirty map as an image model.\label{Fig:ringbydmap} 
The upper panels show a larger area of $500~\mu\rm as$ width, and the lower panels show the enlargements of the central part of the images.
}
\label{temp2}
\end{figure}

We, however, try that here and show what happens.
The left-side panels of Figure~\ref{temp2} show the dirty maps from the data of the first day observations by EHT.
To obtain these maps, we applied
the phase-only calibration by the self-calibration technique using a point source as the image model.
We can see a ring-like structure at the center, and many images which look like the ghosts of this central ring-like structure.

\begin{figure}[p]
\rotatebox{90}{
\begin{minipage}{\textheight}\centering
\epsscale{1.15}
\vspace{15mm}
\plotone{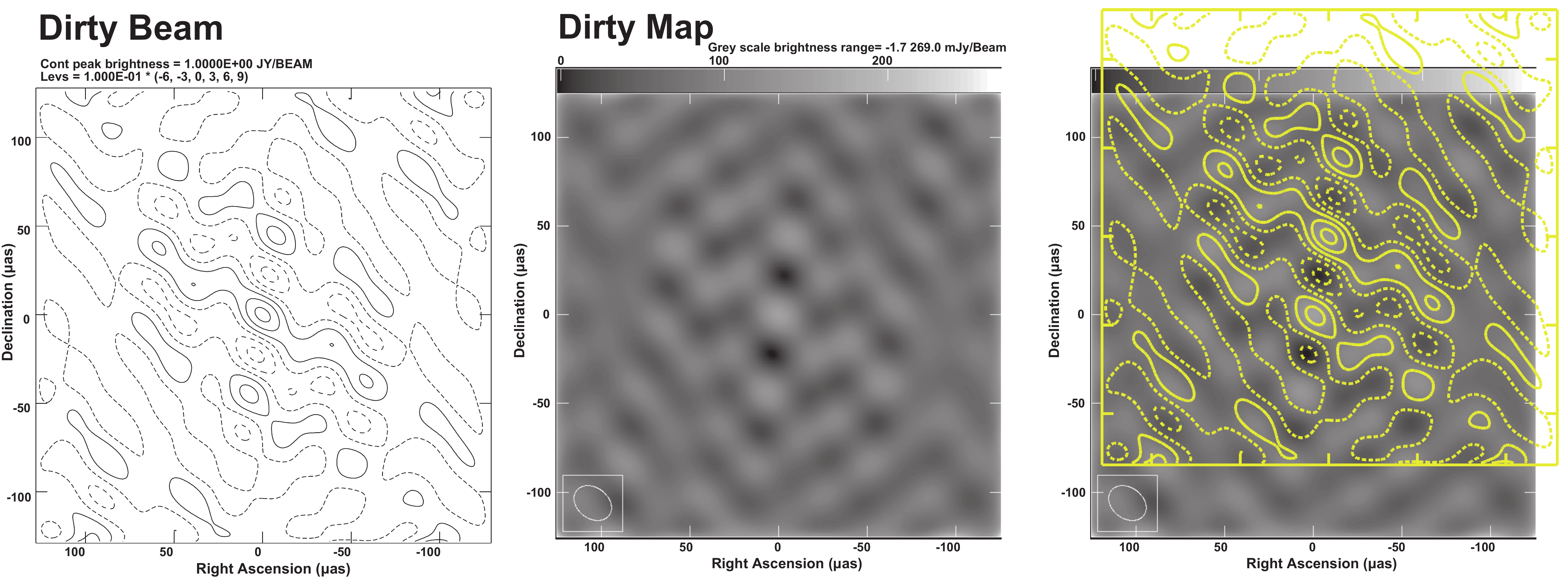}
\caption{Comparison of the dirty beam (PSF) and the dirty map. 
The left panel shows the dirty beam (PSF) and the center panel shows the dirty map. 
The right panel shows a case where the two images were adjusted to fit well together and overlap.
The yellow contours show the dirty beam, and the grayscale show the dirty map. 
The dirty beam (PSF) here is the same as shown in Figure \ref{Fig:uvd}, but the number of contours has been reduced.
The dirty map here is the same as in the left panels of Figure \ref{temp2}, but the grayscale shows the intensity inversion image.}
\label{Fig:ovdbdm}
\end{minipage}}
\end{figure}

This is not a blurred intrinsic image of M\, 87, but a strong reflection of 
the substructure in the dirty beam (PSF) of the EHT array.
Figure~\ref{Fig:ovdbdm} shows the dirty map and the dirty beam. 
Not surprisingly, they agree rather well, and that means the central
ring-like structure we showed in Figure \ref{Fig:ovdbdm}
is just the reflection of the shape of the dirty beam and not a physical reality. 
If we were to believe there should be a ring, it would be very natural to select the partial image from what looks like a ring in the dirty map. 
If we do so, what will happen? To answer this question, we tried the calibration of phase and amplitude using this ring-like structure at the center as the image model.
We calculated the amount of calibration of the amplitude and phase by the self-calibration method.
After the calibration is applied, we made an image using the CLEAN algorithm
assuming that the source is single and compact. 
Here, the CLEAN subtraction area is limited by a circle with $30~\mu\rm as$ radii centered at ($+2~\mu\rm as$, $+22~\mu\rm as$) from the map center, which is the BOX setting as used in the EHTC-DIFMAP pipeline (Section~\ref{Sec:pipeline}).

The obtained ring image is shown in the right panel of Figure \ref{Fig:ringbydmap}. The size of the ring is close to that of the EHTC ring. What we also would like to emphasize here is that in order to create a ring structure, we need a narrow BOX setup. Without a narrow BOX, a ring structure cannot be created. If we change the position or size of the BOX, the ring deforms. When we use BOX offset setting of $22~\mu \rm as$ from the center, we get a ring very similar to the EHTC ring. In other words, the shape strongly depends on the location of the imaging region. As we will show in Section ~\ref{Sec:simData}, the EHTC BOX setting limits the FOV so narrowly that it can produce the EHTC \FTMUAS~ring shape even from simulated data that does not contain the ring shape.

Close inspection reveals that the ring-like structure of the dirty map is at an offset position from the center.
Since the structure of the PSF always shows center symmetry, we can suppose that the offset ring-like structure is not due to the PSF, but really exists at the offset position.
However, the shape of Figure~\ref{Fig:mB} shows that this is not the case.
Figure~\ref{Fig:mB} shows the structure of the dirty beam shape after deconvolution from only the brightest point at the center of the dirty map. There is no ring-like structure here.
We can see that the ring-like structure of the dirty map (Figure~\ref{temp2}) is created by convolution of the dirty beam to the brightest point in the center.
The ring-like structure is offset because the true source image inherent in the data does indeed exhibit non-center symmetry, but it is not an offset ring, but a core-knot structure as shown in Figure~\ref{Fig:sA3}.

\subsection{Ring from the simulated data of different structures\label{Sec:simData}}

In this section, we demonstrate that, with the \UVC coverage of EHT for M\,87, one could ``observe'' a ring even if the physical sources have different structures.
~As the physical source structure, we consider 
a single point with flux density of 1~Jy at the center.
 We made the simulation visibility data by applying the Fourier transformation to the single point image (We used UVMOD, a task in AIPS.).
 The \UVC coverage is exactly the same as that of EHT for M\,87 observations.
As for noise,
we consider two cases, one with relatively large noise (case 1) and the other with no noise (case 2). For case 1, we added thermal noise. The noise level is proportional to the weight of  each data noted in original FITS public data.  The signal-to-noise ratio (S/N) is 0.01 on average.
We then perform experiments of calibration and imaging. For calibration, we obtain solutions from self-calibration with incorrect image models that are different from the true structures of the source, then apply the solutions to the data. We inspect what kinds of images appear from CLEAN imaging.
For the self-calibration, we assume models of two incorrect images.
One is the EHTC ring image (model A) and the other is a pair of two points separated by $45 ~\mu \rm as$ (model B). Model B is a pair of points of 0.5~Jy located at ($0~\mu\rm as$,~$0~\mu\rm as$) and ($0~\mu\rm as$,~$+45~\mu\rm as$) respectively. 
The locations of these two points roughly correspond to the positions of the main peak and the first sidelobe peak in dirty beam.
We get calibration solutions of amplitude and phase from the self-calibrations (We used CALIB, a task in AIPS.).

\begin{figure}[H]
\includegraphics[height=5.5cm]{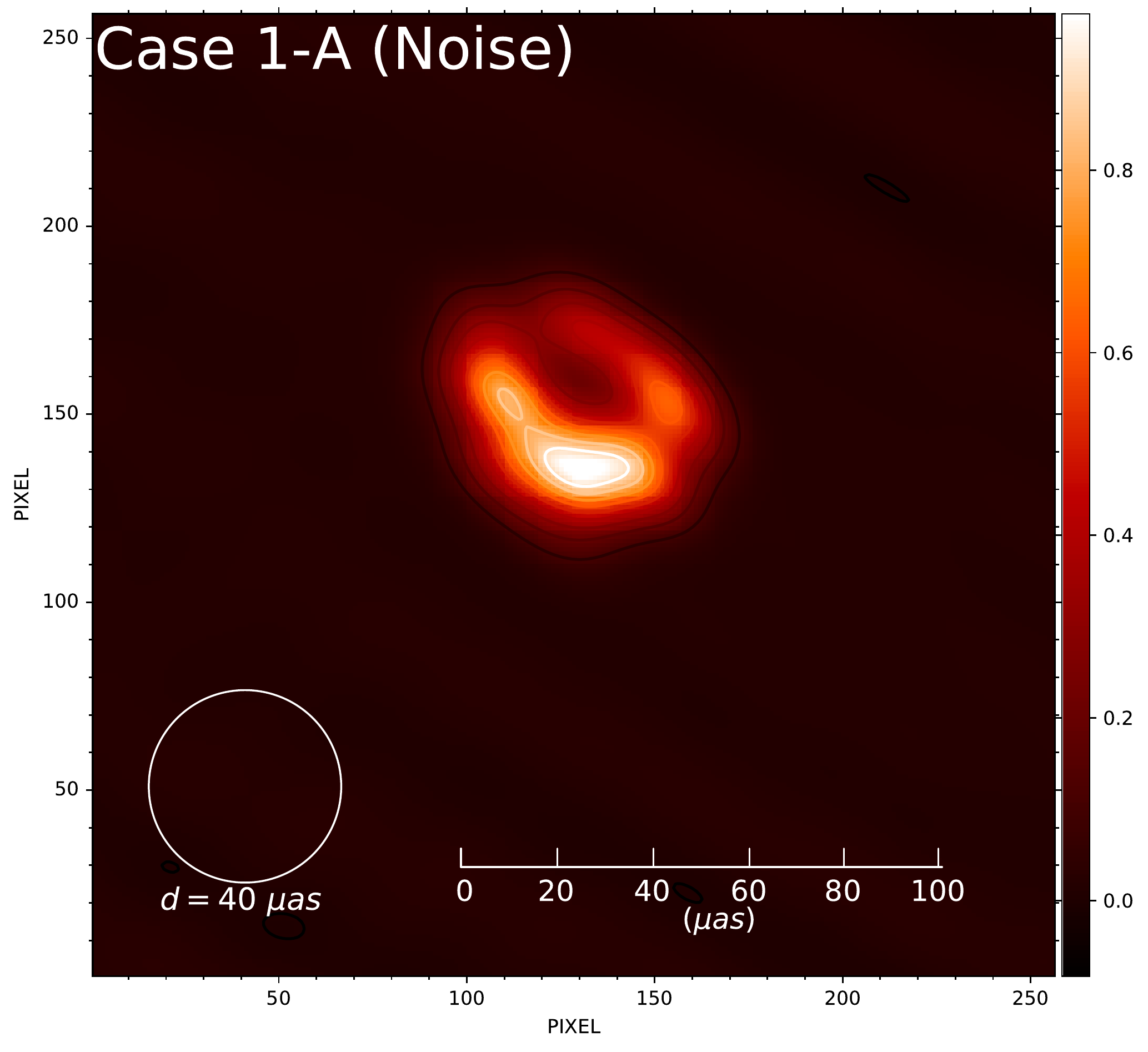}
\includegraphics[height=5.5cm]{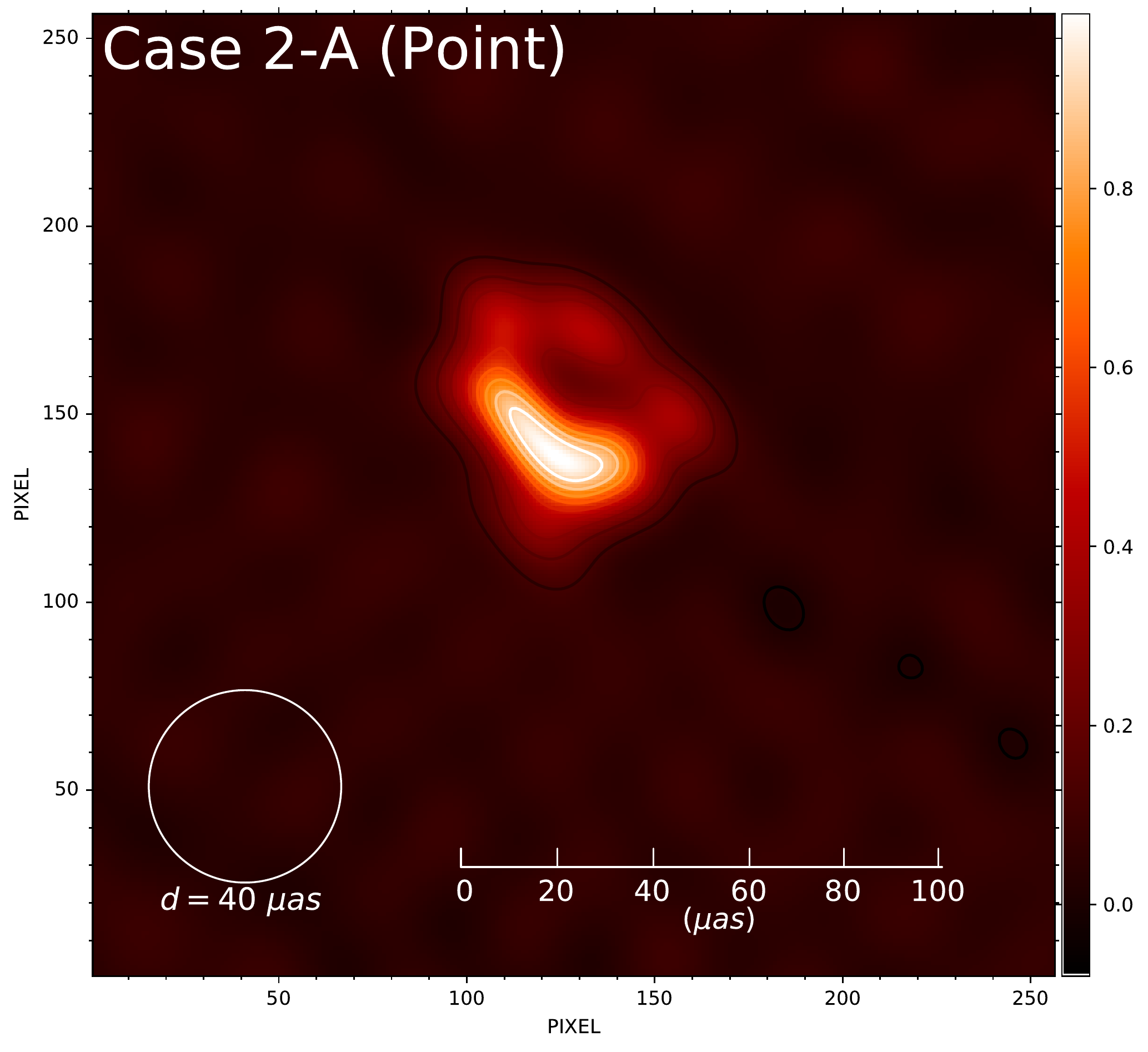}
\includegraphics[height=5.5cm]{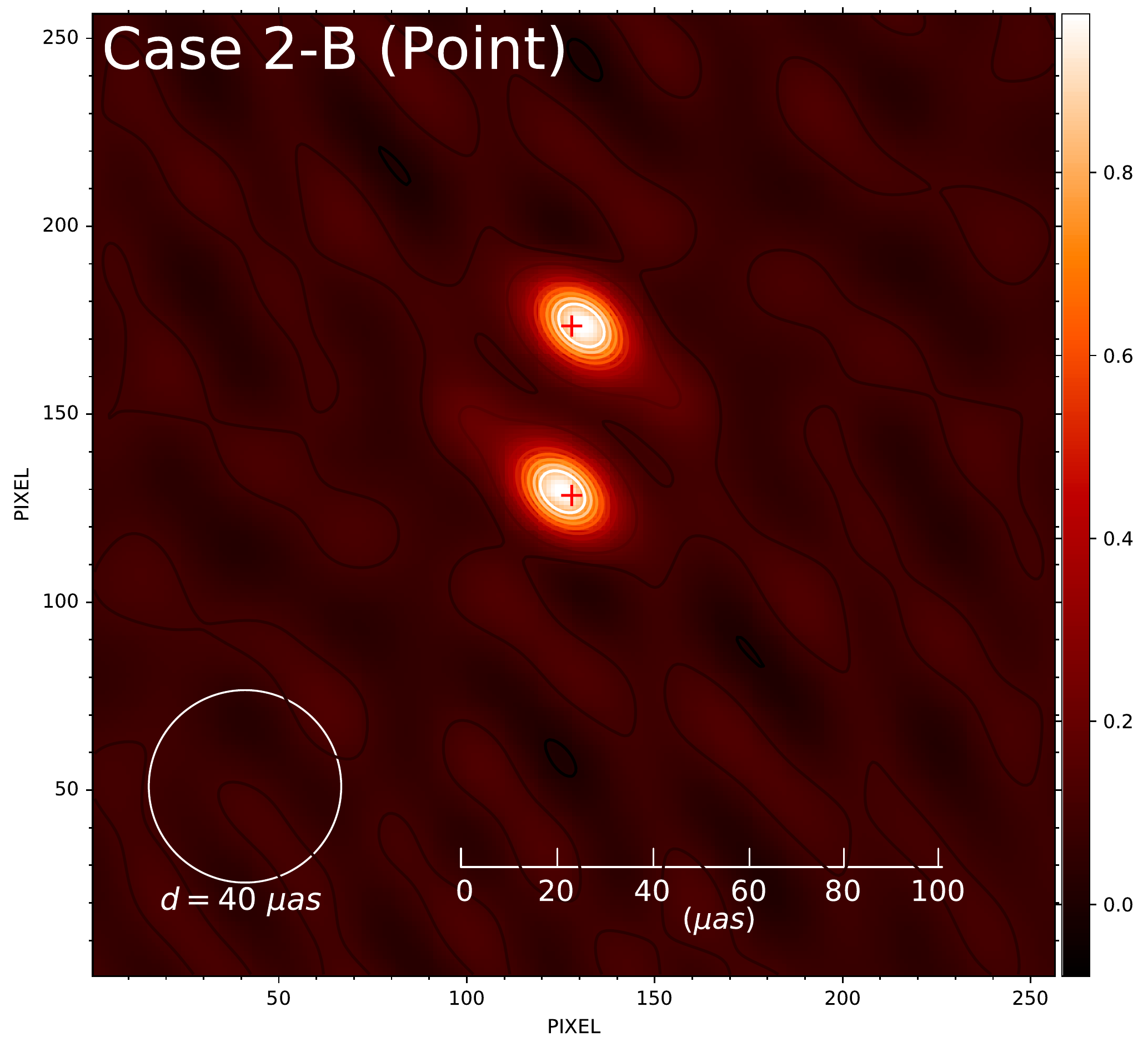}
\caption{
Resultant images from the simulated data sets. 
Case 1-A (left panel) shows the ring shape created from nearly noise data set, after applying the solution by self-calibration using the EHTC ring as the image model.\label{fig:ringbynoise} 
Case 2-A (middle panel) shows the ring shape created from a single point data set with infinite S/R (no noise), after applying the solution by self-calibration using the EHTC ring as the image model.\label{fig:ringby1pt}
Case 2-B (right panel) is a two-point image created by applying the solution by self-calibration with two points as image models to a single point data (no noise data) with infinite S/N. 
The two red crosses indicate the positions of the points in the model image used in self-calibration.\label{fig:2ptby2pt}
}\label{RFDD}
\end{figure}

By using IMAGR, a task in AIPS, we performed the CLEAN imagings. 
We limit the CLEAN subtraction area by a narrow BOX setting; a circle 
with $30~\mu\rm as $ radii centered at ($+2~\mu\rm as $, $+22~\mu\rm as $) from the map center, which is a mimic BOX setting as used in the EHTC-DIFMAP pipeline (Section~\ref{Sec:pipeline}).
%
The left panel in Figure~\ref{RFDD} shows the ring image obtained from the large noise data (case 1). The obtained image from CLEAN is identical to the model image A used in self-calibration.
When the S/N is very low, the CLEAN image after self-calibration can be identical to the model image used.
Therefore, the result of case 1 does not tell much.

The results of case 2 are surprising.
The middle panel in Figure~\ref{RFDD} shows the ring image obtained from no noise data. The obtained image from CLEAN is very close to the model image A used in self-calibration.
Even when there is no noise, if we start with the assumption that there is a ring in the self-calibration image model, with the \UVC coverage of EHT, we obtain a ring similar to that obtained by the EHTC.

 The right panel in Figure~\ref{RFDD} shows the result of using model image B for self-calibration. Though we used the data of a single point with no noise, we obtained a two-point image, which is very close to model B used in self-calibration.
 In addition, the appeared two points show small shifts from the positions in model B. The center point shifted towards the east, while the upper point shifted towards the west. These shifts seem to be the influence of the structure of the dirty beam.
The shift of the center point is along the ridge elongated from the main beam in the dirty beam, while the shift of the upper point is towards the peak position of the first sidelobe.


The results of these experiments show that the reproduced images are very close to the image model used in the self-calibration.  Such a result does not occur when there is  sufficient \UVC coverage, as in the case of VLA and ALMA. Unfortunately, the \UVC coverage of EHT is not sufficient, and it has a special bias that makes it prone to producing shapes (including rings) with a size of \FTMUAS. We have shown that the shape of the PSF (essentially the effect of the sampling bias that makes it easy to create \FTMUAS~ structures) has the power to create \FTMUAS~ring structures even from data with only one central point structure (centrosymmetry).  However, the most powerful factor is not data correction by self-calibration with wrong model images. 
 Here and in Section~\ref{Sec:byDmap}, we showed that the very narrow FOV setting of the BOX has great power to create a \FTMUAS~ring structure. In Section ~\ref{Sec:pipeline}, we will discuss the setting of BOX in the EHT-DIFMAP pipeline.

\subsection{What the EHTC really did in their DIFMAP imaging process\label{Sec:pipeline}}
In this section, we investigate the data processing pipeline that the EHTC used for imaging the EHTC ring with DIFMAP 
\footnote{\url{https://github.com/eventhorizontelescope/2019-D01-02}}. 
Among the three methods the EHTC used for the imaging, DIFMAP is the closest to the usual procedure \footnote{\url{difmap/EHT_Difmap} in the above web page}. Other two are rather new, and difficult to study here. Since all three methods gave essentially the same image, we believe it is sufficient to analyze one of the three methods to find out why the ring structure was formed.
The EHTC-DIFMAP imaging team used the hybrid mapping technique to calibrate the data and obtained the image. 
The hybrid mapping method is the standard method of VLBI calibration and imaging.

Let us first summarize the standard procedure of VLBI data calibration and imaging.
The radio interferometer samples the spatial Fourier components of the brightness distribution of the observed source (often called visibility). Theoretically, the brightness distribution of the observed source can be obtained by collecting the samples and performing the inverse Fourier transform. However, there are actually two problems as follows.
(a) The sampled visibility data contain errors, so they are not the correct spatial Fourier components. Therefore, it is necessary to calibrate them. Self-calibration is a method of obtaining calibration solutions by using an assumed image. If the assumed image is not correct, the correct calibration solution cannot be obtained.
(b) The number of observed samples is limited. The inverse Fourier transform alone does not produce the correct image. The PSF does not become a point. Therefore, the PSF shape must be deconvolved. For this purpose, image processing is performed. As a method to obtain correct calibration solutions and to attain clear imaging results, hybrid mapping, which alternates between the above two tasks, is usually used in VLBI imaging. We use self-calibration for calculating the calibration solution using a tentative image model, and CLEAN for performing deconvolution of the scattered PSF shape.

Hybrid mapping is the following algorithm.
(a) Assume an image model for self-calibration. For the first one, a 1-point model is usually used.  
(b) Calculate tentative calibration solutions from self-calibration using the image model. 
(c) Calibrate visibility data using the tentative calibration solutions.
(d) Obtain the next tentative image using CLEAN from the calibrated visibility data.
CLEAN image is composed of point sources (CLEAN components). 
(e) Make the next image model for self-calibration by picking up reliable CLEAN components from the image.
(f) Go back to (b) and repeat the iteration from (b) to (e) until a satisfactory image is obtained.

The visibility (spatial Fourier component) is a complex quantity and so it has amplitude and phase. It is safe first to repeat self-calibrations only for phase solutions until a nearly satisfactory image is obtained and to perform self-calibration for amplitude and phase by using that image. This is because self-calibration for amplitude tends to give a solution which is too close to the model image used in self-calibration~\citep{CornwellandFomalont1999}. 

However, the repetition of amplitude self-calibration is often performed under the careful check of the practical quality of the data, and then, in many cases, good calibration solutions can be obtained.\\

In addition, the BOX technique is an auxiliary means used for imaging with CLEAN.
(It is essentially unrelated to the hybrid mapping method.)
We can limit the area of CLEAN subtraction intentionally by the BOX setting. The technique may give us a good image, despite an incomplete calibration. As already mentioned, we used the BOX technique in our analysis. Also the EHTC-DIFMAP team did.\\

Now we investigate the EHTC-DIFMAP procedure. The EHTC-DIFMAP imaging team performed the self-calibration only for phase solutions with a single point source model, following the general procedure of the hybrid mapping method.
However, after the first phase self-calibration, they used the BOX technique with a very narrow BOX\footnote{\url{BOXCircMask_r30_x-0.002_y0.022.win} in the web} in the first imaging trial. 
The CLEAN subtraction area was restricted to within the circle of $60~\mu \rm as$ diameter specified by the BOX.
This means that from the beginning of their data analysis, the EHTC-DIFMAP team assumed that the image is a very compact single one, and excluded other possibilities. 

\begin{figure}[H]
\begin{center} 
\epsscale{0.7}
\plotone{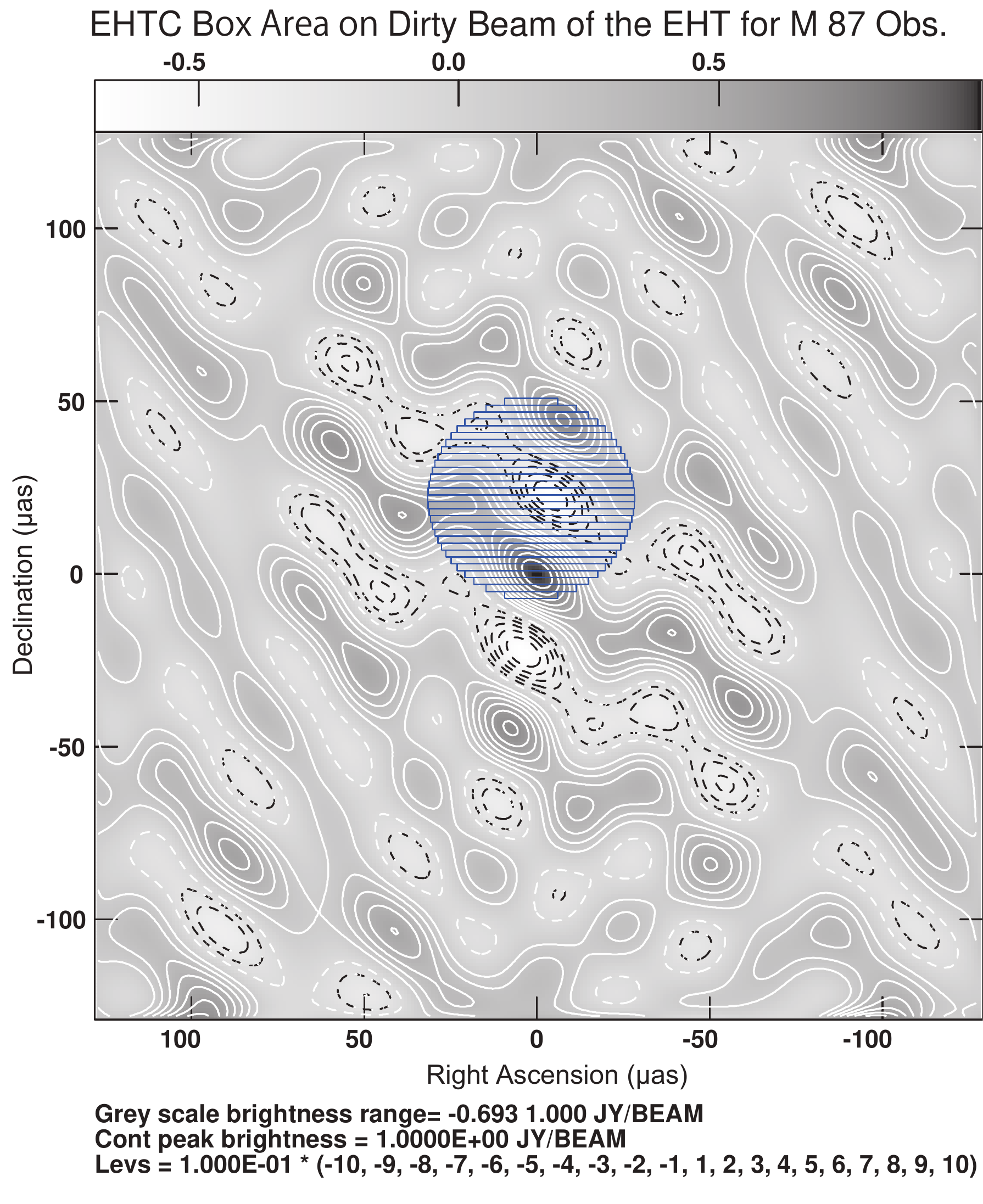}
\end{center} 
\caption{The size and position of the BOX used in their DIFMAP pipeline (blue shaded area). We overlaid the BOX on the dirty beam (PSF) which is already shown in Figure \ref{Fig:uvd}.
We align the two figures at their phase centers.\label{Fig:the EHTCBOX}}
\end{figure}

The BOX is roughly a circle of diameter $60~\mu\rm as$ composed of 30 small rectangles. By comparing the BOX shape and the dirty beam we found that the BOX covers the main beam and the first sidelobe, but leaves out the second and other sidelobes.  
The BOX is not located at the phase center but offset by $+22~\mu\rm as$ in the y-direction ($\delta$ direction).  This offset of $22~\mu\rm as$ coincides with the radius of the EHTC ring.  Figure~\ref{Fig:the EHTCBOX} shows the EHTC BOX position on the dirty beam (PSF) aligned by their phase centers.
The area of the EHTC BOX covers (1) the main beam peak, (2) the ridge extending to the left (east) from the main beam peak, (3) a part of the first sidelobe located north of the main beam, and (4) an area of negative strength near the center of the box area.  The structure of the dirty beam within the EHTC BOX is close to the shape of the EHTC $\sim 40~\mu\rm as$ ring. 

 The PSF structure within the BOX explains not only the diameter of the EHTC ring, which is \FTMUAS, but also the asymmetric structure of the EHTC ring. The EHTC ring has an asymmetric structure with a bright south side. The brighter south side corresponds to the main beam (the brightest in the box), while the darker north side of the ring corresponds to the first sidelobe in the north (the less bright peak in the box).

If such a narrow BOX with the offset is used from the beginning to the end of the hybrid mapping process, 
the effect of the dirty beam can be enhanced as we showed in Section \ref{Sec:byDmap}. 
It is quite clear that the EHTC \FTMUAS~ring is the result of such enhancement of \FTMUAS ~substructures in PSF.

\begin{figure}[H]
\gridline{\fig{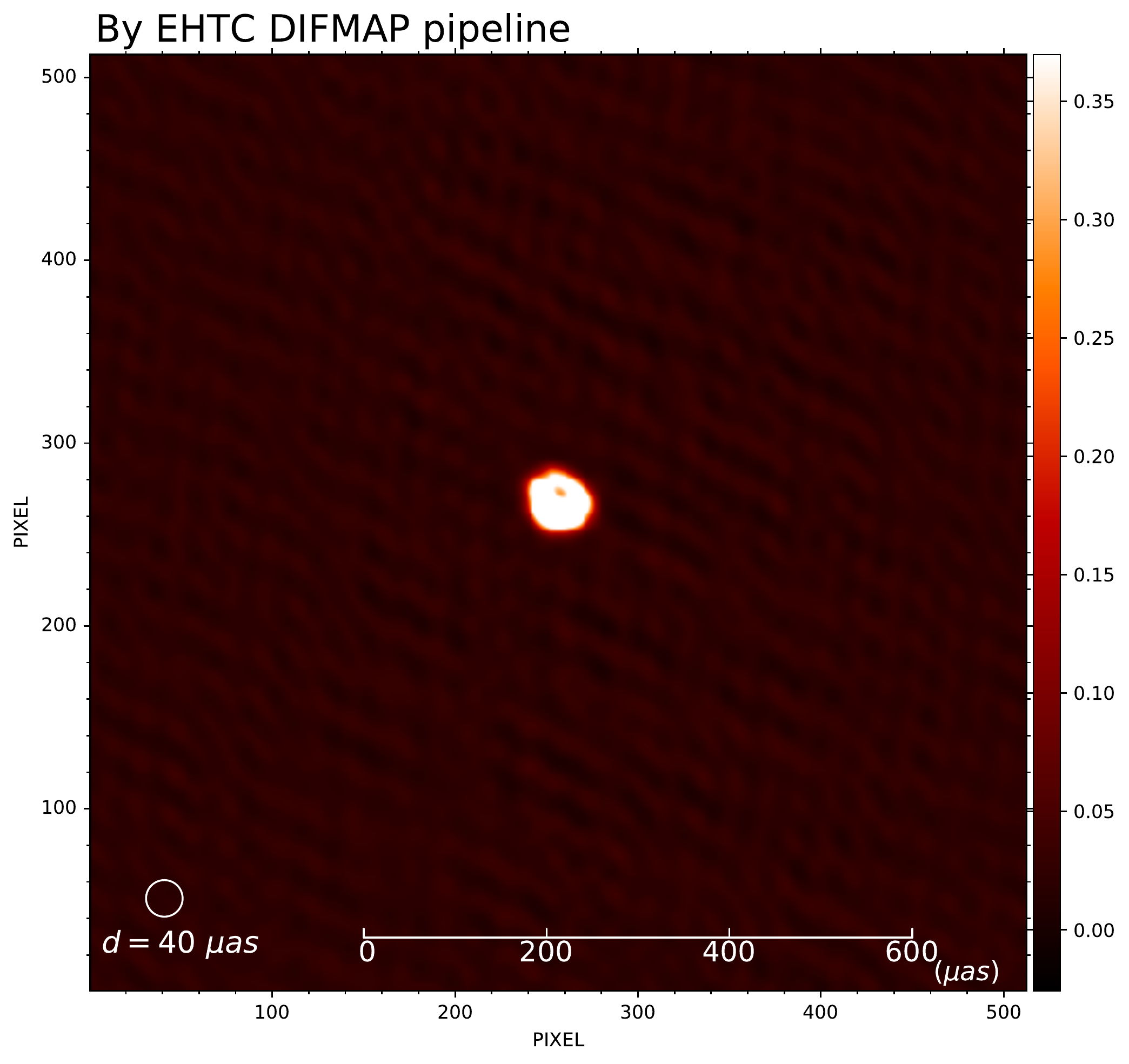}{0.5\textwidth}{}\fig{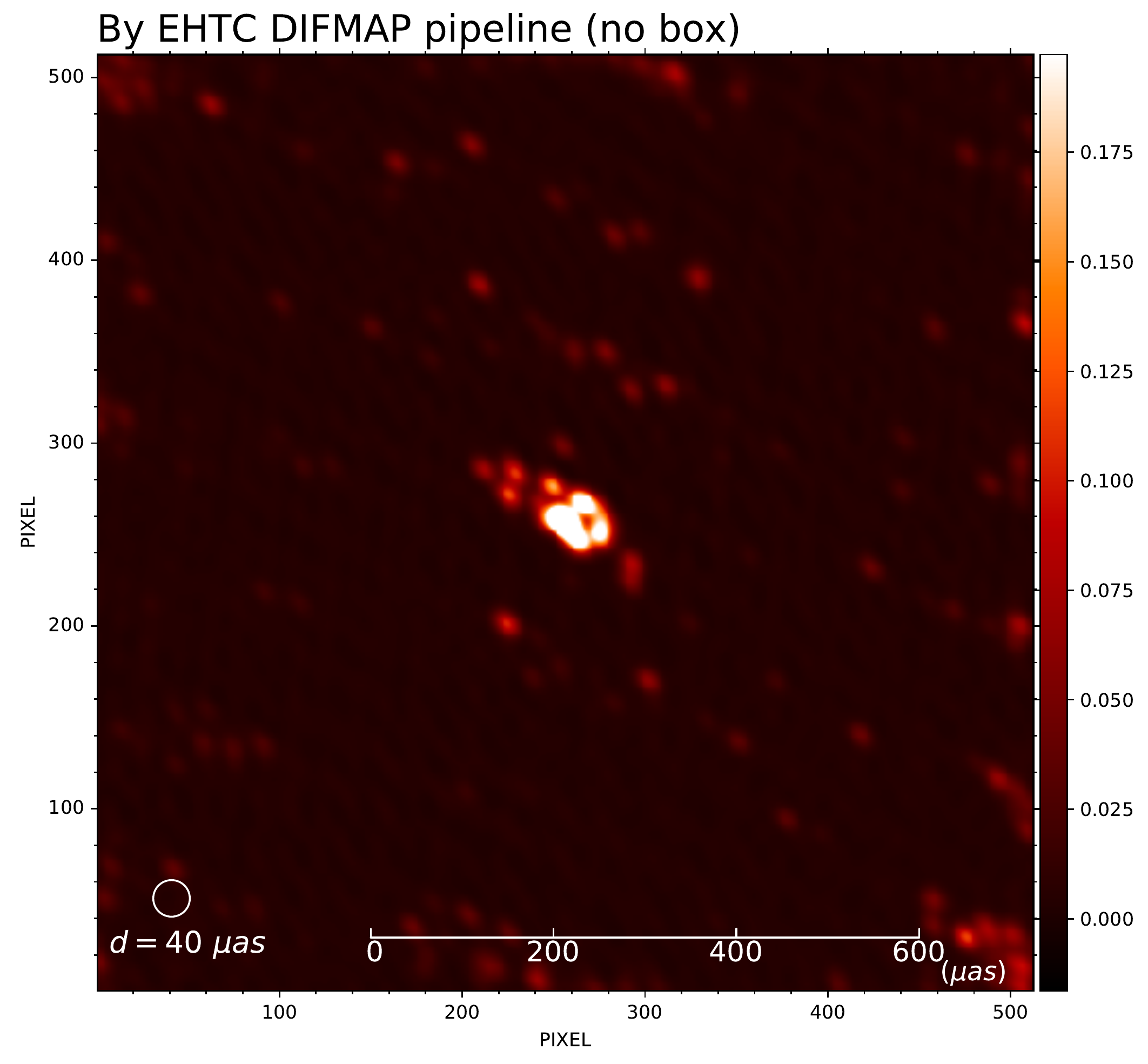}{0.5\textwidth}{}}
\gridline{\fig{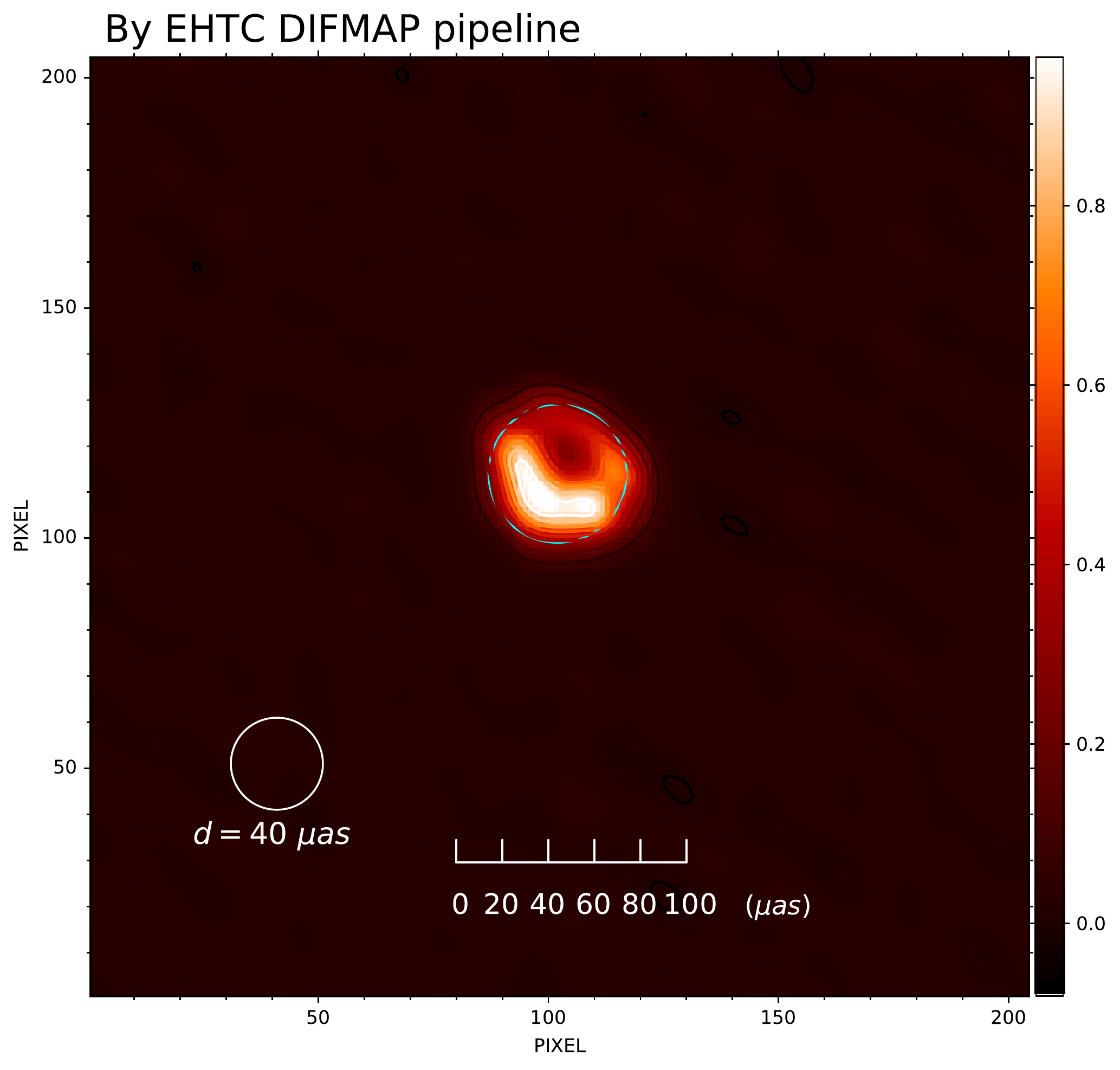}{0.5\textwidth}{}\fig{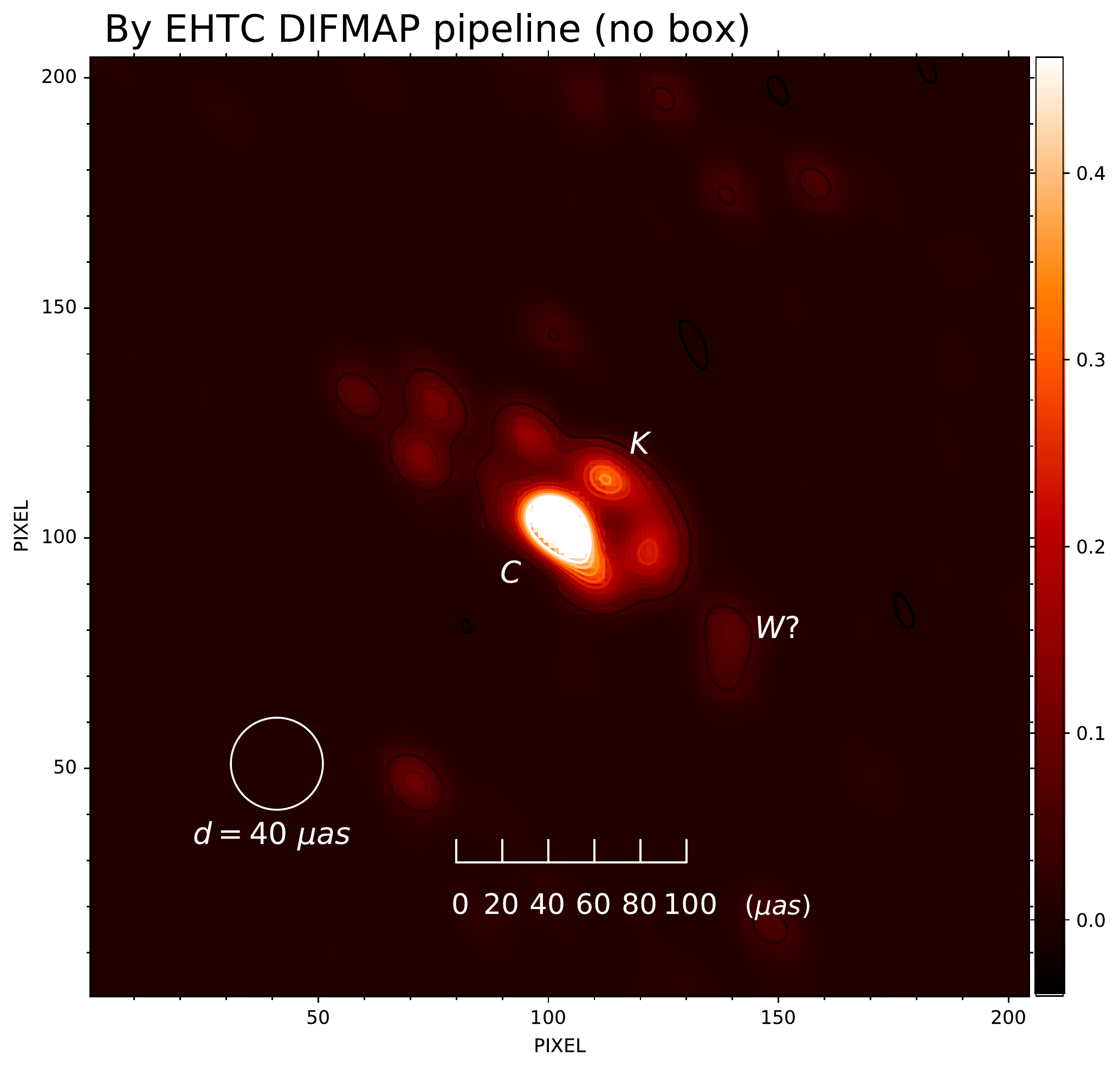}{0.5\textwidth}{}}
\vspace{-0.5cm}
\caption{
Resultant images from the EHTC-DIFMAP pipeline. 
The upper panels show the entire field of view with the default setting of the EHTC-DIFMAP pipeline (1 mas square). 
The lower panels show the enlargement of the central part of the upper images. 
The left panels show the result of running the EHTC-DIFMAP pipeline with the default parameter settings. The light blue circle shows the location and size of the BOX. The ring-shaped image appears in the BOX area. 
The right panels show the resulting image when the EHTC-DIFMAP pipeline is run without the BOX setting. 
In the enlarged image (bottom right), we can see the emission peaks corresponding to the three components we found: core (C), knots (K), and west component (W). Here, we used the data named
SR1\_M87\_2017\_095\_hi\_hops\_netcal\_StokesI.uvfits.
\label{fig:the EHTCDIFMAP1}}
\end{figure}

We actually ran the EHTC-DIFMAP pipeline to investigate its behavior and performance. The left panel of Figure~\ref{fig:the EHTCDIFMAP1} shows the resultant ring image from running the EHTC-DIFMAP pipeline with the default parameter settings.  
On the other hand, the right panel of Figure~\ref{fig:the EHTCDIFMAP1} shows the image obtained by removing the BOX setting (Figure~\ref{Fig:the EHTCBOX}) from the EHTC-DIFMAP pipeline.  The ring in the BOX setting region has been destroyed and the brightness distribution has been moved to outside the BOX setting region.  
The new brightness distribution shows a core-knot structure in the center, similar to the results of our analysis. 
In addition, several emission peaks appear in the 
$PA= 45\DEG$ and $PA= 225 \DEG$ directions from the center. 
These are probably sidelobes, but it is very interesting that one emission peak in the $PA= 225\DEG$ direction corresponds to the feature W we found in our analysis. This result implies that the very narrow BOX setting of the EHTC-DIFMAP pipeline is the main factor that creates the EHTC ring image.\\

\begin{figure}[H]
\gridline{\fig{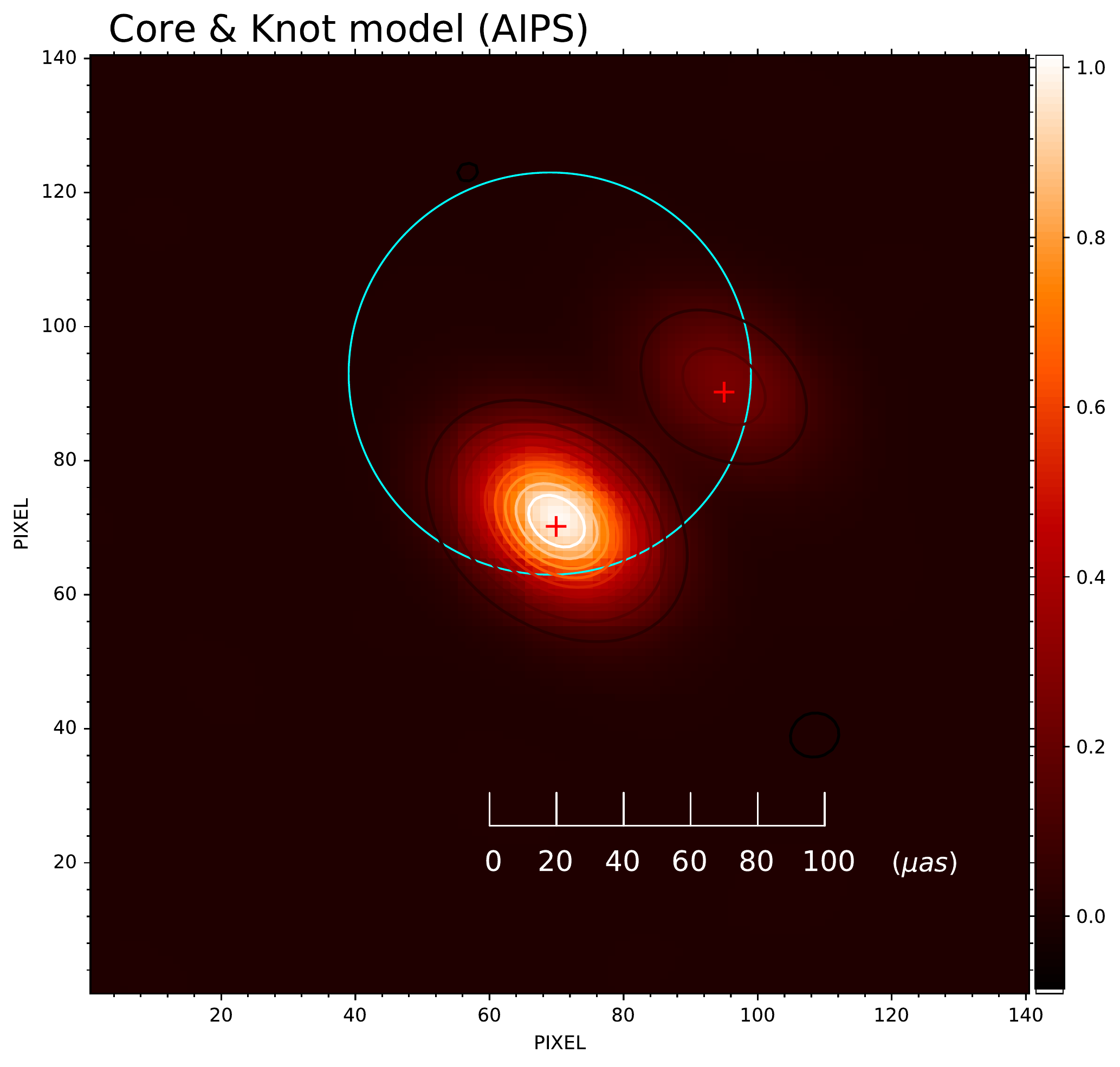}{0.3\textwidth}{}\fig{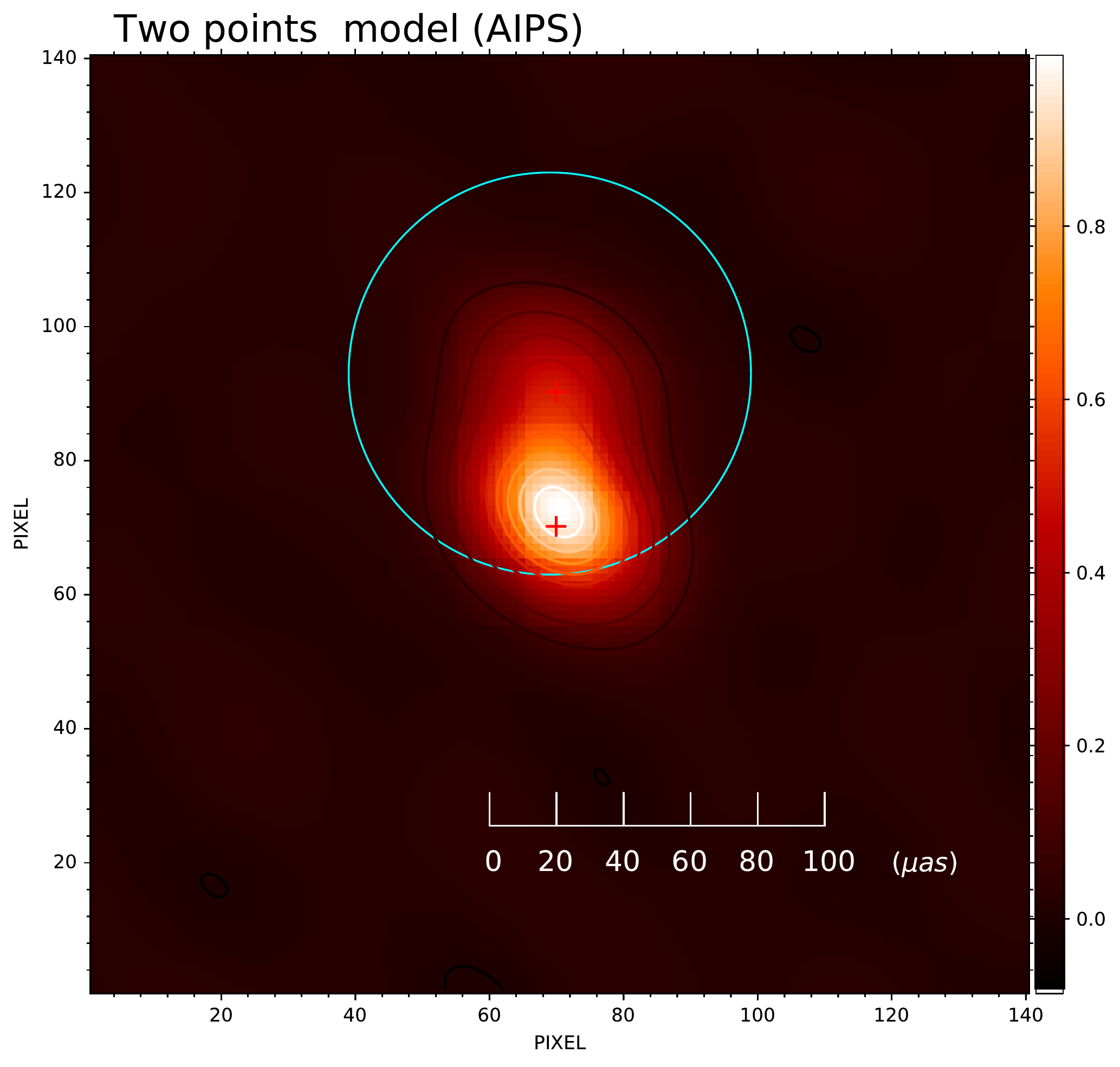}{0.3\textwidth}{}\fig{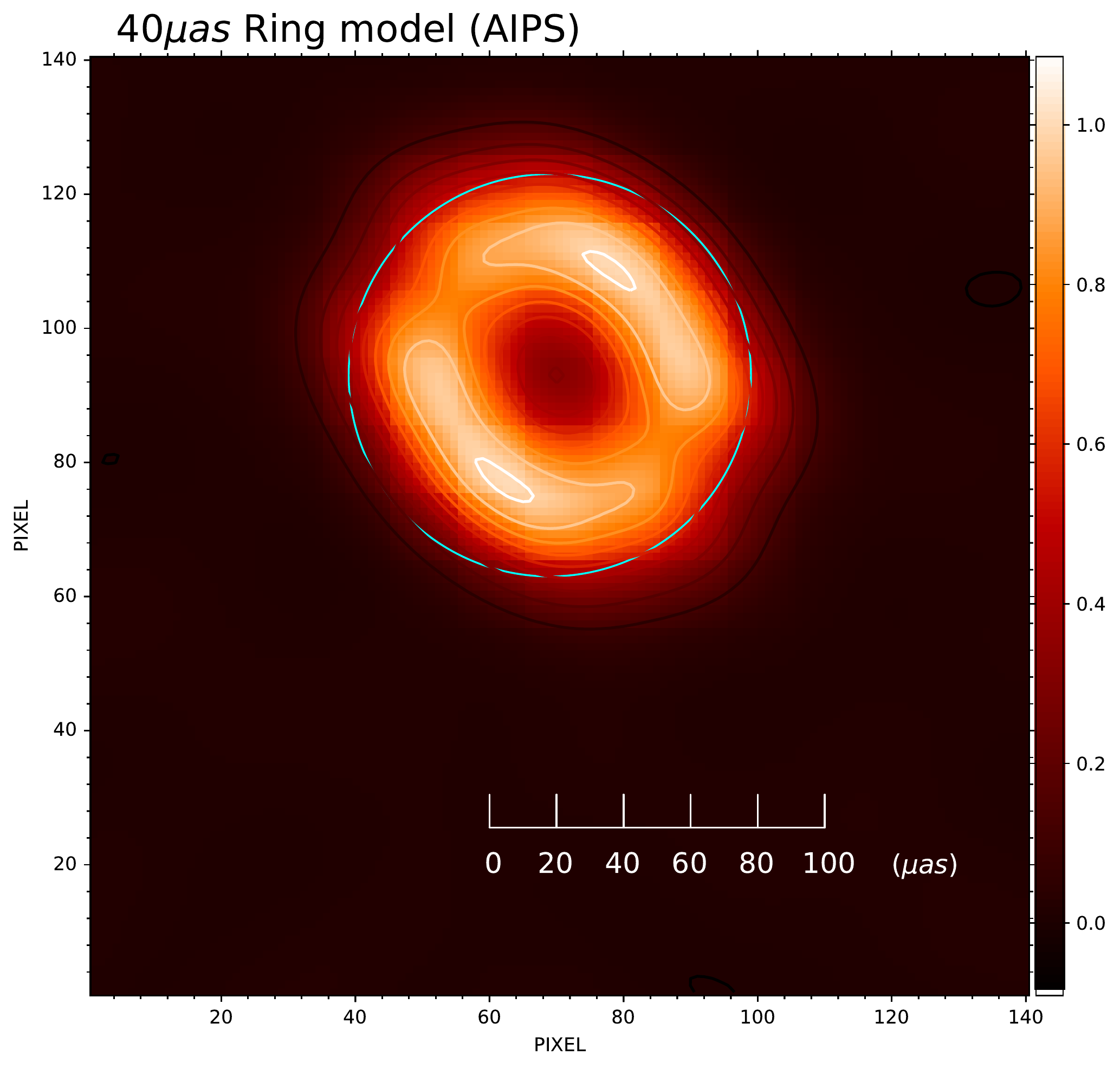}{0.3\textwidth}{}}
\gridline{\fig{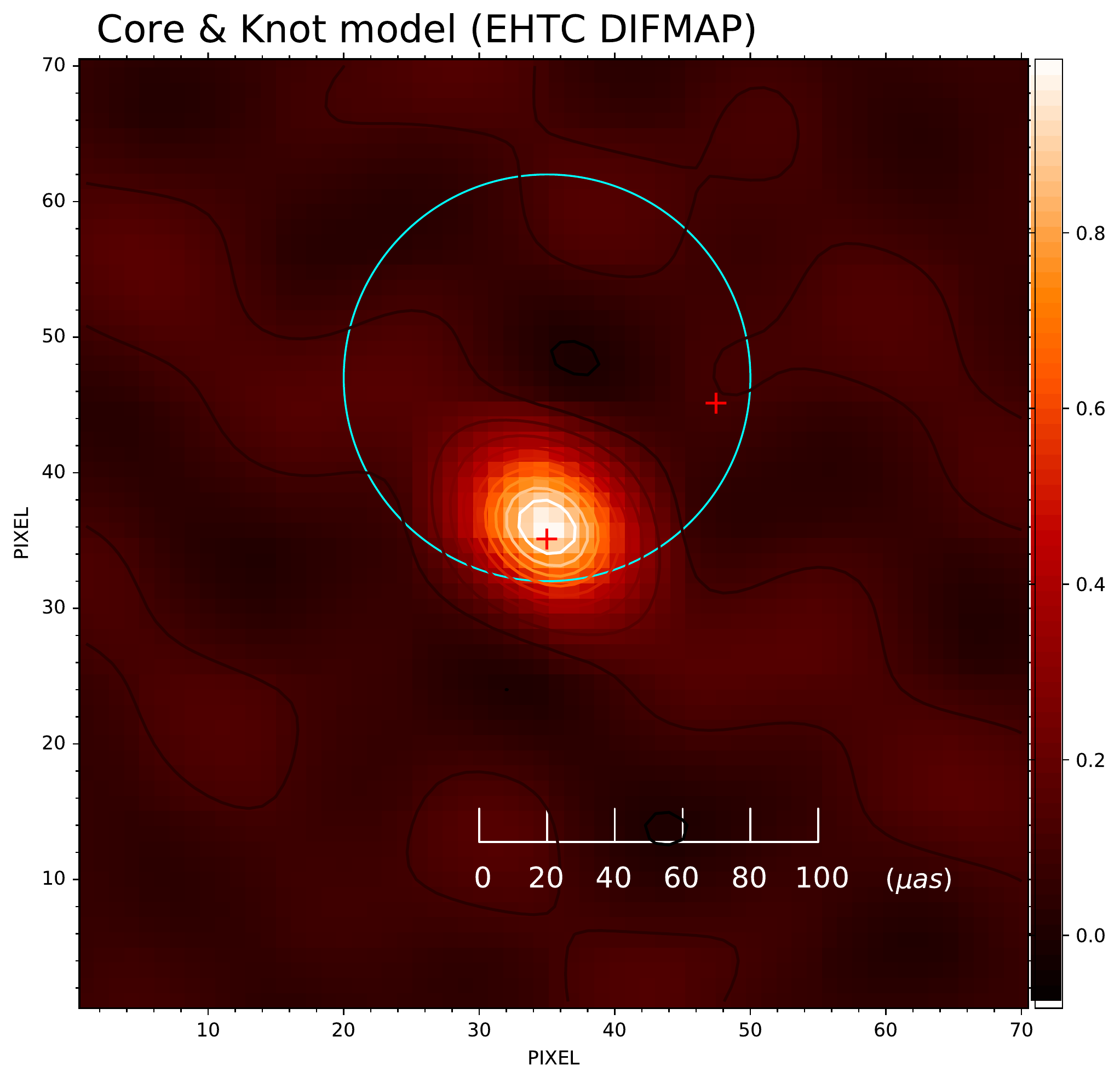}{0.3\textwidth}{}\fig{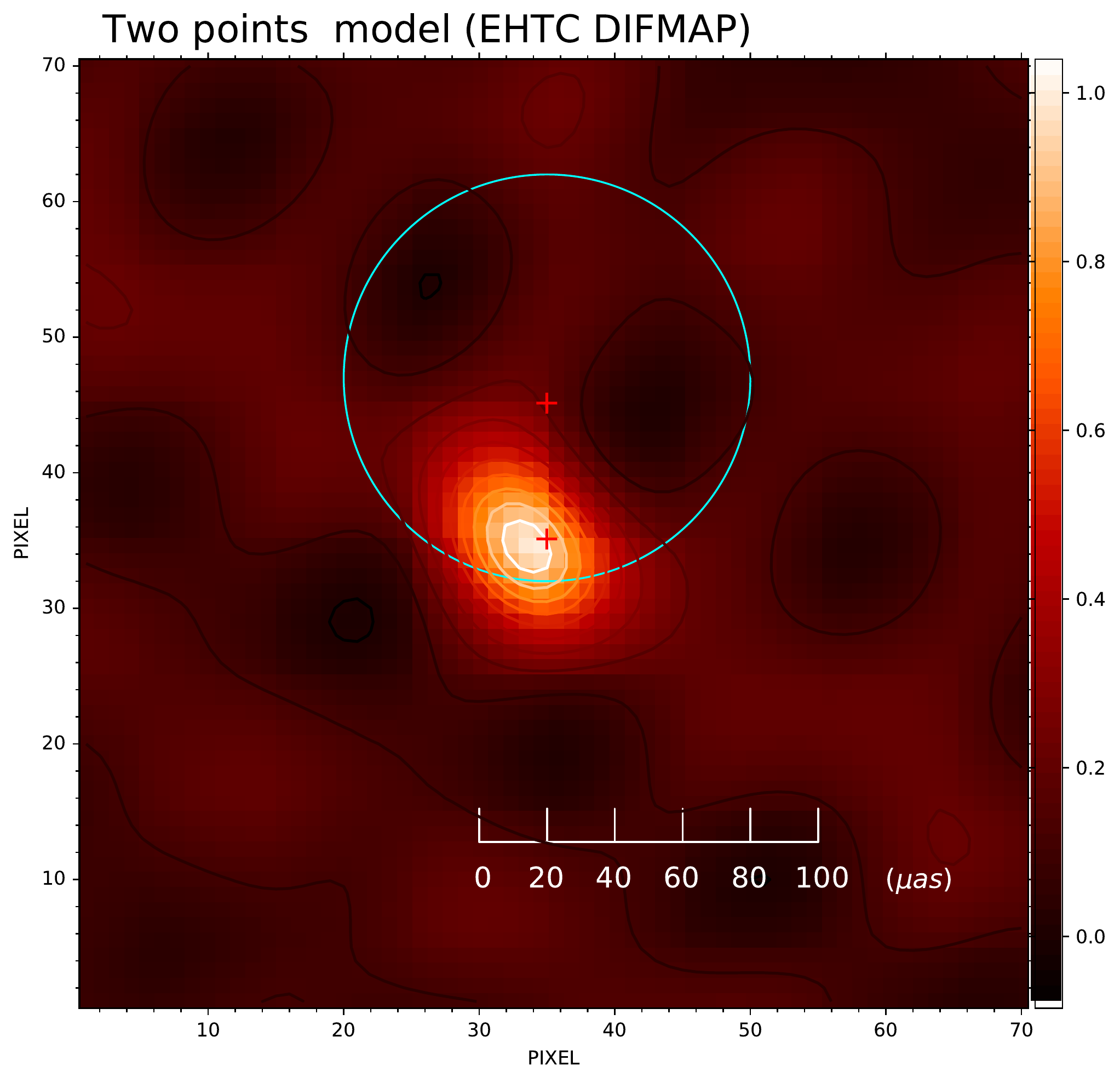}{0.3\textwidth}{}\fig{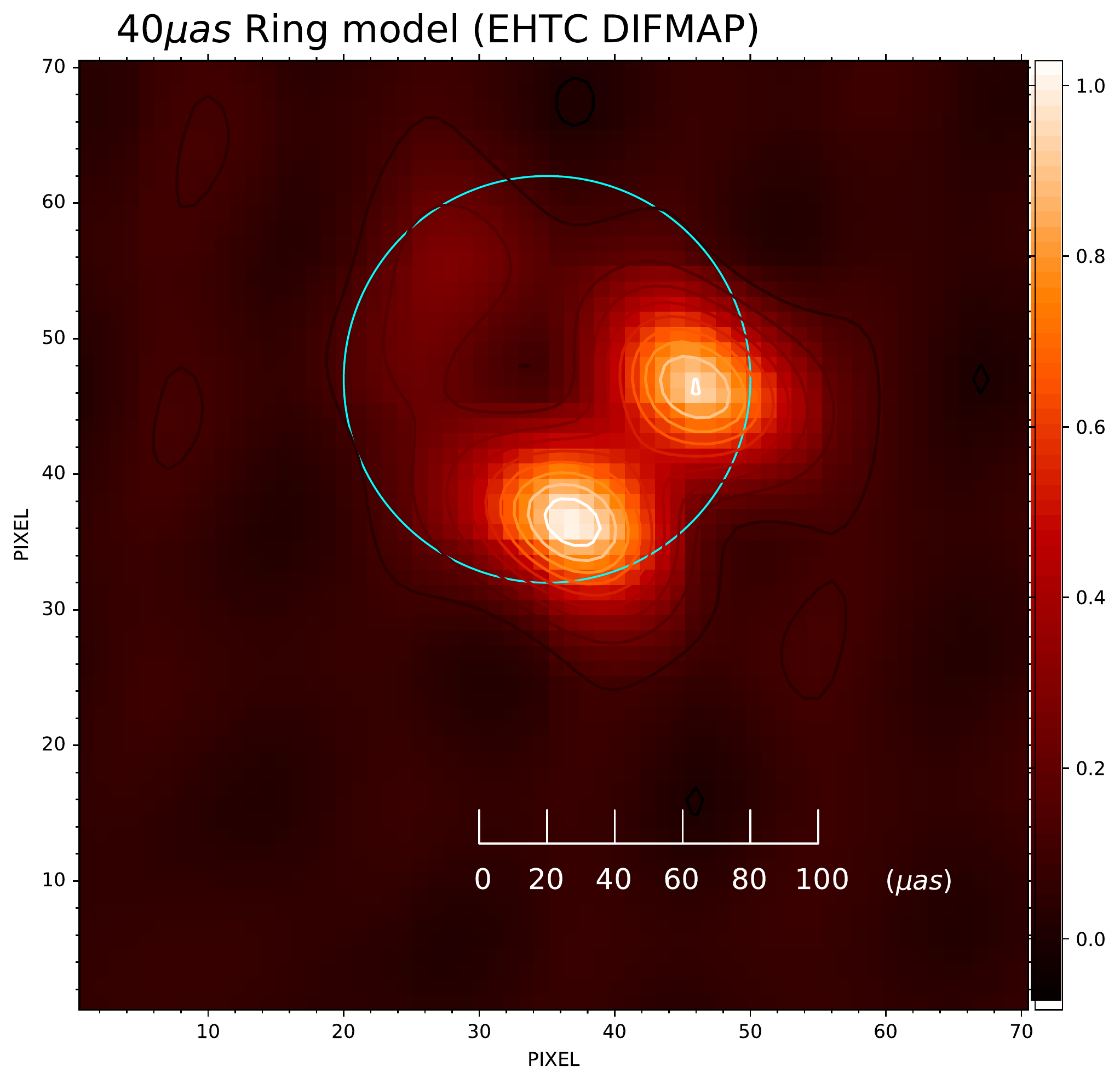}{0.3\textwidth}{}}
\epsscale{1.15}
\caption{
Resulting images of simulation data. 
The upper panels show the imaging results when using IMAGR in AIPS. The lower panels show the imaging results when using the EHTC-DIFMAP pipeline with default parameter settings. The same BOX setting is used for IMAGR as well as the EHTC-DIFMAP pipeline. The light blue circles show the position and size of the BOX setting. 
The three results are presented. 
A core knot model (left panel), 
a two-point model (center panel), and 
a $\sim 40~\mu\rm as$ ring model (right panel). 
The red crosses on the left and in the center indicate the positions of the emission points placed during the simulation data preparation.
\label{fig:the EHTCDIFMAP2}}
\end{figure}

We also created simulated data and applied the EHTC-DIFMAP pipeline to it to investigate what kind of images are obtained.  The purpose of this is to check if the EHTC-DIFMAP pipeline always calibrates the data correctly and reproduces the true picture of the data.  
We created simulated data with the task UVMOD in AIPS and acquired images with both the task IMAGR in AIPS and the EHTC-DIFMAP pipeline for comparison.
Figure~\ref{fig:the EHTCDIFMAP2} shows examples of the simulation results. 
Three cases are shown. 
(1) a core-knot model aligned in about $PA= -38.7\DEG$ direction (left panels), 
(2) a two-point model aligned North-South direction (center panels),
and 
(3) a $40~\mu\rm as$ diameter ring, but the ring width is $2~\mu\rm as$, different from that of the EHTC ring (right panels).

The top panels show images obtained from data simulated using the AIPS IMAGR task with the same box settings as the EHTC-DIFMAP pipeline. The IMAGR produced the expected image. 

The bottom panels of Figure~\ref{fig:the EHTCDIFMAP2} show the results of the EHTC-DIFMAP pipeline. 
The original images of the simulated data could not be reproduced. 
In the case of the core-knot model (lower left panel), the knot component is lost. 
Also, in the case of the two-point model (bottom center), the weak north point has disappeared.
Even more surprisingly, in the case of the $40~\mu\rm as$ diameter ring model (bottom right panel), the image quality is degraded and the ring structure is obscured.

Since these images are compact and fit within the BOX setting of the EHTC-DIFMAP pipeline, this phenomenon is not due to a narrow field of view caused by the BOX setting. 
This is probably because the performance of the hybrid mapping process of the EHTC-DIFMAP pipeline strongly depends on the image structure and noise structure of the input data.
%

Note that our simulation this time is not only that shown 
in Figure~\ref{fig:the EHTCDIFMAP2}.
The summary of the simulation is as follows.
\begin{enumerate}
\item A single point source model of 1 Jy with no noise. The image was reproduced from both IMAGR in AIPS and the EHTC-DIFMAP pipeline.
\item Complete noise model. No specific image was detected from the IMAGR in AIPS, but a single point was obtained from the EHTC-DIFMAP pipeline. This is probably due to the fact that the first self-calibration was performed using the image model of a single point source. This is a common phenomenon that occurs when self-calibration is applied to data sets with very low S/N.
\item  Core knot model. Two points model consisting of a point of 1 Jy at the center ($0~\mu\rm as$, $0 ~\mu\rm as$) and a point of 0.25 Jy at ($-25~\mu\rm as$, $+20~\mu\rm as$). The distance between them is about $32~\mu\rm as$ s and they are located along a line in the direction $PA= -38.7\DEG$. 12 simulation data with noise ranging from zero to 640 Jy/weight were generated. IMAGR in AIPS detected the core-knot structure as expected. In the EHTC-DIFMAP pipeline, the knot component disappeared in 11 of the 12 cases.
\item Ring images with diameters around $40~\mu\rm as$. 
The ring width is $2~\mu\rm as$, and the ring diameters are 30, 35, 40, and 45~$\mu\rm as$. The flux density of the image is 1 Jy. No noise was added.  
Using IMAGR in AIPS, the hole in the center of the ring is obscured when the ring diameter is smaller than $30~\mu\rm as$. When the ring diameter is larger, the ring can be recognized. On the other hand, using the EHTC-DIFMAP pipeline, the structure of the ring became ambiguous in all images, like the example shown in Figure~\ref{fig:the EHTCDIFMAP2}.
\item Two-point model (shift in the north-south direction). Place a point of 1Jy at the center ($0~\mu\rm as$, $0 ~\mu\rm as$)  and a point of 0.5~Jy north of it. There are 
11~intervals: 0, 5, 10, 15, 20, 25, 30, 35, 40, 45, and 50 $~\mu\rm as$. Using IMAGR in AIPS, two points cannot be separated and resolved when the spacing is very narrow.  At a separation of 10$~\mu\rm as$, the image is elongated in the north-south direction. If the separation is larger than 25$~\mu\rm as$, the two points are reproduced separately. Using the EHTC-DIFMAP pipeline, when the separation is larger than 45$~\mu\rm as$, the two points are reproduced correctly, but otherwise the north point is not reproduced in the correct position.
\item Two-point model (east-west shift). A point of 1Jy is placed at the center ($0~\mu\rm as$, $0 ~\mu\rm as$) and a point 0.5-Jy is placed at $\delta =42~\mu\rm as$. The R. A. position of the second point is shifted 
every $5~\mu\rm as$ from -20$~\mu\rm as$ to 20$~\mu\rm as$
to create 9 simulation data. 
The positions of these two points are reproduced from IMAGR in AIPS. the EHTC-DIFMAP pipeline reproduces these two points in four cases. In the other four cases, the position of the north point deviated from the correct position by more than 10~$\mu\rm as$. 
\end{enumerate}
Thus, in most cases, the EHTC-DIFMAP pipeline did not reproduce the intrinsic image of the simulation data.
Figure 10 of~\cite{EHTC4} shows that their procedure successfully reproduces images from simulated data (ring, crescent, disk, and double source), but their results are inconsistent with ours (at least for the EHT DIFMAP case).  
In our simulations, noise is either not added at all, or if added, it is thermal noise. IMAGR in AIPS almost reproduces the input model images, while the EHTC-DIFMAP pipeline does not. The difference between the two methods is the hybrid mapping process. The EHTC-DIFMAP pipeline lacks the performance of the general calibration of the data described in Section ~\ref{Sec:vsimulation}.

\subsection{\textbf{The amounts of calibrations performed in the EHTC imaging process}\label{Sec:EHTSNPLT}}
\begin{figure}[H]
\begin{center} 
\epsscale{0.9}
\plottwo{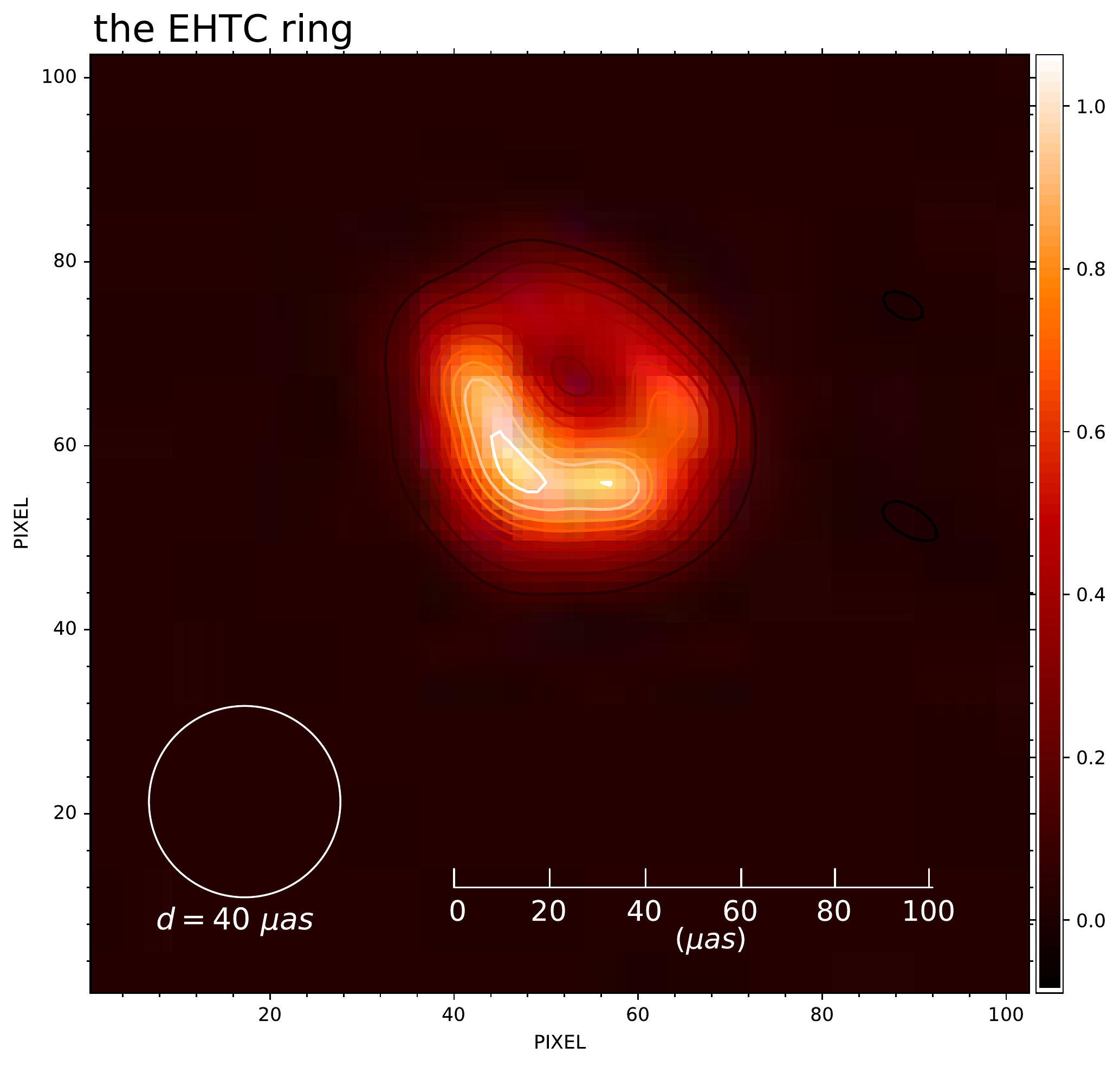}{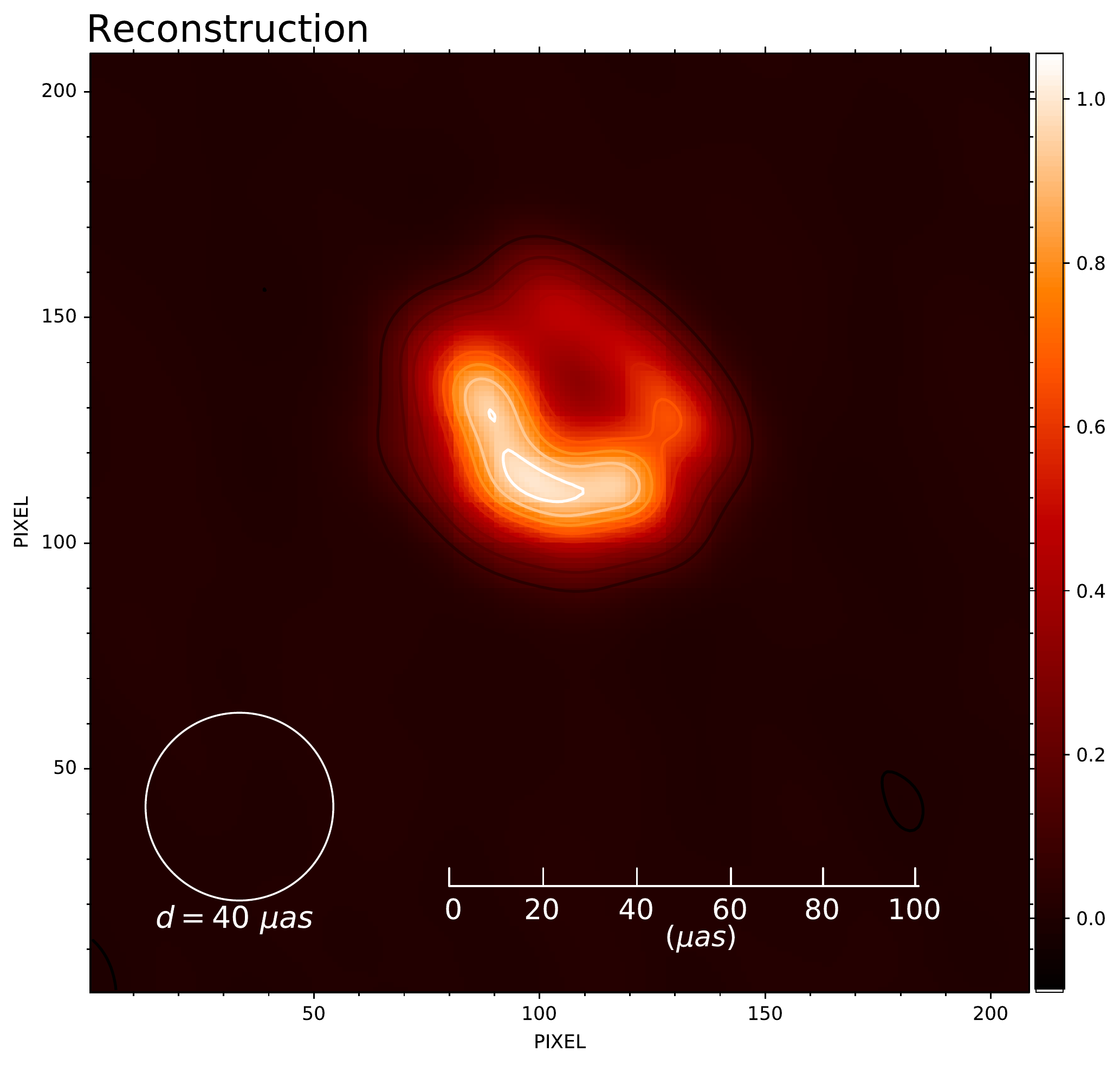}
\end{center} 
\caption{Left panel shows the EHTC ring image of the first day observations. We obtained this image by using their open procedure.
The right panel shows our reconstruction image by applying the self-calibration solutions shown in Figures \ref{Fig:pthe EHTCring} and \ref{Fig:athe EHTCring}. The same BOX setting as that used in the EHTC-DIFMAP pipeline was used.
Also the same restoring beam was applied ($23.1 \times 17.0~\mu \rm as$~with $PA = 44.4\DEG$).
\label{Fig:the EHTCring}
}
\end{figure}

\begin{figure}[H]
\begin{center} 
\includegraphics[width=20cm,height=22cm]{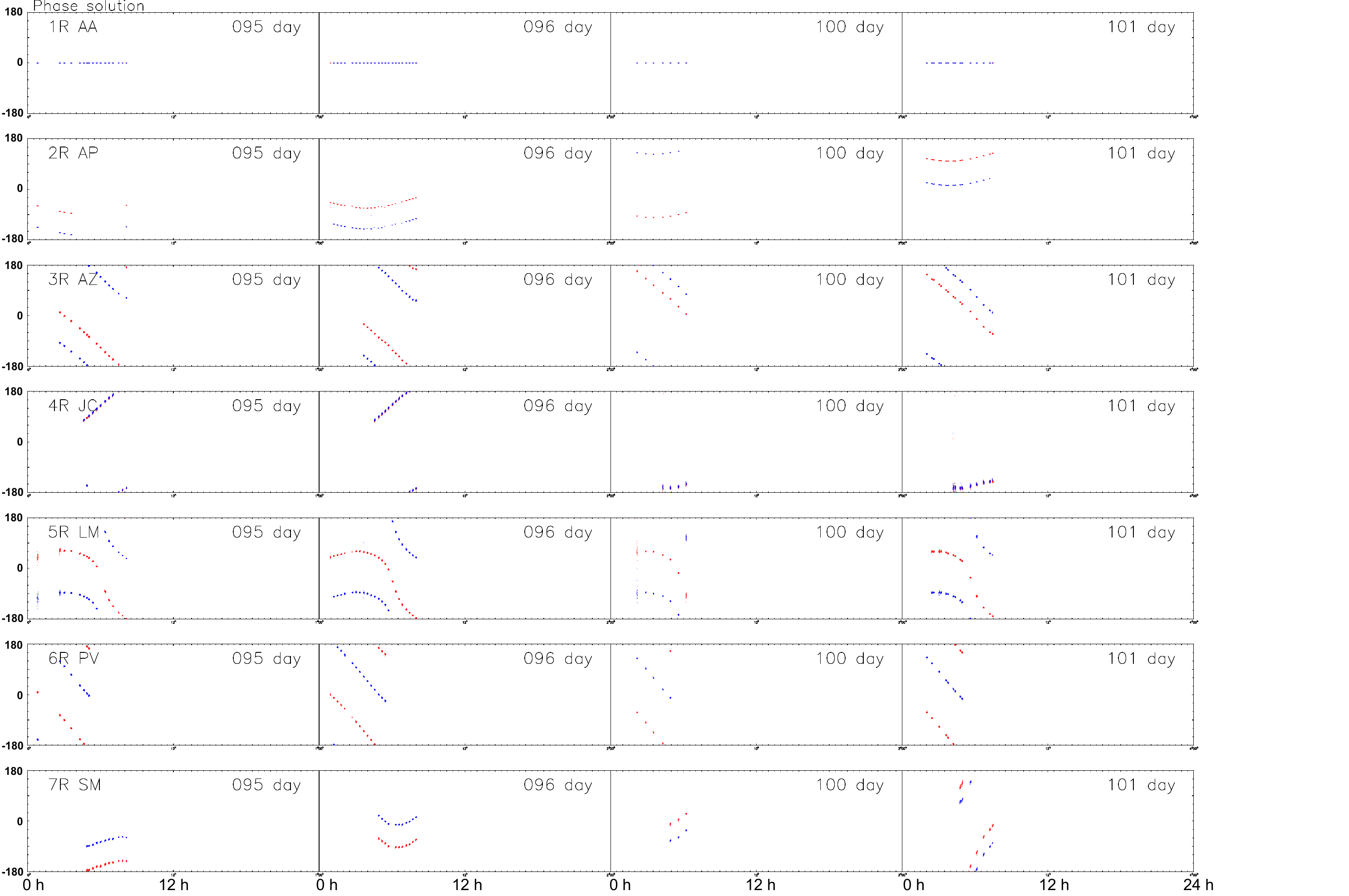}
\end{center} 
\caption{Phase solutions obtained from self-calibration using the EHTC ring image.
The red points show the solutions for IF 1 data, the blue ones show those for IF 2 data.
\label{Fig:pthe EHTCring}}
\end{figure}

\begin{figure}[H]
\begin{center} 
\includegraphics[width=20cm,height=22cm]{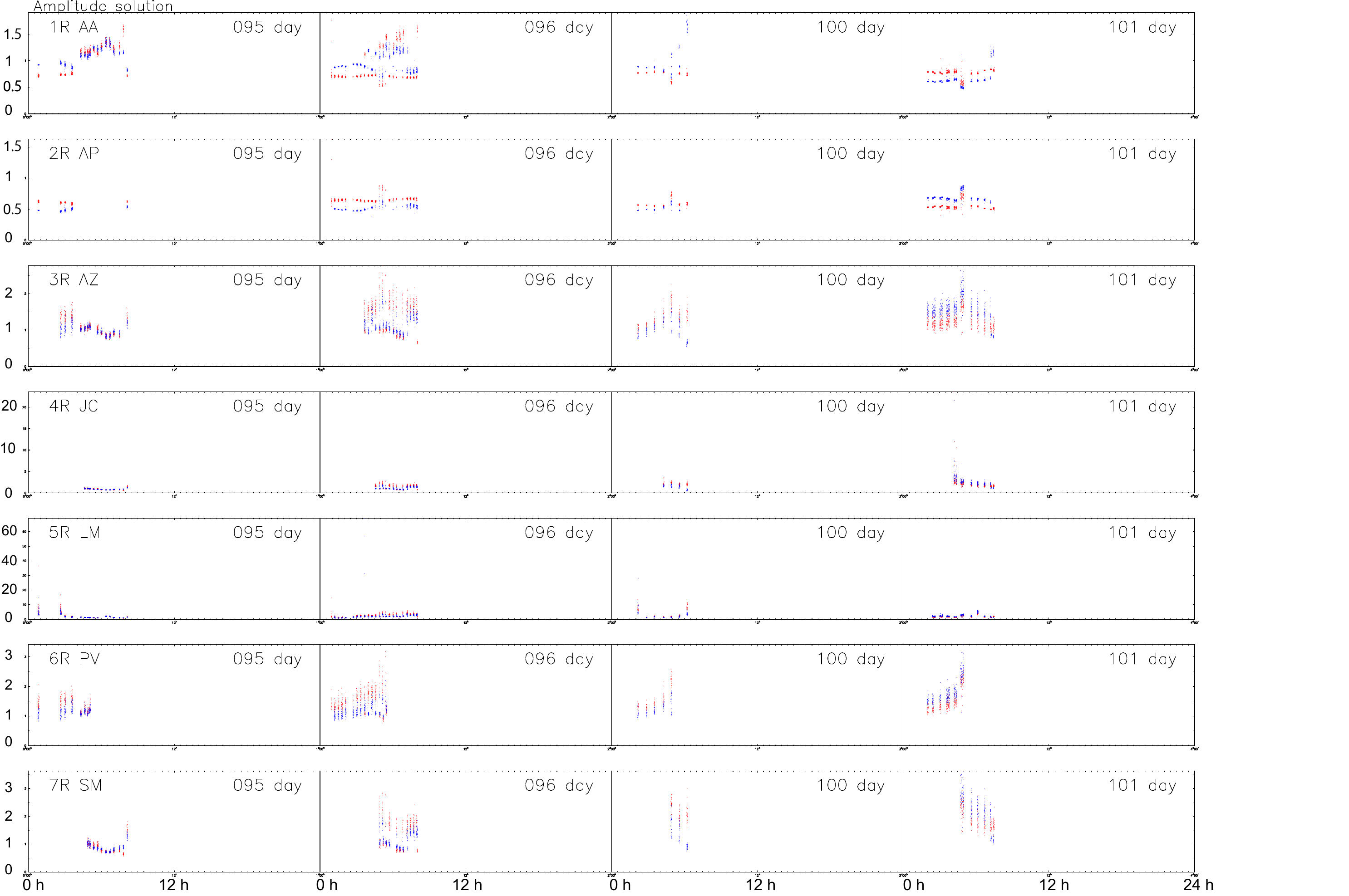}
\end{center} 
\caption{Amplitude solutions obtained from self-calibration using the EHTC ring image.
The red points show the solutions for IF 1 data, the blue ones show those for IF 2 data.
\label{Fig:athe EHTCring}}
\end{figure}

The EHTC papers give the detailed description of the data
calibrations at the pre-stage process, but not much about the
calibrations performed during the imaging process (through the
self-calibration in the hybrid mapping process of the EHTC-DIFMAP imaging
team case). This is an insufficient description of data calibrations because it is rare that only the pre-calibrations are sufficient for obtaining VLBI fine images.
We, therefore, estimate and show the amounts of calibrations performed by the EHTC during their imaging process.
We attempted to reconstruct them in the following procedure.  First,
following the EHTC open procedure, we reproduced the EHTC ring image
of the first observing day (the left panel in Figure
\ref{Fig:the EHTCring}).  Then, using the CLEAN components of the ring
image as the model image, we performed self-calibration by the task
CALIB in AIPS and got solutions for both phase and amplitude.  
The parameters of CALIB are shown in Table~\ref{Tab:final}.
In Figures
\ref{Fig:pthe EHTCring} and \ref{Fig:athe EHTCring}, we show the total amount
of calibrations both of phase and amplitude in the case of the EHTC
ring image model.  Note that we performed CLEAN imaging to verify if
the solutions from 
the self-calibrations are consistent with the original the EHTC ring
image.  
First, we flagged out data points where the amplitude solutions were more than 4 by using the task SNCOR in AIPS. Here we followed the standard VLBI teaching, which states that "discarding bad data will make a better image than keeping them". 
Then, we applied the solutions to the selected data and
obtained an image using the task IMAGR in AIPS for CLEAN. 
The right panel of Figure \ref{Fig:the EHTCring} shows the resultant image whose
quality is similar to that of the original EHTC ring.  Thus, ``our'' solutions of self-calibrations successfully
reproduced the EHTC ring image. 

The phase and amplitude solutions for the EHTC ring image are very
similar to, or worse than our results in Figures \ref{Fig:mD} and \ref{Fig:mE}.
Therefore, if such large amplitudes and their rapid variations found in the self-calibration solutions are negative signs against the resultant image quality, both the results of our final images and the EHTC ring images should be
rejected. This implies that the "calibrated" data released
by the EHTC are not of high quality 
(as is often the case with VLBI archival data). \\
\begin{table}[H]
\begin{center}
\begin{tabular}{lr} \hline
 SOLTYPE &'L1'\\
 SOLMODE &'A\&P' (both phase and amplitude) \\
 REFANT &1 (ALMA)\\
 SOLINT (solution interval) &0.15 (min)\\
 APARM(1)&1\\
 APARM(7) (S/N cut off) &3\\ \hline
\end{tabular}
\end{center}
\caption{Parameters of CALIB for the amplitude and phase self-calibration.} 
\label{Tab:final}
\end{table}
%
%
%
\subsection{Sensitivity of the \texorpdfstring{$\sim 40 ~\mu \rm as$}~~ring to BOX size\label{Sec:staERING}}
In Section \ref{Sec:EHTSNPLT}, we reproduced the ring structure obtained by the EHTC, by using the very narrow box used by the EHTC. 
If the ring image is the definitely correct image, it should only weakly depend on the BOX size.
In fact, we studied the effect of
changing the BOX in Section~\ref{Sec:ROFI}, and found that the
main structure of our final image (features C \& K) does not disappear.
In this section, we investigate how sensitive the ring structure obtained by
the EHTC to the BOX method they used.

We calibrated the data by applying the calibration solutions obtained
from the self-calibration using the EHTC ring as the image model,
which are shown in Figures~\ref{Fig:pthe EHTCring}
and~\ref{Fig:athe EHTCring}.
We performed CLEANs using BOXes of circles with diameters of 
(a) $60~\mu\rm as$, 
(b) $80~\mu\rm as$, 
(c) $100~\mu\rm as$, 
(d) $120~\mu\rm as$, 
(e) $240~\mu\rm as$, and 
(f) $450~\mu\rm as$. 
Their centers are located at the same position as that of the EHTC-DIFMAP case.  
Other parameters of the CLEANs are the same as those used for the right image in Figure~\ref{Fig:the EHTCring}.

\begin{figure}[H]
\begin{center}
\epsscale{1.2}
\plotone{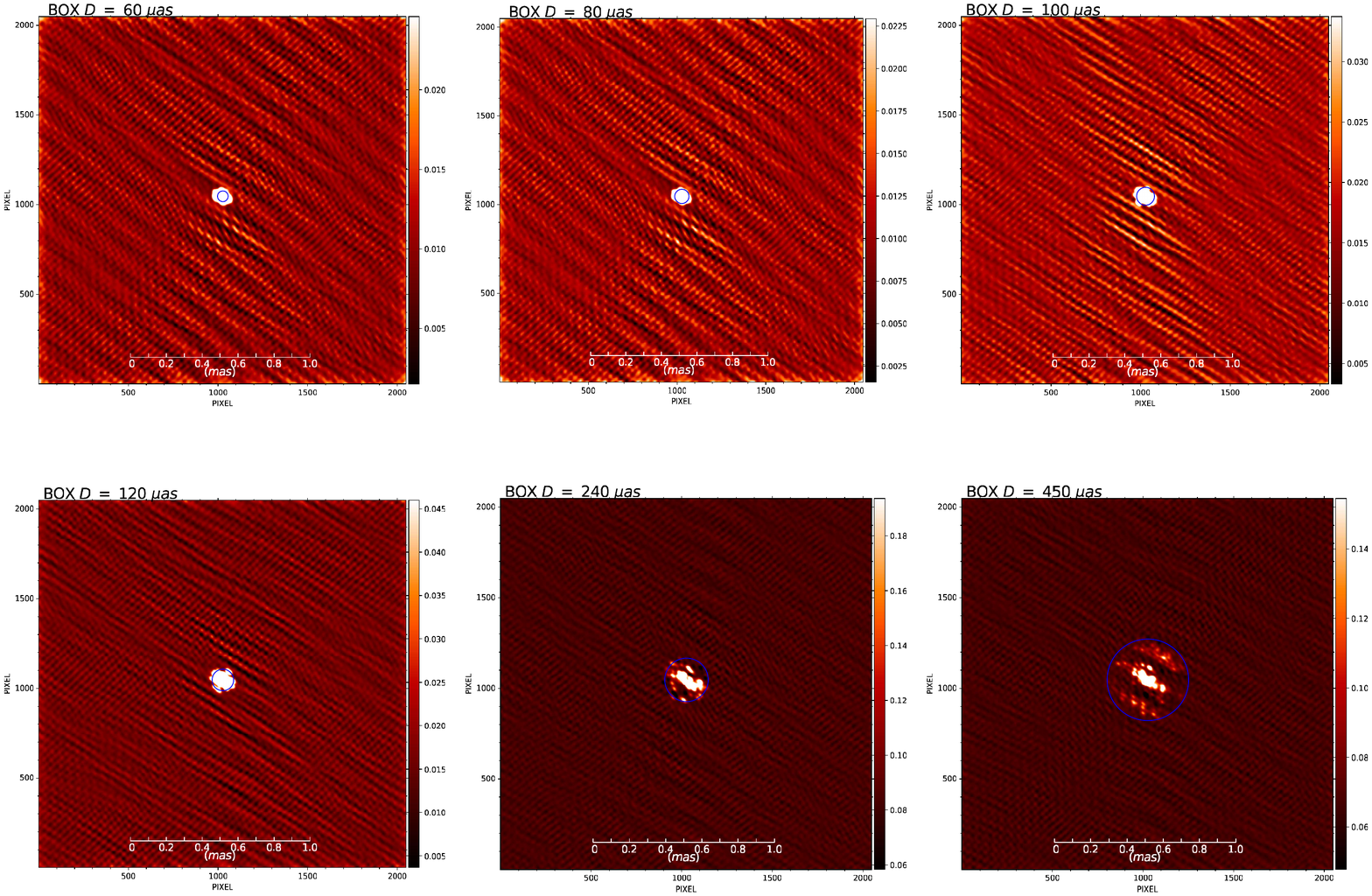}
\end{center}
\caption{Stability of the EHTC ring structure against the BOX sizes: The whole $2~mas$ square images of the CLEAN results by IMAGR (AIPS).
Panels (a) -- (f) show the resultant images in cases of expanding the diameter of the BOX area with $D = 60, 80, 100, 120, 240$, and $450~\mu\rm as$~respectively.
The blue circle shows the respective BOX area.
The center positions of the circles are ($-2~\mu\rm as$, $+22~\mu\rm as$) as followed the BOX the EHTC-DIFMAP team used.
}
\label{fig:add3}
\label{temp1}
\end{figure}

\begin{figure}[H]
\begin{center}
\epsscale{1.2}
\plotone{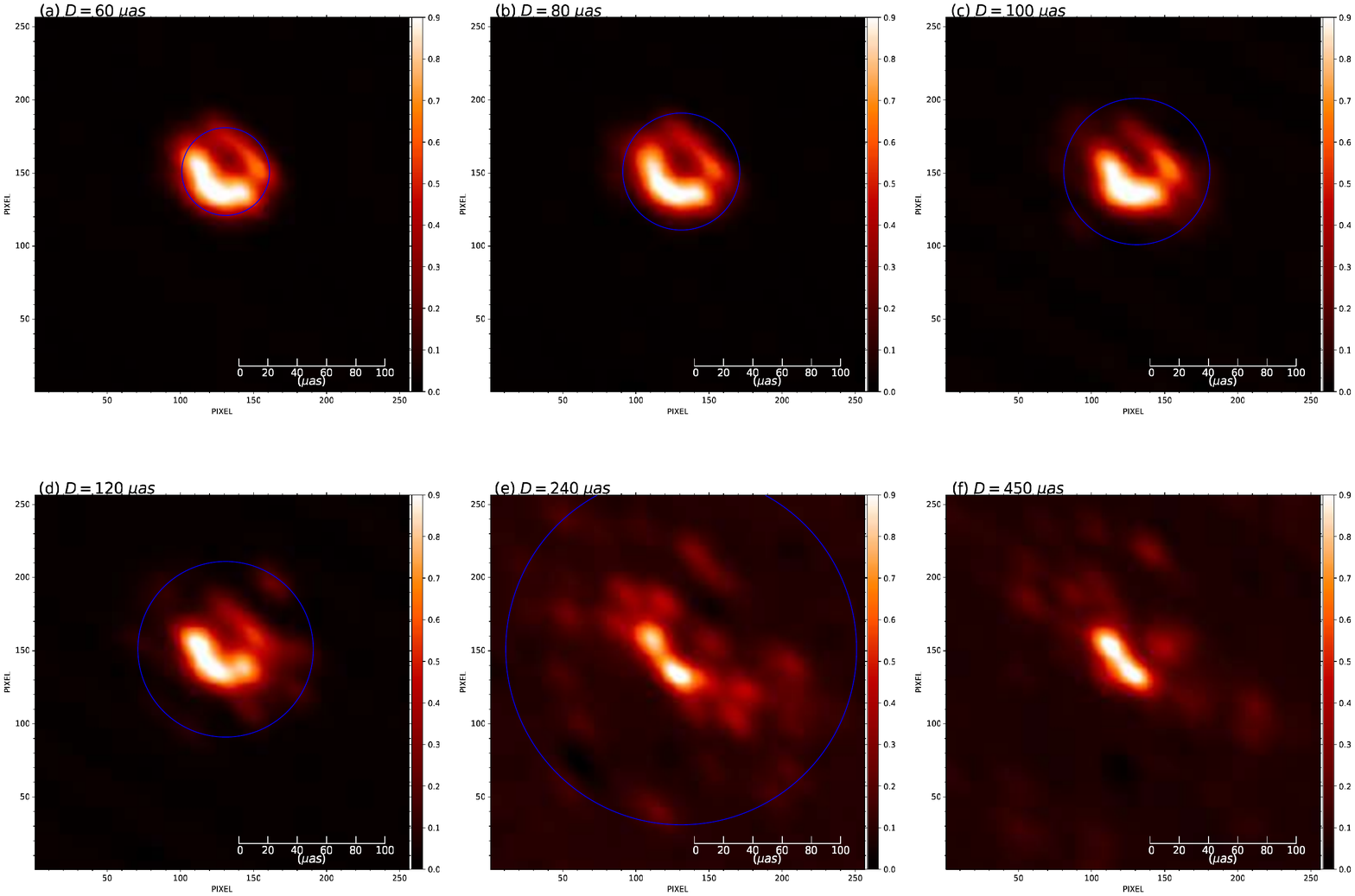}
\end{center}
\caption{Stability of the EHTC ring structure against the BOX sizes: The enlarged views of the central $256~\mu\rm as$ square from the resultant CLEAN images by IMAGR (AIPS).
Panels (a) -- (f) show the resultant images in cases of expanding the diameter of the BOX area with $D = 60, 80, 100, 120, 240$, and $450~\mu\rm as$~respectively. The blue circle shows the respective BOX area. 
The center positions of the circles are ($-2~\mu\rm as$, $+22~\mu\rm as$) as followed the BOX the EHTC-DIFMAP team used.
}
\label{fig:add4}
\label{temp1B}
\end{figure}

The results are shown in Figures~\ref{fig:add3} and~\ref{fig:add4}. 
Figure~\ref{fig:add3} shows views of the entire imaging area of $\rm 2~mas$ square, 
and Figure~\ref{fig:add4} shows enlarged views of the central $256~\mu\rm as$ square. 
The BOXes used are indicated by blue circles.
The ring structure in panel (a) ($D = 60~\mu\rm as$) is the same as
that in the right panel of Figure \ref{Fig:the EHTCring}. Here, we can see
the ring structure similar to that obtained by the EHTC.
The ring extends beyond the BOX area. This is because the obtained CLEAN components are
near the boundary of the BOX, and a $20~\mu\rm as$ restoring beam is convolved to form the CLEAN map.
In the cases of (b) $D = 80~\mu\rm as$ and (c) $D = 100~\mu\rm as$,
the ring is still recognizable. 
However, we can see CLEAN components not on the ring but near the boundary of the BOX when the BOX becomes larger.
In the case of (d) $D = 120~\mu\rm as$, there are several CLEAN
components almost on the boundary of the BOX, though the ring
structure is still recognizable.
In the cases of (e) $D = 240~\mu\rm as$ and (f) $D = 450~\mu\rm as$,
there is no ring. Instead, one elongated bright spot appears.
We also tried larger BOX cases, $D = 900~\mu\rm as$, and $D = 1300~\mu\rm as$, and the resulted image is similar to the case of $D = 450~\mu\rm as$. They show a bright spot at the map center.

If the field of view (FOV) is smaller than the case (d), a ring image will appear, and if FOV is larger than that, the ring image will be destroyed.
In other words, the boundary FOV is around $120~\mu\rm as$.
It is a queer coincidence that the EHTC imaging teams set the FOV to be $128~\mu\rm as$ \citep{EHTC4}, just around the boundary FOV.
 (It is $60 ~\mu\rm as$ in the case of the EHTC-DIFMAP pipeline as already mentioned)
 \footnote{Figure 6 in \cite{EHTC4} shows similar test results to ours, but they only show the results with BOX sizes narrower than $100~\mu\rm as$ in diameter.}.

Thus, we can conclude that the EHTC ring appears only when a very narrow BOX is applied. 
%
%

\subsection{Problems in the EHTC analysis\label{Sec:EHTCL4}}
Here we explain why the EHTC created the image of the ring.
The EHTC described the details of their imaging process in \cite{EHTC4}, in which we found some of the reasons why they created the artifact.
The EHTC conducted various surveys, but their methods were not objective and biased towards their desires from the very beginning of their analysis.
They also failed to perform the basic data checking that VLBI experts always do.
These points are noted below.

\subsubsection{Blind test by the EHTC\label{Sec:btest}}
According to \cite{EHTC4}, "in first stage, four teams, each blind to the others’ work, produced images of M\,87. This stage allowed us to avoid shared human bias and to assess common features among independent reconstructions." They added, "All four images show an asymmetric ring structure. For both RML teams and both CLEAN teams, the ring has a diameter of approximately $40~\mu \rm as$, with brighter emission in the south." 
The EHTC first conducted a blind test and obtained consistent images of each other.

~But their isolated imaging processes were not really blind tests.
The angular size of a black hole shadow can be easily calculated from the distance and mass of the black hole. As every black hole shadow researcher knows, if the mass of the SMBH in M\ 87 is $\sim 6\times 10^9 M_{\odot}$, the expected shadow size is about $40~\mu \rm as$ in diameter.
Based on that calculation, everyone tries to find a ring or arc-shaped image with a diameter of $40~\mu \rm as$. The members of the EHTC imaging team are not exception.

~As shown in the previous Section~\ref{Sec:uvs}, the EHT public data has a sampling bias in that it lacks $\sim 40~\mu \rm as$ spatial Fourier component.
This bias tends to create $\sim 40~\mu \rm as$ scale structures that are not present in the original image.
It is due to this sampling bias effect that the PFS has a $\sim 40~\mu \rm as$ scale structure (Section~\ref{Sec:PSF}).

Not surprisingly, all the EHTC imaging teams, trying to find the black hole shadow with a size of $\sim 40~\mu \rm as$, were able to find the desired illusion in the sampling bias.
The blind imaging test should be conducted by researchers with no prior knowledge of the expected image of the SMBH in M\,87.

\subsubsection{Attention to sampling bias}
It is well known that sampling bias in data can seriously affect data analysis.
For sampling bias, two things should have been done.

~First, PSF (dirty beam) structure should have been checked.
In radio interferometer data analyses, checking the structure of the dirty beam usually provides insight into the effects of insufficient \UVC coverage; 
in a sparse array configuration such as the EHT array, it is more important to check the shape of the dirty beam and utilize it as consideration for false imaging.
\cite{EHTC4} describes the size and shape of the main beam of the dirty beam and the corresponding CLEAN beam, but we do not find any discussion of the overall dirty beam structure in the EHTC papers.
As already shown, the separations between the main beam and the first sidelobes of the dirty beam structure are $\sim 40~\mu \rm as$, which coincidentally is the same as the expected size of the black hole shadow in M\,87. An old-time VLBI expert would have noticed the spacing.

Second, performance of the new imaging method with respect to sampling bias needs to be verified. The CLEAN algorithm, the well-known imaging method in radio interferometry, was originally designed to obtain correct images by deconvolving false structures appearing in dirty beams (the origin of which is the sampling bias). However, in actual use, it is not easy to completely eliminate this effect; most of the relevant researchers know the CLEAN method's performance against sampling bias from its long history of use, and pay attention to its residual structure.
The EHTC used two new imaging methods that are unfamiliar to most radio astronomers, and judging from the case of the CLEAN algorithm, we infer that the performance of the new methods against sampling data bias is not perfect. Its performance needs to be demonstrated.

\subsubsection{The EHTC simulations\label{Sec:vsimulation} }
The EHTC conducted simulations to find the optimal parameters for imaging.
According to \cite{EHTC4}, "In the second stage," the EHTC "reconstructed synthetic data from a large survey of imaging parameters and then compared the results with the corresponding ground truth images. This stage allowed us to select parameters objectively to use when reconstructing images of  M\,87."~However, there are two major problems with this simulation.

First, it seems that the idea of what to be obtained from the simulation was not well considered: the EHTC created synthetic data from the model images and searched for optimal parameters to reproduce the input model images.
But what the EHTC should be obtaining here is a parameter that can correctly calibrate the data and then reconstruct the image correctly.
In their simulations, it seems that no attention was paid to the reproducibility of the input error model.
Optimal parameters are those that need to be satisfied not only for the reproduction of the input image model, but also for the reproduction of the input error model.
As mentioned before, the reproduction of the input image model can be due to sampling bias. In order to reject this possibility, it is useful to check whether the input error model is also reproduced simultaneously.

We tested the performance of the EHTC-DIFMAP with images of similar size to the EHTC image model and found that most of the images were not reproduced correctly (Section~\ref{Sec:pipeline}). The EHTC-DIFMAP parameter settings are of serious concern for the imaging performance of general images other than the EHTC four image models.
We put thermal noise and/or no noise into our simulated visibility. From simple CLEAN by IMAGR in AIPS the input images are reproduced, while from the EHTC-DIFMAP pipeline, they are not reproduced correctly. It means that the EHTC optimal parameters have problems about its calibration performance. We could not find description about the reproduction of the input model errors in the EHTC papers. 
The EHTC conducted vast amount of simulations to search for the optimal imaging parameters, but we worry that what was optimized was only the reproducibility of the input image model, not the input error model.
They should have searched for a parameter that could reproduce not only the input model images, but also input model errors at the same time.

~Second, the large-scale simulation of the EHTC was not so vast as to provide for optimal imaging of M\,87. The number of selected image models and the range of FOV settings are not sufficient for the purpose of the simulation. Therefore, the imaging method using their optimal parameters loses the general performance of producing correct images.
The following are some of the insufficient aspects of their parameter search.

\begin{enumerate}
\item The selected model images (geometric models)\\
~The following four central compact model images have been adopted~\citep{EHTC4}.
~(1) Ring: A delta-function ring with radius $r_{0} = 22~\mu \rm as$, convolved with a circular Gaussian with FWHM of $10 ~\mu \rm as$.
(2) Crescent: crescent composed of a delta-function ring with radius $r_{0} = 22~\mu \rm as$ and a dipolar angular intensity proﬁle.
(3) Disk: A uniform disk with radius $r = 35 ~\mu \rm as$, convolved with a circular Gaussian with FWHM of $10~\mu \rm as$.
(4) Double: Two circular Gaussian components, each with a FWHM of $20~\mu \rm as$. The ﬁrst component is located at the origin and has a total ﬂux density of 0.27~Jy, while the second is located at 
$\Delta_{R.A.} = 30~\mu \rm as$
and $\Delta_{decl.} = -12~\mu \rm as$
and has a total ﬂux density of 0.33~Jy.
These image models were selected based on a plot of visibility amplitude versus \UVC distance (Figure 2 in~\cite{EHTC4}). 
The EHTC claimed that they found two null points in this plot, at $\sim 3.4 \rm~and \sim 8.3~G\lambda$, and identified them as important marks for selecting image models including a ring model.
In the EHTC plot of amplitude versus u-v distance, there appear to be two null points, but it should be noted that the amplitude axis is logarithmic. It means that the null level ($amplitude = 0$) is not shown in the plot area.
It might, naturally, be a good choice to adopt the minimum detection sensitivity as the de facto null level.
The EHTC papers did not describe the detection sensitivity in detail, but according to \cite{EHTC2}, the characteristic sensitivity is 1~mJy for ALMA connected baselines and 10~mJy for other baselines without ALMA.
The lower limits of their amplitudes of the plots are set slightly higher than these characteristic sensitivity levels. 
\footnote{
There are 8 plots of "Amplitude versus u-v distance" for M\,87 in~\cite{RefEHT1-6} papers.
Figures, where the lower limit of the amplitude axis is $3~mJy$, are
~Figure 2 in~\cite{EHTC1}, 
~Figure 10, 13 in~\cite{EHTC3}, 
~Figure 2 in~\cite{EHTC4} , and
~Figure 1 in~\cite{EHTC6} .
Figure, where the lower limit of the amplitude axis is $5~mJy$, is
~Figure 12 in~\cite{EHTC4} .
Figures, where the lower limit of the amplitude axis is $10~mJy$, are
~Figure 13 in~\cite{EHTC5} , and ~Figure 11 in~\cite{EHTC6} .
}
Therefore, the existences of the two null points are not obvious.

~In the old days of VLBI observations, when synthesis imaging was not available, such plots were the only way to get a rough estimate of the structure of the observed source.
One must be aware that there are numerous candidates that satisfy the visibility amplitude plot. Therefore, in principle, it is not possible to identify a unique structure from the plot.
In the EHTC simulations, only four image models were selected, which seems insufficient to search for optimal imaging and calibration parameters.

~Furthermore, as can be seen from the attached error bars, this amplitude measurement contains considerable ambiguity.
Also, as shown in Sections~\ref{Sec:SN138AP} and \ref{Sec:EHTSNPLT}, the amplitude values before the final calibration contain large errors that cannot be neglected.

Especially for interferometers consisting of antennas of various aperture sizes, comparing the amplitudes of individual baselines tends to be confusing.
Therefore, image estimation from such amplitude plots should be done with a large uncertainty in mind.
Since the error is not small, it cannot be easily determined to be a null point, and even if it is, its exact location is not clear.
Therefore, it is not possible to select definitive image models based on a strong recognition of this null point.

~Here we raise another serious concern: of the four image models, two have structures with a diameter of $44~\mu \rm as$. This diameter corresponds to the size of the false structure that can be caused by the sampling bias.
Their optimal imaging parameters are likely to be rather optimized to increase the sampling bias effect.
\item Consideration in large-scale jet model\\
In addition to the compact image model, the EHTC has added an image model for the large scale jet.
The EHTC has adopted three Gaussian shapes as the jet model, based on the results of the previous study of 86~GHz observations~\citep{Kim2018}. The largest Gaussian has a size of $1000\times~600~\mu \rm as$ and the other two have a size of $400\times~200~\mu \rm as$. These mimic the structure surrounding the core and the jet flow with both edges brightened.
However, this choice of jet model does not seem to be a well-considered model for VLBI observations at 230~GHz.

~First, the size of the model jets is too large to simulate observations at 230~GHz. It is not appropriate to use the results of 86~GHz observations; the jets observed at 86~GHz are blurred due to the spatial resolution and do not show the intrinsic size at 230~GHz.
The intrinsic size is probably much smaller.
We can expect to detect a smaller jet component at 230~GHz because of the higher spatial resolution.

~Second, the EHT imaging performance does not have sufficient power to detect large, smoothed structures (such as Gaussian shapes). According to our EHT array performance simulations (Appendix~\ref{Sec:UVsim}), the detectable size is less than $15~\mu \rm as$.
Let us explain what is actually done in this EHTC simulation with the jet model.
This simulation result only says that the large, smooth jet structure does not have any effect on the imaging of the compact structure of the core.
Here, it is very important to note that the simulation is limited to "large, smooth structure".
If the jet consists of compact features as found in our imaging, their simulation cannot rule out the effect of jet features to the image of the core created using small BOX size.
Our imaging results (Section~\ref{Sec:Jet}) 
show that the jet structure is composed of a cluster of small features, which means that the EHTC simulation was performed with an unrealistic assumption.
\item The FOV setting\\
"The choice of FOV must be made with care, as incorrect restrictions can result in false image structure. (page~4 in~\cite{EHTC4})"  --- which is really a very important point. The EHTC set the FOV to less than $128~\mu \rm as$. As we show in Section~\ref{Sec:results}, the actual image distribution far exceeded the FOV size.
~Also, as shown in Section ~\ref {Sec:staERING}, there is a critical FOV at around $FOV = 128~\mu \rm as$.
If the FOV is larger than that, the image of the $\sim 40~\mu \rm as$ ring is destroyed and if the FOV is smaller, the image of the $\sim 40~\mu \rm as$ ring is enhanced.
 The optimal FOV selected by the EHTC coincides with precisely the critical FOV required to create $\sim 40~\mu \rm as$ ring artifact.

M\,87 is a very important target to study the jet structure of AGN. Therefore, previous VLBI observations have mainly pursued the investigation of its jet structure.
In 2017, the EHT array achieved unprecedented sensitivity, and thus we naturally expected to detect weak emission belonging to jet structures that were not detected in previous EHT observations (without ALMA). A wide FOV had to be set to take the jet detection into account.\\
\end{enumerate}
 The EHTC selected their optimal parameters based only on the reproducibility of the image models in simulations. Nothing was written about the reproducibility of the input error models, probably the reproducibility of the input error models was not taken into account.
 Both reproducibilities should be examined when selecting the optimal parameters for data calibration and imaging. This would help to detect false imaging due to the effects of data sampling bias.
 
\subsection{The reasons why the EHTC obtained the artifact image\label{Sec:reasons}}
We here summarize the reasons why the EHTC obtained the artifact image unintentionally.
The fundamental reason is the sampling bias in EHT's \UVC coverage for  M\,87, i.e., the missing  $\sim 40~\mu \rm as$ spatial Fourier components (Section~\ref{Sec:uvs}). 
The sampling bias tends to create an artificial structure of $\sim 40~\mu \rm as$ in imaging.
We confirmed that the artificial $\sim 40~\mu \rm as$ sized structures appear in the dirty beams (PSF) (Section~\ref{Sec:PSF}, Appendix~\ref{Sec:otherDbeams}) and in our CLEAN imaging results (Appendix~\ref{Sec:MOD}).

The CLEAN algorithm was originally designed to remove false structures such as sidelobes that appear in PSF, but it does not always succeed in removing them.
As well, other imaging methods of the EHTC, if no special countermeasures are taken against \UVC sampling bias, will produce artificial structures in the imaging results with sizes of
$\sim 40~\mu \rm as$ of the sampling bias origin.
Also, \UVC sampling bias effects are independent of the need for data calibration. Therefore, even with imaging methods based on observables that are free from systematic errors such as closure phase, we have no choice but to be aware of the effects.
The sampling bias effect does not produce serious errors if careful data analysis is performed. However, the EHTC expected a ring-like structure of $\sim 40~\mu \rm as$ in size from the beginning, and thus performed image synthesis with a very narrow FOV ($128~\mu \rm as$ in size).
A large FOV setting will not enhance the artificial ring image of $\sim 40~\mu \rm as$ size. On the other hand, the very small FOV set by the EHTC enhances the sampling bias effect to create the $\sim 40~\mu \rm as$ ring structure (Section~\ref{Sec:staERING}).
There were large amplitude errors in the observed data possibly originating from atmospheric variations (Section~\ref{Sec:SN138AP}). 
This could have played a role in confusing the initial analysis of the EHTC.
\section{Conclusion\label{Sec:CR}}
Using the public data released by the EHTC,
we obtained images of the central region of M\,87 using the improved calibration obtained using the standard hybrid mapping method. 
As a result, we found the following.
\begin{enumerate}
\item
The core of M\,87 is resolved into a core-knot structure, instead of a ring.
Three features C, K and W are seen. While feature C is definitely a core and feature K is a knot, feature W is not so easy to explain.
Feature W may be a lensing image due to the strong gravity of SMBH. Another possibility is that there are two SMBH system and that feature W is another SMBH in the core of M\,87.
Assuming that W is another knot, the three features could be initial jet structures with an opening angle of $\sim~70\DEG$ at a distance of about 10~R$_{s}$ from the core.
\item
The 230~GHz image has a jet structure consistent with the previous lower-frequency observations. 
It has brightened edges from the core to at a few mas points. The intensity is decreased along the jet axis much rapidly as compared with lower observations. 
\item The $\sim 40~\mu\rm as$ ring that the EHTC reported is an artifact due to the effect of data sampling bias and the very narrow FOV setting that enhances the bias effect. The \UVC coverage of the EHT for M\,87 observations lacks the $\sim 40~\mu\rm as $ spatial Fourier components that produce artifact structures of $\sim 40~\mu\rm as $ size.
\end{enumerate}
\acknowledgments
We would like to dedicate this work to an emeritus professor Yoshiharu Eriguchi of the Tokyo University and Dr. Jose Ishituka in Peru, both passed away in 2020, also to a professor Naruhisa Takato of SUBARU observatory in Hawaii suddenly passed away in 2021.
We thank Jun Fukue, Hiroyuki Takahashi, Hisaaki Shinkai, Hiroshi Imai, Hiroshi Sudou, Masao Saito, Hiromi Saida, Yasusada Nanbu,and Masaaki Takahashi for their kind comments. Takahiro Tsutsumi and Shunya Takekawa kindly provided new tools for checking the data.

\vspace{5mm}

\facilities{EHT}

\software{AIPS~\citep{Greisen2003},
DIFMAP~\citep{Shepherd1997}
}


\appendix
\section{Imaging simulations for investigating the \UVC performance of the EHT array\label{Sec:UVsim}}
~In VLBI imaging with sparse \UVC coverage, sometimes the reconstructed images do not show the real brightness distribution correctly. 
In particular, when the size of the source is significantly larger than that of the synthesized beam, the source image disappears, namely ''resolved-out'' occurs. 
Here, we investigated such an effect for the EHT \UVC coverage for the M\,87 observations.

\begin{figure}[H]
\begin{center} 
\plottwo{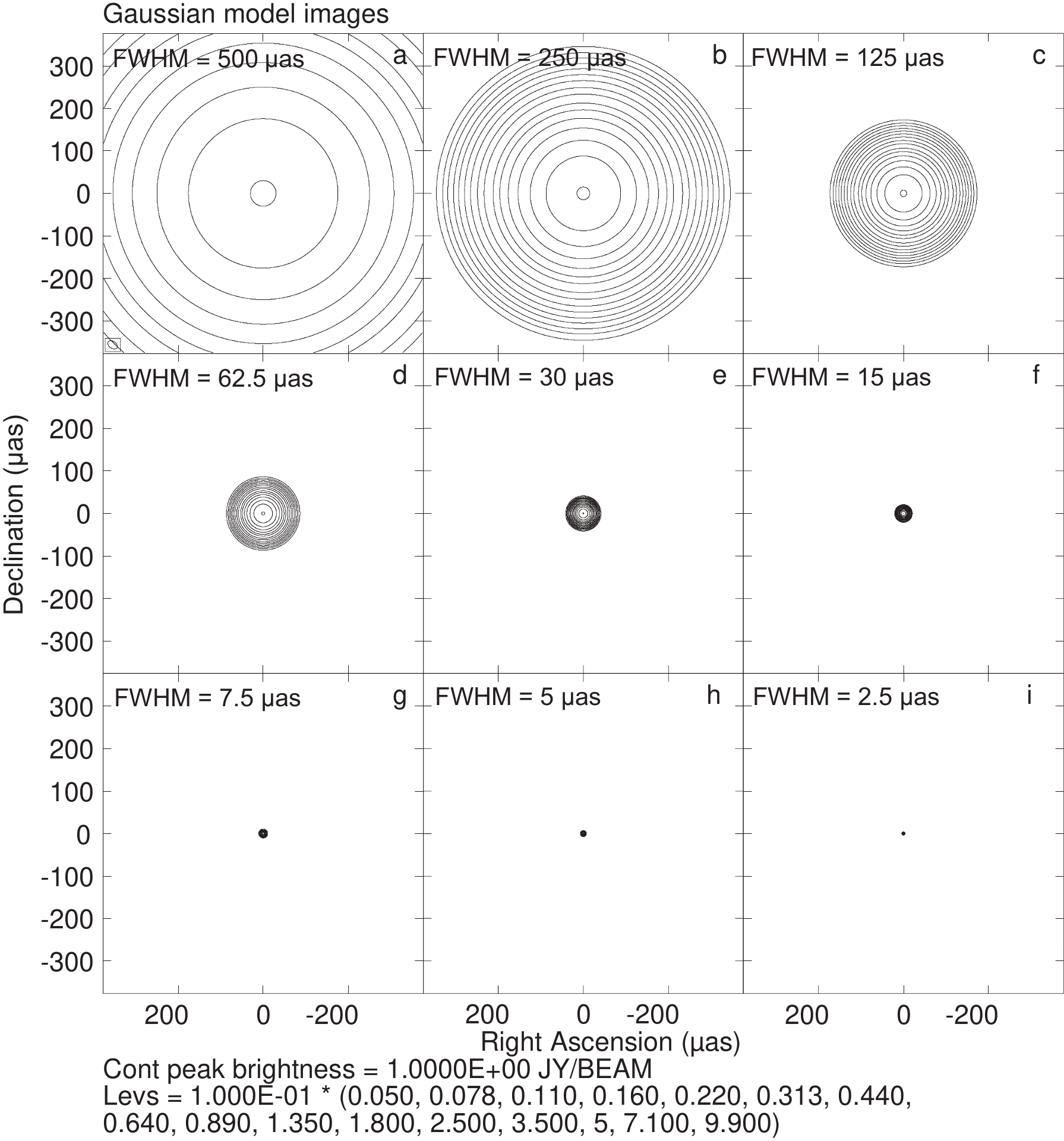}{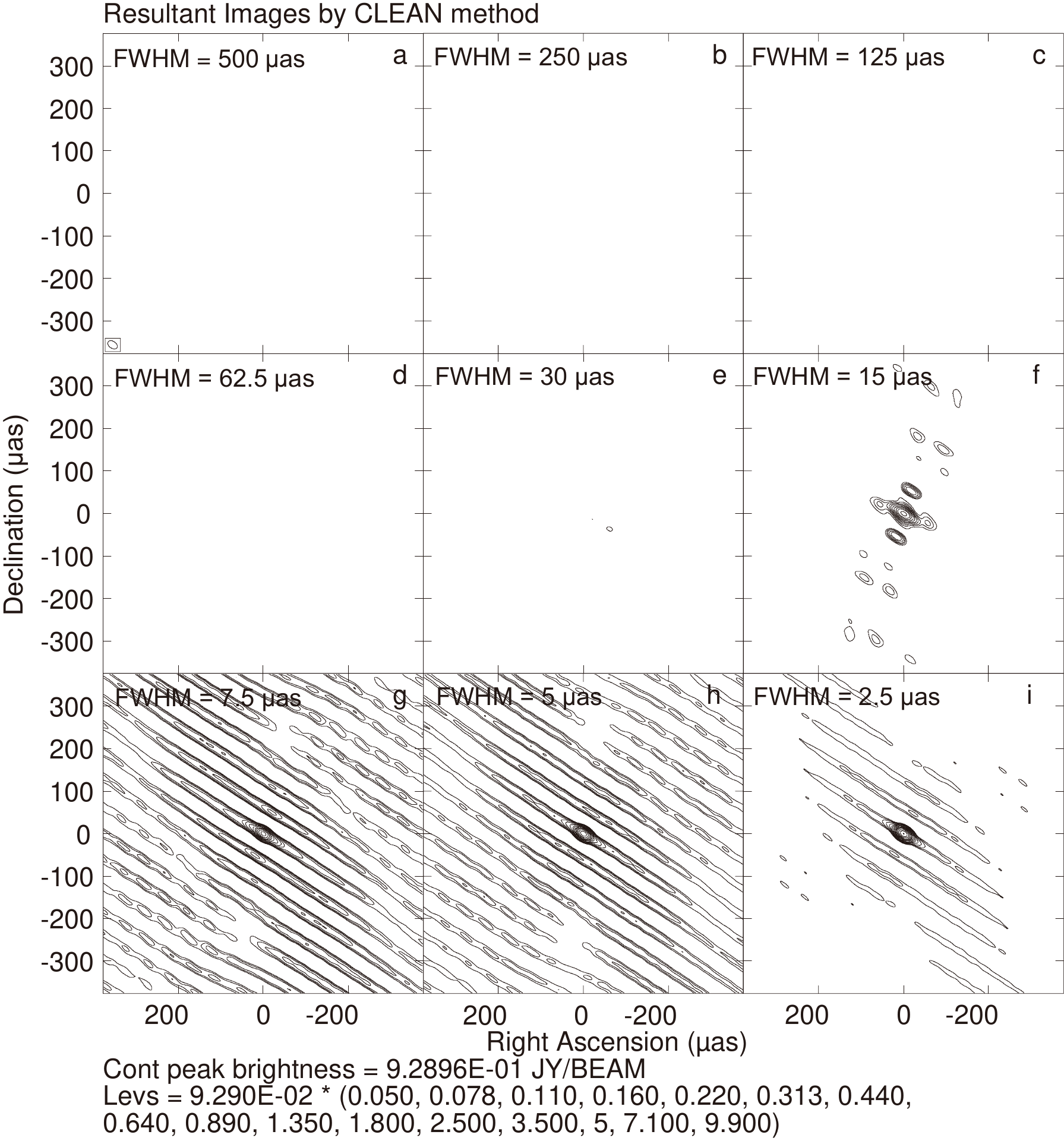}
\end{center} 
\caption{\UVC simulations using circular Gaussian shapes with various FWHMs.
Larger structures than that of $FWHM = 30~\mu\rm as$ cannot be detected by the EHT \UVC coverage.
The contour levels are 0.5, 0.78, 1.1, 1.6, 2.2, 3.13, 4.4, 6.4, 8.9, 13.5, 18, 25, 35, 50, 71, and 99~$\%$ of the peak. }
\label{Fig:mF}
\end{figure}

Figure~\ref{Fig:mF} shows the simulation results for the reconstruction of Gaussian brightness distributions using the EHT array configuration. 
We used model images of circular Gaussian shapes with a total flux density of 1~Jy. The sizes are~500,~250,~125,~62.5,~30,~15,~7.5,~5, and~2.5$~\mu\rm as $ in FWHM  (the left panels in Figure~\ref{Fig:mF}). 
Here, we assume that the EHT array has infinite sensitivity \textbf{and that the data calibration is perfect}, that is, we did not add any thermal and systematic noise error into the simulated visibility data.

By CLEAN, we obtained the corresponding reconstructed images, as shown in the right panels of Figure~\ref{Fig:mF}.
When the model image sizes are larger ($FWHM =~500,~250,~125,~62.5~\mu\rm as $), they are completely resolved-out and cannot be detected even with infinite sensitivity.
In the case of $FWHM = 30~\mu\rm as$, it was almost decomposed but was detected as a small dark spot. 
In other words, the brightness distribution that can be modeled as a Gaussian shape larger than $30~\mu\rm as$ in FWHM cannot be detected by the EHT array configuration.
Although the smaller sizes with $FWHM = 15~\mu\rm as$ or less can be detected, they are not correctly reproduced as the original Gaussian shapes.
In addition, the reconstructed images show streaks running in the NE-SW direction at lower levels. 
This also reflects the shape of the dirty beam (PSF) produced by the space \UVC coverage.

 In conclusion, the detectable structure of M\,87 by the EHT array configuration is limited to less than $30~\mu\rm as$ in size, which corresponds to 2.43$\times 10^{-3}$ pc (500~au) or 4.2~$R_{\rm S}$. Any larger extended structure cannot be detected. We must consider the possibility of the existences of more extended structures.

\section{Dirty beam shapes with different weightings\label{Sec:otherDbeams}}
\begin{figure}[H]
\begin{center} 
\plotone{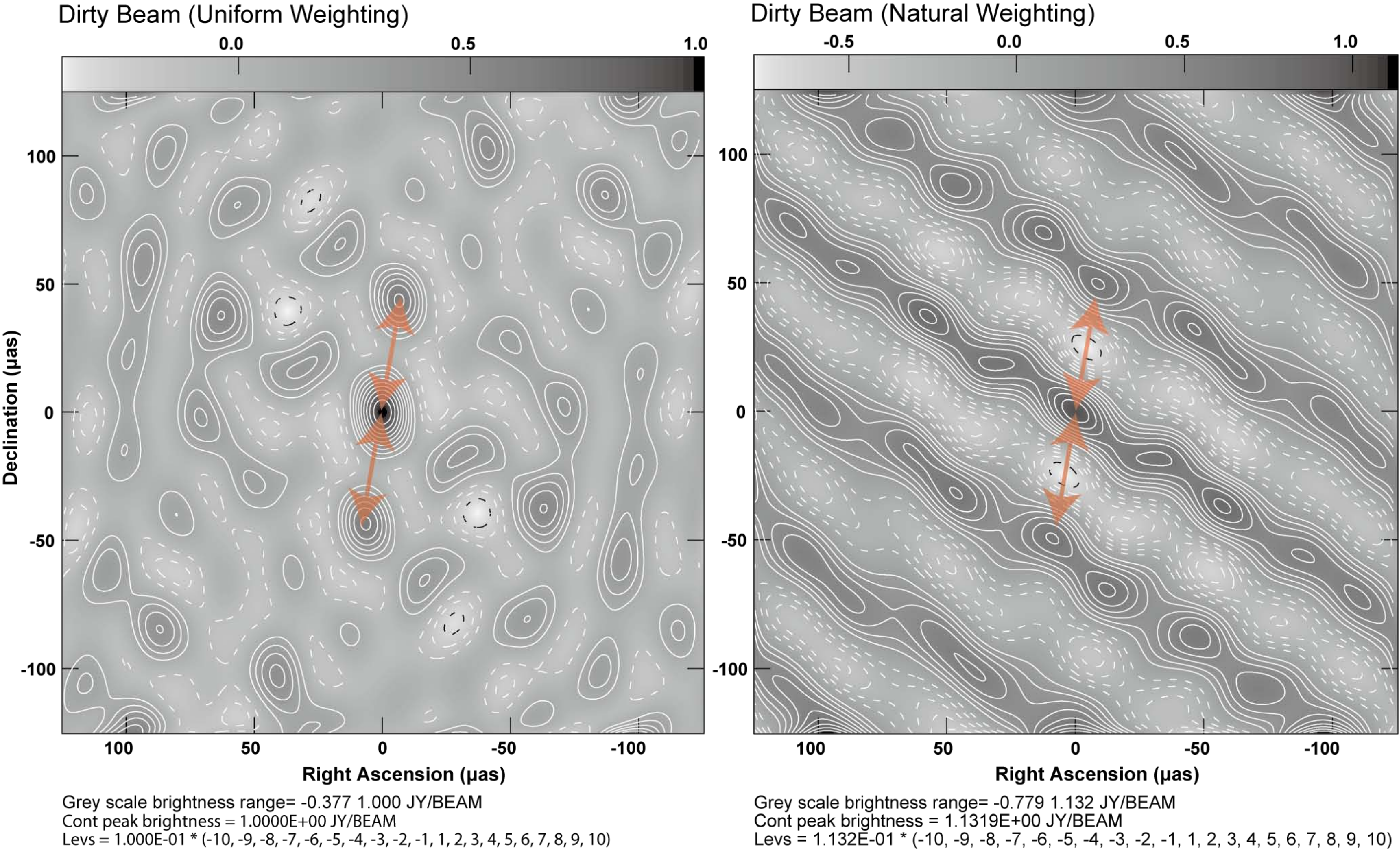}
\end{center} 
\caption{
Other different dirty beam shapes (PSFs) of the EHT during the first day of observations of M\,87.
The left panel shows the dirty beam shape by uniform weighting.
The right panel shows that by natural weighting.
Contour lines are shown in 10~\% increments from -100 to 100~\% of the peak. Positive levels are indicated by white lines and negative levels by dotted lines.
The gray scale shows the range from the minimum to the maximum intensity of the dirty beam image.
Both cases show that the first sidelobe levels reach more than 60~\% of the height of the main beam.
As shown in Figure~\ref{Fig:uvd}, red arrows with a length of $45~\mu \rm as$ are placed at the same position and angle.
\label{Fig:otherDBs}
}
\end{figure}

In synthesis imaging using a radio interferometer, the shape of the PSF (dirty beam) can be changed by adjusting the weighting of \UVC points.
The normal PSF shape when EHT observes M\,87 is shown in Fig~\ref{Fig:uvd}, but here we show the famous PSF shapes often used in synthetic images, i.e.,~the uniform weighting case and the natural weighting case.
These PSFs were obtained by changing the parameter ROBUST in AIPS 
\footnote{
Refer to the NRAO site, \url{http://www.aips.nrao.edu/cgi-bin/ZXHLP2.PL?ROBUST}
}.\\
Figure~\ref{Fig:otherDBs} shows them. 
We set $ROBUST = -5$ for uniform weighting, 
and $ROBUST = +5$ for natural weighting.
As a whole, the shapes of each are not similar. In particular, the shape of the main beam is different.
\footnote{
The sidelobe level is extremely high even when changing the \UVC weights.
Both cases show that the first sidelobe levels reach more than 60~\% of the height of the main beam.
}
However, they both have the same spacing substructure of about $40-50~\mu \rm as$ in common. In particular, the spacing between the main beam and the first sidelobe and the direction of the line connecting them, i.e., the position angle, are almost the same. This means that even if the weights of the \UVC points are adjusted, the effect of spatial Fourier components that were not sampled during the observation cannot be essentially removed.

\section{Structure by the sampling bias in our CLEAN images\label{Sec:MOD}}
The EHT data sampling is biased in that the spatial Fourier components of \FTMUAS~are missing. Because of this, the imaging results tend to contain false \FTMUAS~-sized structures that do not actually exist.
 In our images, which were obtained by independently analyzing the same EHT data, there is no \FTMUAS~size ring, but the effect of data sampling bias could still appear. Here, we investigate the spatial distribution of the CLEAN components in one of the CLEAN maps obtained from the hybrid mapping process and found the effect of data sampling bias.
 The CLEAN algorithm, assuming that the observed structure consists of a large number of point sources, finds the points from the visibility data. Accordingly,  the result of CLEAN is obtained as a set of CLEAN components (position and intensity).
 We examined the spacing distribution of all pairs of CLEAN components.
 Raw CLEAN components can be in the same position as other components. In such cases, we used CCMRG, an AIPS task, to merge the raw CLEAN components in the same position into one.
The number of CLEAN components after merging is $N_{cc} = 2579$, and the number of pairs of them is $N_{pair} = 3234331$.

\begin{figure}[H]
\begin{center} 
\plotone{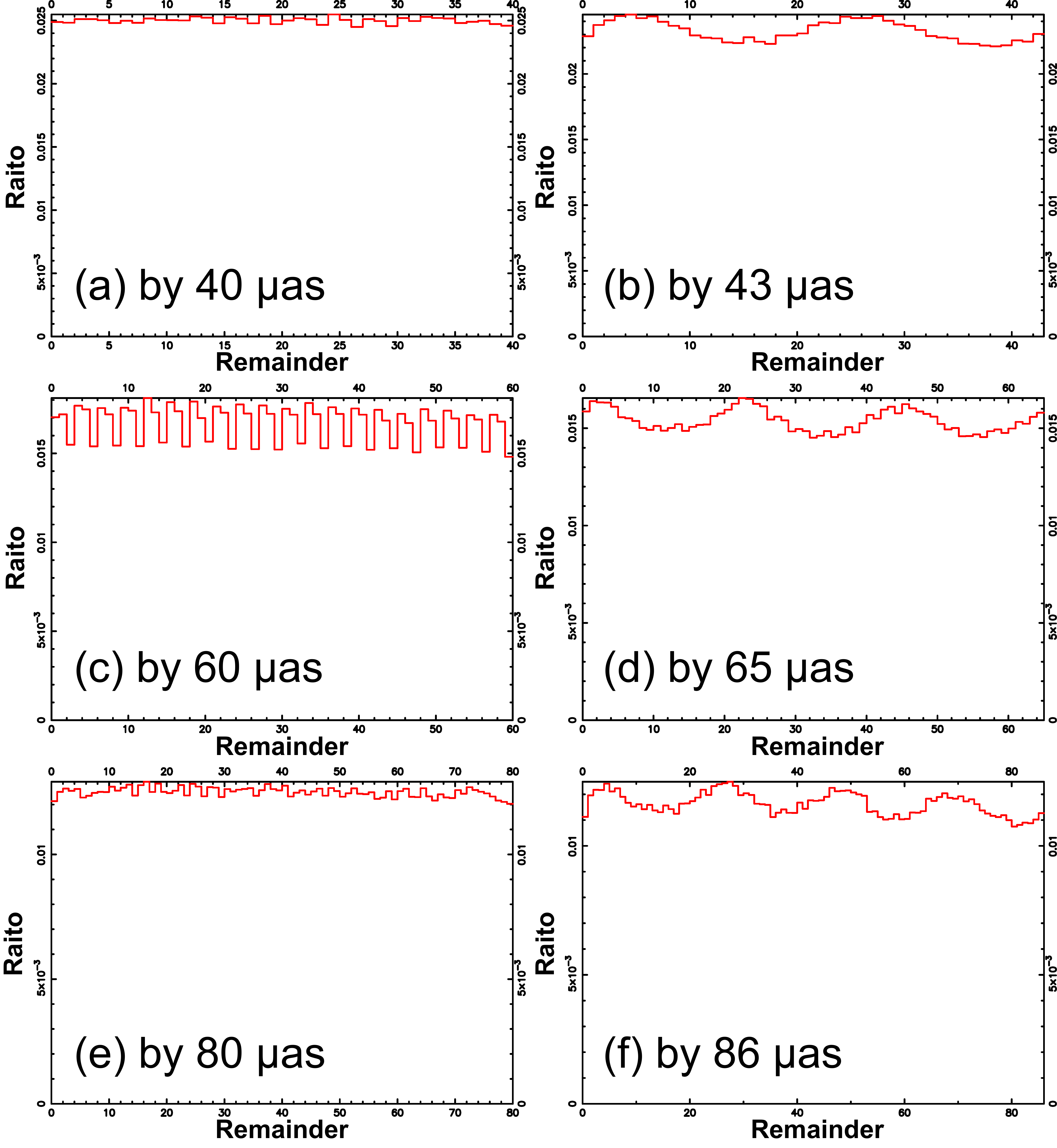}
\end{center} 
\caption{
Histogram of intervals for CLEAN component pairs. The horizontal axis shows the remainder of the interval after dividing by $d~\mu \rm as$. 
The vertical axis shows the ratio of pairs in a bin.
The width of the bin is $1~\mu \rm as$. 
Panel (a) shows $d = 40~\mu \rm as$,
(b) $d = 43~\mu \rm as$,
(c) $d = 60~\mu \rm as$,
(d) $d = 65~\mu \rm as$,
(e) $d = 80~\mu \rm as$, and 
(f) $d = 86~\mu \rm as$.
\label{Fig:MOD1}
}
\end{figure}

\begin{figure}[H]
\begin{center} 
\plotone{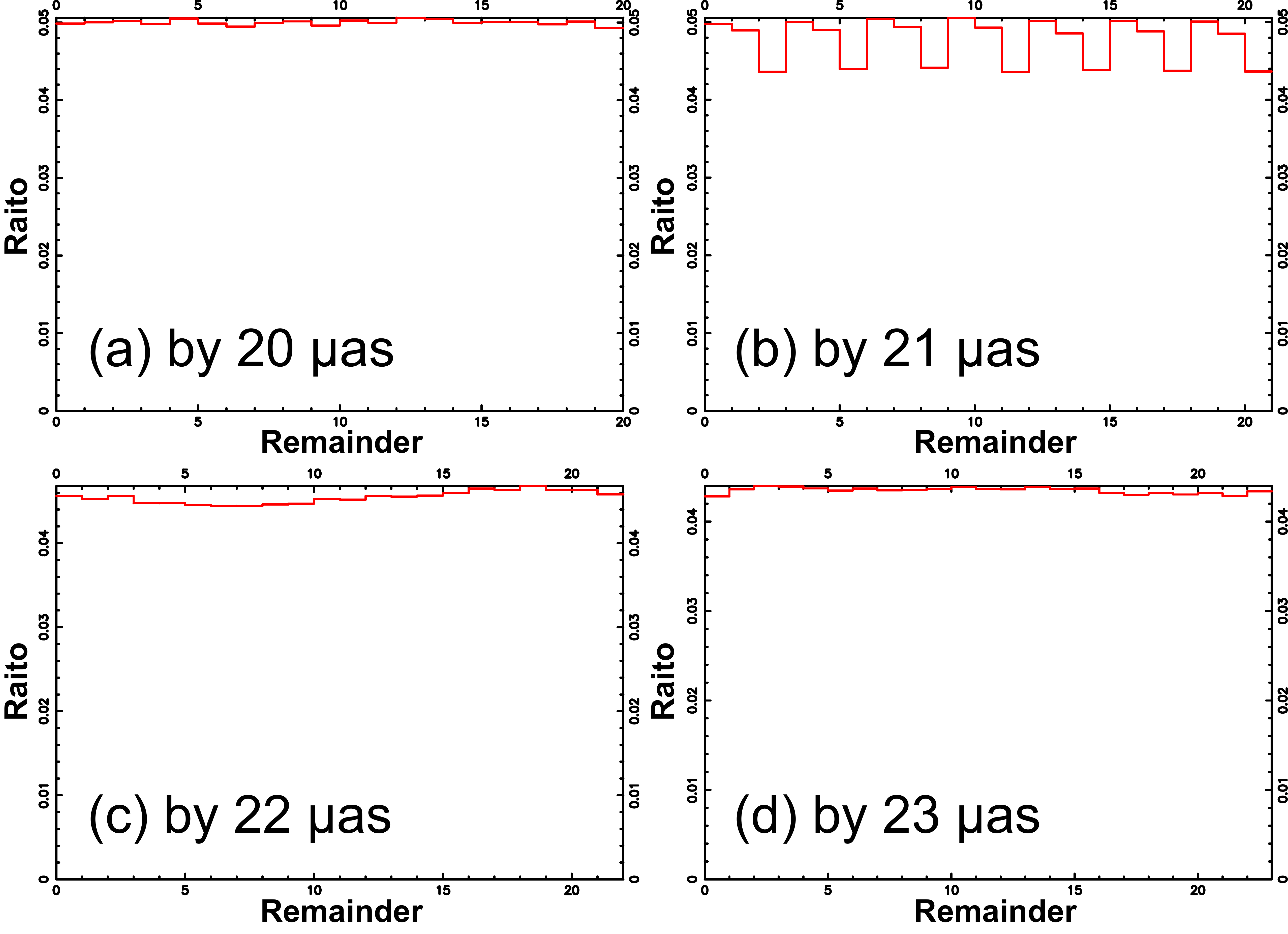}
\end{center} 
\caption{
Histogram of intervals for CLEAN component pairs. The horizontal axis shows the remainder of the interval after dividing by $d~\mu \rm as$. 
The vertical axis shows the ratio of pairs in a bin.
The width of the bin is $1~\mu \rm as$. 
Panel (a) shows $d = 20~\mu \rm as$,
(b) $d = 21~\mu \rm as$,
(c) $d = 22~\mu \rm as$, and
(d) $d = 23~\mu \rm as$.
\label{Fig:MOD2}
}
\end{figure}

Figures~\ref{Fig:MOD1} and~\ref{Fig:MOD2} show the distribution of spacings for all pairs of CLEAN components.
The distance between a CLEAN component pair is divided by a certain interval d ($\mu \rm as$), and the distribution of the remainders are shown in the figures.
If there is no characteristic structure at interval d, the remainder distribution will be flat. The panels (a), (c), and (e) of Figure~\ref{Fig:MOD1} show such cases. That is, the three cases of $d = 40,~60,~80 ~\mu \rm as$ show a flat distribution.
Fine saw-tooth shapes appear in panel (c). This is due to the effect of the grid spacing during CLEAN imaging. In this case, $d = 60~\mu \rm as$ is divisible by the grid spacing ($1.5~\mu \rm as$).

However, at $d = 43~\mu \rm as$, an arch shape appears and two peaks are formed. The spacing between these peaks is about $20~\mu \rm as$. 
A similar arch-like distribution is seen from $d = 42.5~\mu \rm as$ to $d = 44.25~\mu \rm as$.
For $d = 65~(= 43 \times 1.5)~\mu \rm as$, three arches can be seen. 
For $d = 86~(= 43~\times 2)~\mu \rm as$, four arches appear.
Interestingly, the spacing between all the peaks is $20 ~\mu \rm as$.
The position of these peaks is an integer multiple of $43~\mu \rm as$. 
This can be explained as the sum of the sampling bias effect and an integer multiple of the spatial resolution of the EHT array ($20 ~\mu \rm as$).
The first peak is not located at $residual = 0$ and shows a slight shift for $d = 43$ and~$86~\mu \rm as$. 
The reason for this is difficult to explain.

Note that the basis of the characteristic interval is not $d \sim 20~\mu \rm as$, but $d\sim43~\mu \rm as$. 
 If the basis of the characteristic interval is 20 µas, then a single arche type shape should appear in the distribution for $d \sim 20~\mu \rm as$.
 As shown in Figure~\ref{Fig:MOD2}, the distributions for $d \sim 20~\mu \rm as$ have flat structures.

Not only the EHTC image, but also our CLEAN image is affected by the sampling bias of the data. As a result, we can see that the structure of the interval d 
$\sim 43~\mu \rm as$ ($d = 42.5\sim~44.25~\mu \rm as$) is emphasized.

Such a trend was not observed in our final images. This is probably because the final image is a composite of several CLEAN images (the first two days' images were created from six CLEAN images and the last two days' images were created from nine CLEAN images. )



\end{document}